\documentclass{article}

\usepackage{graphicx}
\usepackage{amsmath,amssymb,amsfonts}
\usepackage{amsthm}
\usepackage{mathtools}
\usepackage{booktabs}
\usepackage{authblk}
\usepackage{makecell}
\usepackage{subcaption}
\usepackage{bm}
\usepackage{algorithm}
\usepackage{algpseudocode}
\usepackage{hyperref}

\graphicspath{{imgs/}{imgs/app/}}

\newcommand{\vrho}{\boldsymbol{\rho}}
\newcommand{\A}{\mathcal{A}}

\theoremstyle{plain}

\theoremstyle{definition}
\newtheorem{definition}{Definition}
\theoremstyle{remark}
\newtheorem{remark}{Remark}

\begin{document}

\title{Bayesian genome-wide clustering and variable selection of transcriptomic data via rank-based mixtures}
\date{}

\author[1]{Emilie Eliseussen\thanks{Emilie Eliseussen, Haakon Muggerud, and Luca Coraggio contributed equally to this work.}}
\author[2]{Haakon Muggerud}
\author[3]{Luca Coraggio\thanks{Corresponding author: luca.coraggio@unina.it}}
\author[2,4]{Ida Scheel}
\author[5]{Thomas Fleischer}
\author[1,4]{Valeria Vitelli\thanks{Corresponding author: valeria.vitelli@medisin.uio.no}}

\affil[1]{Oslo Centre for Biostatistics and Epidemiology, Department of Biostatistics, University of Oslo, Oslo, Norway}
\affil[2]{Department of Mathematics, University of Oslo, Oslo, Norway}
\affil[3]{Department of Economics and Statistics, University of Naples Federico II and CSEF, Napoli, Italy}
\affil[4]{The Norwegian Centre for Knowledge-driven Machine Learning, Integreat, Norway}
\affil[5]{Department of Cancer Genetics, Oslo University Hospital, Oslo, Norway}

\maketitle

\begin{abstract}
With the increasing availability of ranking data, there has been a growing demand for appropriate unsupervised rank-based inferential frameworks capable of handling high-dimensional datasets and providing uncertainty quantification for all estimates.
Rank-based methods have also seen a growing popularity in -omics pipelines, as ranking continuous measurements provides a robust means of handling non-normally distributed data.
The Bayesian Mallows model (BMM) has emerged as a promising choice because of its adaptability to various types of ranking data and its flexible framework, integrating cluster-wise rank aggregation with inference at the individual level.
However, the scalability of BMM to ultra-high-dimensional settings, such as -omics analyses, has remained limited.
The present paper addresses this issue by introducing the first rank-based model generalizing BMM to jointly handle clustering and variable selection, namely the lower-dimensional Bayesian Mallows Model Mixture (lowBM3).
The proposed method provides a novel Bayesian framework that simultaneously handles heterogeneity in the sample, unsupervised parameter estimation, and model selection in a scalable manner for ultra-high-dimensional data.
Additionally, a companion postprocessing framework is introduced to provide posterior summaries of the discrete posterior distributions of both the consensus ranking and the variable selector.
Simulation studies are performed to assess the performance of the method.
The usefulness of the method is also shown in an application to signature discovery for cancer genomics, where RNA-seq bulk gene expression data obtained from breast cancer patients are clustered genome-wide.
\end{abstract}


\section{Introduction}\label{sec:intro}

In recent years, the availability of data in the form of rankings has increased significantly due to the wider availability of technology to collect, share, and explore information, particularly on online platforms.
We here refer to the umbrella term ``ranking data'' for data arising from preference assessment types of experiments, where a set of assessors (judges, users) expresses preference statements on a set of items that take the form of (potentially partial) rankings. In this context, rank-based methods are the natural choice as they can typically handle more sparse types of preference information, such as ratings, pairwise preferences, and clicks.
However, rank-based methods have also seen a growing popularity in the statistical analysis of -omics data (see e.g., ~\cite{deng2014,asfari2014, badgley2015, shang2020}), as transforming the vector of continuous measurements of each patient into a ranking provides a robust mean for handling non-normally distributed data, which are common in this context.
Such ranking transformation is obviously suited for all -omics layers that can successfully be understood in terms of the respective ordering of the observed features (e.g., a gene being over- or under-expressed as compared to another gene), such as for RNA-seq, proteomics, and metabolomics. Despite partially losing the informative content of the data, rankings can be helpful, for instance, in the presence of confounding batch effects in the original measurements.

Several probabilistic methods suitable for ranking data have been studied, among which the Plackett-Luce~\cite{luce1959, plackett1975} and the Mallows~\cite{mallows1957} model are probably the most solid and popular choices.
Each probabilistic model brings its own advantages: the Plackett-Luce model offers the interpretability of a continuous model parameter, whereas the Mallows model affords greater flexibility through the choice of the distance function governing the model.
The Plackett-Luce model explores much less the complexity of the permutation space, due to its restrictive score-based data generation scheme, and thus for instance provides less diverse recommendations~\cite{liu2019}, while the Mallows model has the disadvantage of being computationally intensive and does not currently scale to large applications.
In -omics settings, when rankings arise from continuous measurements, the Mallows model is particularly suitable as, being a model of the exponential family, it naturally models the dispersion of ranked observations around a consensus ranking.
In this paper, we will build on frameworks that use the Mallows model to address the current limitations in scalability, and propose their use in the context of ultra-high dimensional heterogeneous ranking data.

The Mallows model expresses the probability of an observed ranking (i.e., a permutation) as a function of its distance to the consensus ranking, the model location parameter, and the main objective of inference is to identify the consensus ranking that most accurately represents the observed rankings.
Such probabilistic model has been successfully applied in various fields, including sports, voting schemes, and recommender systems, and for diverse tasks, such as rank aggregation, clustering, and individual-level prediction. Prediction of individuals' latent rankings from partial preference information is for example impossible when using a Plackett-Luce model. However, one challenge associated with the Mallows model is the computationally demanding task of calculating the partition function, which is used for normalizing the model.
Since the model partition function is defined on the entire permutation space, research has been mostly limited to using distance functions that permit analytical computation of the partition function, such as the Kendall distance, which has been studied extensively~\cite{fligner_verducci1986,meila_chen2010, meila_bao2010,lu_boutilier2014}.
The first probabilistic framework capable of performing inference in the Mallows model using any right-invariant distance has been the Bayesian Mallows model (BMM)~\cite{vitelli2018}, an extensive inferential endeavor that can perform both clustering and individual-level preference prediction, both for complete and sparse preference data, allowing proper uncertainty quantification of all unknowns.
However, BMM does not scale to high- and ultra-high-dimensional settings, such as -omics settings. Applications of BMM to -omics data have been pioneered by an extension of BMM called the lower-dimensional Bayesian Mallows Model (lowBMM)~\cite{eliseussen2022}, which handles cases in which the pool of items is considerably large, by embedding a variable selection step into the modeling framework.
Indeed, lowBMM assumes that only a minor portion of the items observed in the data can truly be modeled by a rank-based model, while the rest is simply noise. Thus, lowBMM includes this set of so-called ``relevant items'' among the model parameters, together with the other usual Mallows parameters, all estimated with uncertainty.
An application to transcriptomic data is shown in~\cite{eliseussen2022} to demonstrate the usefulness of lowBMM in the context of -omics applications, 
particularly in transcriptomics, as it is biologically reasonable to assume that only a subset of genes is expressed in a specific biological process, and thus relevant for prediction or subtyping~\cite{alexandrov2013, hoadley2014}.
However, lowBMM cannot handle the quite standard setting in which the data show additional heterogeneity due to the presence of several groups in the sample, as is often the case in cancer patient data, due to the presence of distinct disease subtypes~\cite{sorlie2001}.

In this paper, we address this issue by proposing a generalization of lowBMM to accommodate the existence of multiple clusters within a dataset, and we name the new method lower-dimensional Bayesian Mallows Model Mixture (lowBM3).
Extending lowBMM to handle clustering yields a robust clustering method for ultra-high-dimensional heterogeneous ranking data,
which combines the clustering techniques used in BMM with the theoretical framework and assumptions of lowBMM. Specifically, the variable selection parameter in lowBM3 is cluster-specific, so that the method allows for enhanced cluster interpretability (as each cluster is characterized by a reduced set of features), and for greater model flexibility than usual Bayesian clustering and variable selection approaches (see the excellent review \cite{fop2018variable}), where a common set of features is selected for all clusters.

Bayesian clustering methods tailored to ultra-high-dimensional data are scarce, possibly due to the computational complexities associated with high-dimensional variable selection.
Bayesian methods also require coherent prior specifications and tailored postprocessing of the posterior to interpret parameters, which adds to their complexity~\cite{wade2018, balocchi2025understanding}.
Two notable methods used in the context of -omics data applications are iClusterBayes~\cite{iClusterBayes} and the Multi-Omics Factor Analysis model (MOFA)~\cite{MOFA}.
iClusterBayes is a Bayesian latent variable model facilitating joint modeling of diverse -omics data types for clustering.
Similarly, MOFA offers a low-dimensional embedding solution for multi-omics data using variational inference, and is grounded in the Bayesian Group Factor Analysis framework, introducing sparsity through sparse PCA.
However, a considerable limitation of both iClusterBayes and MOFA is the lack of a variable selection and regularization step embedded \emph{within} the clustering scheme as proposed in this paper.
In fact, an integrated approach ensures both a more accurate clustering, providing a better identification of the variables that are critical for characterizing each group, and a more parsimonious model, by more effectively selecting those variables that are heterogeneous across groups.
This line of research was initiated by the seminal work of Tadesse et al.~\cite{tadesse2005}, who for the first time introduced an infinite mixture model that incorporates a variable selection step, where inference is performed using reversible-jump MCMC\@.
More recent noteworthy methods include several works by Witten and coauthors, e.g. \cite{witten2010,gao2022,chen2023}, all recognizing the importance of incorporating variable selection within the clustering framework. Bayesian bi-clustering approaches~\cite{moran2021spike} have been recently proposed, even though these are based on a very specific factor-modeling assumption (i.e., that the data are well-described by a sum of rank-1 matrices), and often assume Gaussianity, thus limiting the model applicability to diverse -omics data types.
A notable (non-Bayesian) method influenced by~\cite{chen2023} is detailed in~\cite{li2023}, where a sparse negative binomial mixture model is employed to perform joint variable selection and clustering with an application to RNA-seq count data.
The method showcases superior performance compared to alternative approaches, excelling in clustering accuracy, gene selection, and biological interpretation via pathway enrichment analysis; the authors rightly emphasize the crucial role of employing more sophisticated models in enhancing the accuracy of results when dealing with RNA-seq data. However, the method does not provide uncertainty quantification for the model unknowns and, moreover, is based on specific distributional choices that only apply to RNA-seq data. We claim that lowBM3 can obtain state-of-the-art results by simply ranking the RNA-seq measurements and analyzing the resulting rankings. Additionally, lowBM3 gains the needed robustness when dealing with heterogeneous cohorts, batch effects, and incomparable data scales.
To our knowledge, there exists no model for rankings that jointly performs both item selection and clustering, and that scales to such data dimensions.

The paper is structured as follows. Section~\ref{sec:method} introduces lowBM3 for clutsering and items selection in the Mallows model, after describing the necessary methodological preliminaries connected to lowBMM. The Bayesian inferential setup used to estimate the model parameters is outlined in Section~\ref{sec:mcmc}.
To evaluate the performance of the method, we conduct several simulation studies, which are described in Section~\ref{sec:simulations}.
In Section~\ref{sec:application}, we present an application of the modeling approach to genome-wide RNA-seq gene expression data from breast cancer patients.
Finally, we summarize our findings and discuss potential future extensions of the model in Section~\ref{sec:discussion}.

\section{Lower-dimensional Bayesian Mallows Model Mixture}\label{sec:method}
We briefly present the lower-dimensional Bayesian Mallows Model (lowBMM) for variable selection in ultra-high-dimensions (as introduced in~\cite{eliseussen2022}) in Section~\ref{subsec:lowbmm_mod}, and then delineate our novel generalization to handle joint clustering and variable selection in Section~\ref{subsec:lowbmm_mix_mod}.

\subsection{Previous work: the Lower-dimensional Bayesian Mallows Model}\label{subsec:lowbmm_mod}
The lower-dimensional Bayesian Mallows Model (lowBMM)~\cite{eliseussen2022} is a generalization of the Bayesian Mallows Model (BMM)~\cite{vitelli2018,sorensen2019} for situations where the number of items is large or \emph{ultra-large}.
In these situations, it is unrealistic to assume that a full consensus ranking on the space of all items can be estimated from the data, while
a better modeling strategy would assume that only a subset of the items are ranked in the data, the so-called \emph{relevant} items, while those not worth ranking are simply noisy. lowBMM relies on the assumption that the data generating process for relevant items follows a Mallows distribution, whereas noise items are uniformly distributed.
The purpose of the model is to infer with uncertainty the set $\A^{*}$ that contains all relevant items, and the consensus ranking of such relevant items, $\rho$, i.e.,\ the ranking of the items in $\A^{*}$.

Consider a finite set of $n$ items denoted as $\A=\{A_1,A_2,\ldots,A_n\}$, and assume that all $n$ items are ranked according to a specified feature by $N$ assessors, thus providing complete rankings $\mathbf{R}_j= \{\mathbf{R}_{1j},\mathbf{R}_{2j},\ldots, \mathbf{R}_{nj}\}$, $j=1,\ldots,N$.
Note that the assumption of observing complete rankings can be straightforwardly relaxed using the same strategies employed in the original Bayes Mallows proposal~\cite{vitelli2018}.
For the formulation of lowBMM, let us define $\A^{*}=\{A_{i_1},\ldots, A_{i_{n^{*}}}\}$ as an $n^{*}-$dimensional subset of the original set of items $\A$, with $n^{*} \ll n$ and $\A^{*} \subset \A.$ The data likelihood under lowBMM is defined via a Mallows model only on the lower-dimensional permutation space of dimension $n^{*}$, and it takes the form:
\begin{align*}
  P(\mathbf{R}_1, \ldots, \mathbf{R}_N | \bm{\rho}, \A^{*}) = \frac{1}{Z_{n^{*}}(\alpha)} \exp \left\{ -\frac{\alpha}{n^{*}} \sum_{j=1}^N d_{\A^{*}}(\mathbf{R}_j, \bm{\rho})\right\} \prod_{j=1}^N \left\{ U_{\mathcal{P}_{n-n^{*}}} (\mathbf{R}_j|_{\A\setminus\A^{*}}) 1_{\mathcal{P}_n}(\mathbf{R}_j) \right\}
\end{align*}
where $\alpha>0$ is a positive scale parameter, $\bm{\rho} \in \mathcal{P}_{n^{*}}$ is the latent consensus ranking defined on the low-dimensional permutation space, $Z_{n^{*}}(\alpha)$ is the model partition function, $1_S(\cdot)$ is the indicator function of the set $S$, and finally $U_S(\cdot)$ is the uniform distribution over the domain $S$. Let $d_n(\cdot, \cdot):\mathcal{P}_n\times\mathcal{P}_n\rightarrow[0,+\infty)$ be a right-invariant\footnote{A right-invariant metric $d(\cdot,\cdot)$ is such that, for any $r_1, r_2 \in \mathcal{P}_n$, it holds: $d(r_1,r_2)=d(r_1 r_2^{-1}, \bm{1}_n)$, $\bm{1}_n= \{1,2,\ldots,n\}$ \cite{diaconis1988}} distance function between two rankings, and let $d_{\A^{*}}(\mathbf{R}_j, \bm{\rho}) \coloneqq d_{n^{*}}(\mathbf{R}_j|_{\A^{*}},\bm{\rho})$ be the restriction of the same distance function to the $n^{*}-$dimensional set of items included in $\A^{*}$.

Several possibilities exist for choosing the distance function $d_n(\cdot,\cdot)$, such as the footrule, the Spearman, the Kendall, and the Cayley distances~\cite{diaconis1988}.
In this paper, we choose to use the footrule distance, the equivalent of an $\ell^1$ measure between rankings and defined as $d_n(\bm{R}, \bm{\rho})=\sum_{i=1}^n |R_i - \rho_i|$, because of its robustness, and to be coherent with the popular choice of $\ell^1$ measures for the penalization terms of high-dimensional regression models, especially in the analysis of molecular data.
Furthermore, we keep the scale parameter $\alpha$ fixed as in~\cite{eliseussen2022}, but this restriction is also easy to relax.
The vector $\mathbf{R}_j|_{\A^{*}}$ is the restriction of $\mathbf{R}_j$ to the set $\A^{*}$, thus defining a partial ordering from which an $n^{*}$-dimensional permutation can be derived.
For instance, if the items $(A_1, A_2, A_3)$ are ranked $( 9, 5, 12)$ in the $n$-dimensional permutation space, then their ranking in the $n^{*}$-dimensional permutation space, with $n^{*}=3$, is $(2, 1, 3)$.
This emphasizes the importance of the items' relative rankings in determining their inclusion in the set $\A^{*}$: consistency in the relative rankings among the items is prioritized over an individual item's original ranking.
This also explains why lowBMM can estimate any relative ranking and not only select a top-ranking lower-dimensional solution.

The parameter $n^{*}$ is not estimated, but rather selected based on prior information about the number of items believed to be relevant for the data at hand.
A good rule of thumb is to set this parameter as large as possible among plausible values for the current application, as this increases the chances of assigning all truly relevant items a large posterior probability of inclusion in the set $\A^{*}$.
As empirically proved in~\cite{eliseussen2022} under several simulation scenarios, setting $n^*$ larger than the true value is not necessarily a problem, as it is also possible to tune this parameter post-hoc by examining the marginal posterior distribution of $\bm{\rho}$: the marginal posterior probabilities assigned to irrelevant items' ranks converge to a uniform, 
which helps in assessing the true $n^{*}$.

Since the inferential approach is Bayesian, we must decide priors for all parameters.
We set a uniform prior on $\bm{\rho}$, however restricted to the $n^{*}-$dimensional space of permutations of elements of $\A^{*}$: $\pi(\bm{\rho}|\A^{*})=\frac{1}{n^{*}!
  }1_{\mathcal{P}_{n^{*}}}(\bm{\rho})$.
We also set a uniform prior for $\mathcal{A^{*}}$ over $\mathcal{C}$, $\pi(\A^{*})=\frac{1}{|\mathcal{C}|} 1_{\mathcal{C}}(\mathcal{A^{*}})$, where we define $\mathcal{C}$ as the collection of all $\binom{n}{n^{*}}$ possible sets of items of dimension $n^{*}$ chosen from a set of dimension $n$.
The posterior distribution for lowBMM can then be written as
\begin{align}
  \label{eq:post}
  P(\bm{\rho}, \A^{*} | \mathbf{R}_1, \ldots, \mathbf{R}_N)  \propto & \ \pi(\A^{*})\pi(\bm{\rho}|\A^{*}) \frac{1}{Z_{n^{*}}(\alpha)} \exp\left\{ -\frac{\alpha}{n^{*}} \sum_{j=1}^N d_{\A^{*}}(\mathbf{R}_j, \bm{\rho})\right\} \cdot \nonumber \\
  & \cdot \prod_{j=1}^N \left\{ U_{\mathcal{P}_{n-n^{*}}}(R_j |_{\A\setminus\A^{*}}) 1_{\mathcal{P}_n}(\mathbf{R}_j) \right\},
\end{align}
which, by removing the terms not depending on any of the model parameters and by assuming the data are full rankings, can be simplified to
\begin{equation}
  \label{eq:postRed}
  P(\bm{\rho}, \A^{*} | \mathbf{R}_1, \ldots, \mathbf{R}_N)  \propto \exp\left\{ -\frac{\alpha}{n^{*}} \sum_{j=1}^N d_{\A^{*}}(\mathbf{R}_j, \bm{\rho})\right\} 1_{\mathcal{P}_{n^{*}}}(\bm{\rho}) 1_{\mathcal{C}}(\mathcal{A^{*}}).
\end{equation}

Note that, from the posterior distribution in \eqref{eq:postRed}, marginal posterior distributions of both model parameters, $\bm{\rho}$ and $\A^{*},$ can be easily derived.
Also note that while BMM aims to find a consensus ranking for the full set of items included in $\A$, lowBMM aims to select and rank items in $\A$ that are \textit{relevant} for explaining the data structure.
This latter model can thus be used to jointly perform ranking and variable selection tasks for high- and ultra-high-dimensional data.

\subsection{Lower-dimensional Bayesian Mallows Model Mixture (lowBM3)}\label{subsec:lowbmm_mix_mod}

The main drawback of lowBMM is that, differently from the original BMM, it assumes that a shared single consensus ranking can well describe the data structure alone. This is an unrealistic assumption in many situations, as one usually expects some heterogeneity of the observed data around few different consensus rankings.
To properly handle such heterogeneous data, we build a finite mixture of lowBMM models, where each of the $C$ mixture components (for $c=1,\ldots,C$) is characterized by a set of relevant items $\A_c^{*}$ and by a low-dimensional consensus parameter $\vrho_c \in \mathcal{P}_{n^{*}}$, defined on a lower-dimensional cluster-specific permutation space of dimension $n^{*}$. The complete-data likelihood of the resulting finite mixture model, which is named Lower-dimensional Bayesian Mallows Model Mixtures (lowBM3), takes the form
\begin{align*}
  P(\textbf{R}_1,.
  ..,\textbf{R}_N | \{\A_c^{*},\vrho_c,\tau_c\}_{c=1}^C; z_1,\ldots,z_N) = & \prod_{j=1}^N \left\{ \frac{1}{Z_{n^{*}}(\alpha_{z_j})} \exp\{-\frac{\alpha_{z_j}}{n^{*}}d_{\A^{*}_{z_j}}(\textbf{R}_j, \vrho_{z_j})\} \right. \\
  &\left.  \cdot U_{\mathcal{P}_{n-n^{*}}}(\textbf{R}_j|_{A\backslash \A^{*}_{z_j}}) 1_{\mathcal{P}_n}(\textbf{R}_j) \right\},
\end{align*}
where $\tau_1,\ldots, \tau_C$ are mixture parameters, and $z_1,\ldots,z_N$ are latent variables taking values in the set $\{1,\ldots,C\}$ and defining the cluster assignments of the $N$ assessors. $d_{\A_c^{*}}(\textbf{R}_j, \vrho) \coloneqq d_{n^{*}}(\textbf{R}_j|_{\A^{*}_c}, \vrho)$ is the distance function restricted to the $n^{*}$-dimensional set of items $\A^{*}_c$, where the subscript $c$ here refers to the fact that the set of selected \emph{relevant} items is cluster-specific.
Furthermore, $\mathcal{U}_S$ is the uniform distribution over the domain S, and hence the term $\mathcal{U}_{\mathcal{P}_{n-n^{*}}}(\textbf{R}_j|_{\A \backslash \A^{*}_c})$ refers to the assumption that $\textbf{R}_j|_{\A \backslash \A^{*}_c}$ includes only noisy unranked data.
Note that we fix the scale parameter $\alpha_c, c=1,\ldots,C$ as in~\cite{eliseussen2022}.

As in lowBMM, lowBM3 also assumes uniform priors for the model parameters $\vrho_c$ and $\A^{*}_c$.
The priors related to the mixture parameters and cluster assignments do not depend on the introduction of the $n^{*}$-dimensional space, so we keep the same choices as used in BMM:
\begin{align}
  \pi(z_1,.
  ..,z_N|\tau_1,\ldots,\tau_C) = \prod_{j=1}^N \tau_{z_j} = \prod_{c=1}^C \tau_c^{n_c},\quad \pi(\tau_1,\ldots,\tau_C)=\Gamma(\psi C)\Gamma(\psi^{-C})\prod^C_{c=1}\tau_c^{\psi-1},
\end{align}
where $n_c=\sum_{j=1}^N 1_{j:z_j=c},$ i.e., $n_c$ is the dimension of cluster $c.$
This leads to the posterior of lowBM3 taking the form:
\begin{align*}
  P(\{ \A_c^{*},&\vrho_c,\tau_c\}_{c=1}^C;z_1,\ldots,z_N| \textbf{R}_1,\ldots,\textbf{R}_N)\\
  \propto  &P(\textbf{R}_1,\ldots,\textbf{R}_N | \{\vrho_c, \A_c^{*},\tau_c\}_{c=1}^C; z_1,\ldots,z_N)
  \pi(\{\vrho_{c}\}_{c=1}^C|\{\A^{*}_{c}\}_{c=1}^C; z_1,\ldots,z_N)                                                                                                                                                                                                                \\
  & \cdot \pi(\{\A^{*}_{c}\}_{c=1}^C|z_1,\ldots,z_N) \pi(z_1,\ldots,z_N|\tau_1,\ldots,\tau_C) \pi(\tau_1,\ldots,\tau_c)                                                    \\
  \propto                                                                                                             & \prod_{j=1}^N \left[ P(\textbf{R}_j | \vrho_{z_j}, \A_{z_j}^{*}, \tau_{z_j}; z_j)
                                                                                                                                        \pi(\vrho_{z_j}|\A^{*}_{z_j}, z_j) \pi(\A^{*}_{z_j}|z_j) \right]                                                                                                                                                                                                          \\
  & \cdot \pi(z_1,\ldots,z_N|\tau_1,\ldots,\tau_C) \pi(\tau_1,\ldots,\tau_c)                                                                                                      \\
  \propto                                                                                                             &
  \prod_{j=1}^N \left[\frac{1}{Z_{n^{*}}(\alpha_{z_j})} \exp \left\{-\frac{\alpha_{z_j}}{n^{*}}d_{\A^{*}_{z_j}}(\textbf{R}_j, \vrho_{z_j}) \right\} \mathcal{U}_{\mathcal{P}_{n-n*}}(\textbf{R}_j|_{\A \backslash \A^{*}_{z_j}})1_{\mathcal{P}_n}(\textbf{R}_j)\right] \\
  & \cdot \prod_{c=1}^C \left[\frac{1}{n^{*}!
  } 1_{\mathcal{P}_{n^{*}}}(\vrho_{c})\frac{1}{|\mathcal{C}|} 1_{\mathcal{C}}(\A^{*}_{c}) \tau_c^{n_c + \psi -1}\right]                                                                                                                                                              \\
\end{align*}
Under the assumption that the data are full rankings, and removing terms not depending on any of the model parameters, this can be simplified to
\begin{multline}\label{eq:lowBM3posterior}
  P(\{\A_c^{*}, \vrho_c,\tau_c\}_{c=1}^C;z_1,\ldots,z_N|\textbf{R}_1,\ldots,\textbf{R}_N)
  \\
  \propto \prod_{j=1}^N \exp \left\{-\frac{\alpha_{z_j}}{n^{*}}d_{\A^{*}_{z_j}}(\textbf{R}_j, \vrho_{z_j}) \right\} \prod_{c=1}^C  \left\{ \tau_c^{n_c + \psi -1}1_{\mathcal{P}_{n^{*}}}(\bm{\rho}_c) 1_\mathcal{C}(\A^{*}_c)\right\}.
\end{multline}
The inferential strategy to tackle the posterior distribution \eqref{eq:lowBM3posterior} is based on a Markov Chain Monte Carlo (MCMC) scheme detailed in Section \ref{sec:mcmc}. Marginal posterior distributions for the model parameters $\bm{\rho}_c$, $\A_c^{*}$ and $\tau_c,$ for $c=1,\ldots,C,$ and of the latent variables $z_1, \ldots, z_N$, can be derived via posterior postprocessing as outlined in Section~\ref{subsec:mh_postprocess}.

\section{MCMC algorithm for inference in lowBM3}\label{sec:mcmc}

The MCMC scheme for tackling the posterior distribution of lowBM3 in \eqref{eq:lowBM3posterior} iterates between four steps: (i) the update of the cluster-specific consensus ranking $\vrho_c$, $c=1,\ldots, C$; (ii) the update of the cluster-specific set of relevant items $\A^{*}_c$, $c=1,\ldots, C$; (iii) the update of the cluster label assignments for each assessor $z_j$, $j=1,.
  .., N$; and (iv) the update of the mixture parameters, i.e., the cluster probabilities $\tau_c$, $c=1,\ldots, C$.
The cluster-specific consensus $\vrho_c$ is updated through a Metropolis-Hastings (M-H) step given the current cluster assignment labels $z_1, \ldots, z_N$ and the current set $\A^{*}_c$.
The set $\A^{*}_c$ is also updated through an M-H step, given the labels $z_1, \ldots, z_N$ and the new consensus ranking $\vrho_c$.
The cluster assignment labels $z_1, \ldots, z_N$ are updated in parallel to one another via an assessor-specific Gibbs sampling step given the full set of updated consensus rankings $\{\vrho_1,\ldots,\vrho_C \}$ and relevant items sets $\{\A^{*}_1,\ldots, \A^{*}_C\}$.
Finally, the mixture parameters $\{\tau_1,\ldots,\tau_C\}$ are also updated via a Gibbs sampling step given the cluster label assignments $z_1, \ldots, z_N$.

For step (i), for each cluster $c=1,\ldots,C$, we propose a new consensus ranking $\vrho'_c\in \mathcal{P}_{n^{*}}$ using the Leap-and-Shift (L\&S) proposal distribution~\cite{vitelli2018}.
This results in the following acceptance probability
\begin{equation*}
  \min \left\{1, \frac{P_l(\vrho_c|\vrho'_c)\pi(\vrho'_c|\A^{*}_c)\pi(\A^{*}_c)P(\textbf{R}_1,\ldots, \textbf{R}_N|\vrho'_c, \A^{*}_c)}{P_l(\vrho'_c|\vrho_c)\pi(\vrho_c|\A^{*}_c)\pi(\A^{*}_c)P(\textbf{R}_1,\ldots,\textbf{R}_N|\vrho_c, \A^{*}_c)} \right\}.
\end{equation*}
Assuming that the rankings are permutations in $\mathcal{P}_{n^{*}}$, and given the uniform priors used for $\vrho_c$ and $\A^{*}_c$, the acceptance probability above can be simplified to the following form:
\begin{equation}
  \label{eq:CLBMM 1}
  \min\left\{ 1, \frac{P_l(\vrho_c|\vrho'_c)}{P_l(\vrho'_c|\vrho_c)}\exp\left[ -\frac{\alpha_c}{n^{*}}\left(\sum\limits_{j\in \{1,\ldots, N\}: z_j = c} d_{\A^{*}_c}(\textbf{R}_j,\vrho'_c)-\sum\limits_{j\in \{1,\ldots, N\}: z_j = c} d_{\A^{*}_c}(\textbf{R}_j, \vrho_c) \right) \right] \right\}.
\end{equation}
In \eqref{eq:CLBMM 1}, $P_l(\vrho'_c|\vrho_c)$ refers to the probability mass function associated with moving from $\vrho_c$ to $\vrho'_c$ in an $n^{*}$-dimensional space according to the L\&S proposal distribution.
As with BMM and lowBMM, $l$ denotes the number of items perturbed in the consensus ranking $\bm{\rho}_c$ to get a new proposal $\bm{\rho}'_c$, and is used to tune the acceptance probability.

Step (ii) is also an M-H step to update the set of relevant items $\A^{*}_c$ for each cluster $c=1,\ldots,C$.
A new set is proposed by uniformly drawing $L_c$ items from the current set $\A^{*}_c,$ and by swapping the chosen items with $L_c$ items uniformly drawn from the set $\A\backslash \A^{*}_c$.
The number of items swapped in the set at each iteration of the MCMC is also selected uniformly with an upper limit of $L$, i.e., $L_c\sim\mathcal{U}(1,L)$.
The transition probability mass associated with the change can formally be written as $q(\A^{*}_{c,\textnormal{prop}}|\A^{*}_c) =|D_{L_c}^{in}|^{-1}|D_{L_c}^{out}|^{-1}1_{D_{L_c}^{in}}(\A^{*}_{c,\textnormal{prop}}\backslash \A^{*}_c)1_{D_{L_c}^{out}}(\A^{*}_c\backslash \A^{*}_{c,\textnormal{prop}})$, where $D_{L_c}^{in}$ is defined as the collection of all $\binom{n-n*}{L_c}$ possible sets of items of dimension $L_c$ chosen from a set of dimension $n-n^{*}$, to be brought into the reduced set $\A^{*}_c$ to obtain $\A^{*}_{c,\textnormal{prop}}$.
Similarly, $D_{L_c}^{out}$ is defined as the collection of all $ \binom{n*}{L_c}$ possible sets of dimension $L_c$ chosen from a set of dimension $n^{*}$, to be brought out of $\A^{*}_c$ to obtain $\A^{*}_{c, \textnormal{prop}}$.

Note that proposing a new set $\A^{*}_{c, \textnormal{prop}}$ in our modeling framework implicitly implies updating $\bm{\rho}_c$ to $\bm{\rho}_{c, \textnormal{prop}},$ as the consensus ranking must be defined on the new set of items. However, since this step is the update of $\A^{*}_c,$ the update of $\bm{\rho}_c$ to $\bm{\rho}_{c, \textnormal{prop}}$ must not add any further information, and therefore is performed such that $\bm{\rho}_{c, \textnormal{prop}}|_{\mathcal{I}_c}\equiv\bm{\rho}_c|_{\mathcal{I}_c}$ for $\mathcal{I}_c := \A^{*}_{c,\textnormal{prop}}\cap \A^{*}_c.$ Items that enter $\A^{*}_{c,\textnormal{prop}}$ from outside of $\A^{*}_c$ are randomly assigned the ranks of the items that are left out, i.e., $\rho_{c,\textnormal{prop}}|_{\A^{*}_{c,\textnormal{prop}}\backslash \A^{*}_c}\sim \mathcal{U}(\bm{\rho}_{c}|_{\A^{*}_c\backslash \A^{*}_{c,\textnormal{prop}}}).$
The probability of accepting the proposed $\A^{*}_{c,\textnormal{prop}}$ is then
\begin{equation*}
  \min\left\{ 1, \frac{\pi(\A^{*}_{c,\textnormal{prop}},\bm{\rho}_{c, \textnormal{prop}})P(\textbf{R}_1,\ldots,\textbf{R}_N|\A^{*}_{c,\textnormal{prop}},\vrho_{c, \textnormal{prop}})q(\A^{*}_c,\vrho_c|\A^{*}_{c,\textnormal{prop}},\vrho_{c, \textnormal{prop}})}{\pi(\A^{*}_c,\bm{\rho}_c)P(\textbf{R}_1,\ldots,\textbf{R}_N|\A^{*}_c,\vrho_c)q(\A^{*}_{c,\textnormal{prop}},\vrho_{c, \textnormal{prop}}|\A^{*}_c,\vrho_c)} \right\}.
\end{equation*}

In the above, the priors factorize in $\pi(\A^{*}_{c, \textnormal{prop}},\bm{\rho}_{c, \textnormal{prop}})= \pi(A^*_{c, \textnormal{prop}})\pi(\vrho_{c, \textnormal{prop}})$ and $\pi(\A^{*}_c,\bm{\rho}_c)= \pi(A^*_c)\pi(\vrho_c)$, and simplify as they are both assumed to be uniform. For the transition probabilities, noting that $q(\A^{*}|\cdot)$ cannot depend on the consensus, we have that:
$$
\frac{q(\A^{*}_c,\vrho_c|\A^{*}_{c,\textnormal{prop}},\vrho_{c, \textnormal{prop}})}{q(\A^{*}_{c,\textnormal{prop}},\vrho_{c, \textnormal{prop}}|\A^{*}_c,\vrho_c)} = \frac{q(\vrho_c|\A^{*}_c,\A^{*}_{c,\textnormal{prop}},\vrho_{c, \textnormal{prop}})\cdot q(\A^{*}_c|\A^{*}_{c,\textnormal{prop}})}{q(\vrho_{c, \textnormal{prop}}|\A^{*}_{c,\textnormal{prop}},\A^{*}_c,\vrho_c)\cdot q(\A^{*}_{c,\textnormal{prop}}|\A^{*}_c)}
$$
which is simply equal to 1 as both transition kernels are symmetrical, since all involved distributions are uniform.
In conclusion, the acceptance probability takes the following simplified form
\begin{equation}
  \label{eq:CLBMM 2}
  \min\left\{ 1, \exp\left[ -\frac{\alpha_c}{n^{*}}\left(\sum\limits_{j\in \{1,\ldots, N\}: z_j = c} d_{\A^{*}_{c,\textnormal{prop}}}(\textbf{R}_j,\vrho_{c, \textnormal{prop}})-\sum\limits_{j\in \{1,\ldots, N\}: z_j = c}d_{\A^{*}_c}(\textbf{R}_j, \vrho_c) \right) \right] \right\}.
\end{equation}

For Step (iii), i.e.,\ the update of the cluster label assignments $z_1,\ldots, z_N$, we can use the Gibbs sampler, as the full conditional distribution associated with the $j$-th cluster label can be explicitly written as $P(z_j|\mathbf{z}_{-j};\{ \A^{*}_c, \rho_c, \tau_c\}_{c=1}^C)\sim \mathcal{M}(p_{1j},\ldots, p_{Cj})$, where $\mathbf{z}_{-j}$ is the vector including all cluster label assignments excluding the $j$-th, and with
\begin{equation}
  \label{eq:z_gibbs}
  p_{cj} = \frac{\tau_{c}}{Z_{n^{*}}(\alpha_c)}\exp \left[ -\frac{\alpha_c}{n^{*}} d_{\A^{*}_c}(\textbf{R}_j, \vrho_c)\right]\ \forall\ c=1,\ldots,C.
\end{equation}

Finally, the parameters $\tau_1,\ldots,\tau_C$ are updated in Step (iv) via a Gibbs sampling step making use of the conjugacy property.
The algorithm described above is summarized in Algorithm~\ref{alg:clbmm}, and at convergence of the chain produces samples from the posterior distribution \eqref{eq:lowBM3posterior}.
\begin{algorithm}
  \caption{MCMC algorithm for lowBMM with mixtures}
  \label{alg:clbmm}
  \begin{algorithmic}[1]
    \Require $\mathbf{R}_1,\ldots,\mathbf{R}_N$, $\alpha_1,\ldots,\alpha_C$, $d(\cdot,\cdot)$, $l$, $L$, $M$, $C$, $\psi$, $Z_{n^*}(\alpha_1), \ldots, Z_{n^*}(\alpha_C)$, $n^{*}$
    \Ensure posterior distribution of $\bm{\rho}_1,\ldots,\bm{\rho}_C$, $\A^{*}_1,\ldots,\A^{*}_C$, $\tau_1,\ldots,\tau_C$, $z_1,\ldots,z_N$
    \State Randomly initialize $\bm{\rho}_{1,0},\ldots,\bm{\rho}_{C,0}$, $\A^{*}_{1,0},\ldots,\A^{*}_{C,0}$, $\tau_{1,0},\ldots,\tau_{C,0}$, and $z_{1,0},\ldots,z_{N,0}$
    \For{$m \gets 1$ to $M$}
    \State \textbf{Gibbs step:} update $\tau_1,\ldots,\tau_C$
    \State Compute $n_c = \sum_{j=1}^{N}1_c(z_{j,m-1})$ for $c=1,\ldots,C$
    \State Sample $\tau_{1,m},\ldots,\tau_{C,m} \sim \mathcal{D}(\psi + n_1,\ldots,\psi + n_C)$
    \For{$c \gets 1$ to $C$}
    \State \textbf{M-H step:} update $\vrho_c$
    \State Sample $\vrho'_c \sim L\&S(\vrho_{c,m-1},l)$ restricted on $\A^{*}_{c,m-1}$, and $u\sim \mathcal{U}(0,1)$
    \State Compute $\mathit{ratio}$ from Equation~\eqref{eq:CLBMM 1} with $\vrho_c \gets \vrho_{c,m-1}$ and $\A^{*}_c \gets \A^{*}_{c,m-1}$
    \If{$u < \mathit{ratio}$}
    \State $\vrho_{c,m}\gets \vrho'_c$
    \Else
    \State $\vrho_{c,m} \gets \vrho_{c,m-1}$
    \EndIf
    \State \textbf{M-H step:} update $\A^{*}_c$
    \State Sample $L_c\sim \mathcal{U}(0,L)$, $(\A^{*}_{prop},\vrho_{prop}) \sim q(\A^{*}_{prop},\vrho_{prop} \mid \A^{*}_{c,m-1},\vrho_{c,m},L_c)$, and $u \sim \mathcal{U}(0,1)$
    \State Compute $\mathit{ratio}$ from Equation~\eqref{eq:CLBMM 2} with $\vrho_c \gets \vrho_{c,m}$, $\A^{*}_c \gets \A^{*}_{c,m-1}$, and $z_j\gets z_{j,m-1}$
    \If{$u < \mathit{ratio}$}
    \State $\A^{*}_{c,m}\gets \A^{*}_{prop}$
    \Else
    \State $\A^{*}_{c,m}\gets \A^{*}_{c,m-1}$
    \EndIf
    \EndFor
    \State \textbf{Gibbs step:} update $z_1,\ldots,z_N$
    \For{$j \gets 1$ to $N$}
    \State Compute $p_{cj} = \frac{ \tau_{c,m}}{Z_{n^*}(\alpha_c)}\exp \left[-\frac{\alpha_c}{n^{*}}d_{\A^{*}_{c,m}}(\mathbf{R}_j, \vrho_{c,m}) \right]$ for $c=1,\ldots,C$
    \State Sample $z_{j,m} \sim \mathcal{M}(p_{1j},\ldots,p_{Cj})$
    \EndFor
    \EndFor
  \end{algorithmic}
\end{algorithm}

\subsection{Postprocessing of the MCMC results}\label{subsec:mh_postprocess}
Since all model unknowns are discrete parameters, summarizing the associated posterior distributions is not a trivial task. We here outline our innovative strategies for deriving posterior summaries of the consensus of the relevant items, $\hat{\vrho}_c$, and of the set of relevant items, $\hat{\A}^{*}_c$, by adapting MCMC postprocessing techniques introduced in~\cite{eliseussen2022} to the case of a mixture model.
In a nutshell, the proposed approach to computing posterior summaries proceeds in three steps:
(i) Define the Highest Probability Set (HPS), $\A'_c$, as the set of items in $\mathcal{A}$ that have the largest posterior probability of belonging to $\A^{*}_c$; note that $|\A'_c|$ is larger than $n^*$, to ensure broader inclusion of candidate relevant items.
(ii) Estimate $A^{*}_c$ with the set $\hat{\A}^{*}_c$, selecting the $n^*$ items in $\A'_c$ featuring the smallest average posterior ranking.
(iii) Compute a summary of the posterior distribution of the consensus ranking $\vrho_c$ by taking $\boldsymbol{\hat{\rho}}_c$, which ranks the items in the set $\hat{\A}^{*}_c$ by their average posterior rank.

Specifically, suppose to collect $M$ samples from the posterior distribution of the two parameters of interest, denoted by $\{\vrho_{c,m},\A^{*}_{c,m}\}_{m=1}^M$ for $c=1,.
  ..,C$, with $\vrho_{c,m}=\{\rho^{c,m}_{i_1},\ldots,\rho^{c,m}_{i_{n^{*}}}\}$ and $\A^{*}_{c,m}=\{A^{c,m}_{i_1},\ldots,A^{c,m}_{i_{n^{*}}}\}$.
Here $\rho^{c,m}_{i_t}$ is the ranking assigned to the item $A^{c,m}_{i_t}$ in the consensus ranking of the $c$-th cluster at the $m$-th iteration.
Let $W^c\in \mathbb{R}^{M\times n}$ for $c=1,\ldots,C$ be a matrix such that the $m,i$ cell is an indicator of item $A_i$ being selected and ranked in the $m$-th posterior sample for the $c$-th cluster: $W^c_{mi}=1_{\A^{*}_{c,m}}(A_i)$, for items $A_i$, $i=1,\ldots,n$.
We introduce a few definitions to infer $\hat{\A}^{*}_c$ and $\hat{\bm{\rho}}_c$.
\begin{definition}
  Given a vector of real numbers $(x_1,\ldots, x_n)\in \mathbb{R}^n,$ its corresponding rank vector is obtained as follows:\\

  $rank(x_1, \ldots, x_n) = (r_1,\ldots, r_n), \text{ such that}\quad r_i = \sum\limits_{j=1}^{n}\delta (x_j - x_i)$ for $i = 1, \ldots, n$, \\
  where $\delta(x) = \left \{
    \begin{aligned}
                         & 1, &  & \text{if } x\geq 0 \\
                         & 0, &  & \text{if } x < 0
                       \end{aligned}
    \right.$
\end{definition}

\begin{definition}
  \label{def:hps1}
  For $k \in \left\{n^{*},n^{*}+1,\ldots, n\right\}$, the Highest Probability Set (HPS) of $\A_c^{*}$ is the set of the $k$ items that appear most frequently in the posterior sample sets $\A_{c,m}^{*}$, for $m \in 1,\ldots, M$.
  Formally: \\

  $\A_c' = \{A_i, i=1,\ldots,n \, | \, rank(\bm{\bar{w}}^c)_i \leq k\}$, \\

  where $\bm{\bar{w}}^c=(\bar{w}_{1}^c, \ldots,\bar{w}_{n}^c)$ and $\bar{w}^c_{i} = \frac{1}{M}\sum_{m=1}^{M}
    W^c_{mi},$ $i=1,\ldots,n$.
\end{definition}
We note that, rather than fixing the value $k$ (i.e., selecting a predetermined number of items to be included in the HPS), an alternative and equivalent approach is to threshold the vector $\bm{\bar{w}}^c$ at a specified value $p\in[0,1]$.
This entails including in the HPS only those items whose frequency of inclusion in the posterior samples $\A^{*}_{c,m}$ is larger than $p$.
For any given $k$, a value of $p$ leading to the same $\A_c'$ can always be found, and viceversa.

\begin{definition}
  \label{def:Astar}
  Given the HPS $\A'_c$ as in Definition \ref{def:hps1}, the posterior summary of the set of relevant items $\mathcal{\hat{A}}_c^{*}$ corresponding to the $c$-th cluster is computed as the set of $n^{*}$ objects in $\A'_c$ with the lowest posterior average rank.
  Formally:
  \[
  \hat{\A^{*}_c} = \left\{A_i \in \A'_c \, | \, rank(-\Bar{\mathbf{x}}_c)_i \leq n^{*}\right\}
  \]
  where $\Bar{\mathbf{x}}_c \in \mathbb{R}^{|\A'_c|}$, $\Bar{x}_{c,i}=\frac{\sum\limits_{m=1}^M\rho^{c,m}_{i}{1_{\A^{*}_{c,m}}}(A_i)}{\sum\limits_{m=1}^M 1_{\A^{*}_{c,m}}(A_i)}\ \forall\ A_i \in \A'_c$.
\end{definition}
\noindent Note that the $rank(\cdot)$ of the opposite of $\Bar{\mathbf{x}}_c$ is taken as the lowest ranked objects on average are the most relevant. Also note that item $A_i$ receives a posterior rank only when it is selected in the posterior sample set $\A^{*}_{c,m}$. Consequently, the posterior average rank $\bar{x}_{c,i}$ is computed by averaging the ranks of the available posterior samples alone.

Finally, we can quantify the posterior summary of the consensus ranking $\bm{\hat{\rho}}_c$ as follows:
\begin{definition}
  \label{def:rho}
  Given $\mathcal{\hat{A}}_c^{*}$ as in Definition \ref{def:Astar}, the posterior summary of the consensus ranking $\bm{\hat{\rho}}_c$ is
  \[
  \bm{\hat{\rho}}_c = rank(-\Bar{\mathbf{x}}_c)|_{\mathcal{\hat{A}}^{*}_c}
  \]
  where $\Bar{\mathbf{x}}_c$ has also been introduced in Definition~\ref{def:Astar}.
\end{definition}

Summarizing the proposed strategy, to obtain the posterior summaries of $\A^{*}_c$ and $\vrho_c$ for $c=1,\ldots, C$, we first obtain the highest probability sets $\A'_c$ according to Definition~\ref{def:hps1}.
We then quantify the posterior summary $\hat{\A}^{*}_c$ by Definition~\ref{def:Astar}, based on $\A'_c$, and finally quantify the posterior summary $\bm{\hat{\rho}}_c$ by Definition~\ref{def:rho}.
The cluster label assignment associated with each assessor, and used in the calculation of the posteriors summaries above, is the maximum a posteriori of $z_j$ for $j=1,\ldots, N$.
\begin{remark}
  The joint evaluation of the posterior distributions of $\bm{\rho}_c$ and $\A^{*}_c$ will induce an ordering on the complete set of items when setting $k=n$ in Definition~\ref{def:hps1}, as $\Bar{\mathbf{x}}_c$ is then defined on $\mathbb{R}^n$, since $|\A_c'|=n$.
  However, even in this extreme case, $\bm{\rho}_c\in \mathcal{P}_{n^{*}}$ by construction, as $|\mathcal{\hat{A}^{*}}_c|=n^{*}$.
\end{remark}



\section{Simulation studies}\label{sec:simulations}

In this section, we provide a comprehensive evaluation of the performance of lowBM3.
The model performance is assessed by several performance metrics, such as the cluster assignment accuracy and the accuracy in the estimation of model parameters.
We also conduct an experiment to test the ability of the method to detect the correct number of clusters $C$, and its sensitivity to the misspecification of  $\alpha$, given that  these parameters are fixed in our current model implementation.
The experiments were conducted using simulated data sets generated according to different data-generating processes to test the model's versatility in handling various types of data that may arise in real-life scenarios.

To compare and evaluate the methods, we computed several performance measures.
Let $\A^{*}_c$ be the true set of relevant items in cluster $c$, and $\hat{\A}^{*}_c$ be its estimate obtained via posterior postprocessing as described in Section \ref{subsec:mh_postprocess}, with $|\hat{\A}^{*}_c|=|\A^{*}_c|=n^{*}$.
Similarly, let $\bm{\rho}_c$ be the true consensus ranking of the relevant items in cluster $c$, and $\bm{\hat{\rho}}_c$ be its estimate.
Also, let $z_1,\ldots, z_N$ be the true cluster labels and let $\hat{z}_1,\ldots, \hat{z}_N$ be the estimated cluster labels, defined as the MAP over the posterior cluster assignments.
We then consider the following measures of performance:
\begin{itemize}
  \itemsep0em
  \item The proportion of correctly assigned assessors: $\hat{p}_z = \frac{1}{N}\sum_{j=1}^N 1_{\hat{z}_j = z_j}.$
  \item The average (over assessors) posterior probability of assigning each assessor to the correct cluster: $Z_\textnormal{post} = \frac{1}{N}\sum_{j=1}^N P(\hat{z}_j = z_j|\textbf{R}_1,\ldots,\textbf{R}_N).$
  \item The average proportion of correctly selected items over the $C$ clusters: $\hat{p}_{\A^{*}} = \frac{1}{C} \sum_{c=1}^C n_\textnormal{c,corr}/n^{*}$, where $n_\textnormal{c,corr}=|\A_c^{*}\bigcap\hat{\A}_c^{*}|$ is the number of correctly selected items in cluster $c$.
  \item The normalized within-cluster mean footrule distance between the true $\bm{\rho}_c$ and the estimated $\hat{\bm{\rho}}_c$ consensus: $d(\bm{\rho},\hat{\bm{\rho}}) = \frac{1}{C} \sum_{c=1}^C d(\bm{\rho}_c,\hat{\bm{\rho}}_c)\big/\bar{d}_{n^{*}}$, where $\bar{d}_x$ is the maximum of the footrule distance between full rankings of $x$ items.
\end{itemize}

\subsection{Data-generating processes: top-rank and consistency}\label{subsec:sim_data_processes}

In what follows, we describe two data-generating processes (DGPs) aimed at producing data that emulate real-life scenarios, which are then used to generate several different simulated setting.
For clarity of exposition, we refer to these DGPs as \textit{top-rank} and \textit{consistency}.

In both DGPs, a total of $N$ assessors is initially randomly split into $C$ clusters, i.e, the random variables $z_j\sim\mathcal{U}(\{1,\ldots,C\})$ are generated independently of one another for $j=1,\ldots,N$. We also define $N_c:=\sum_{j=1}^N 1_{z_j=c}$ for $c=1,\ldots,C.$
Then, for each cluster, a $n^{*}$-dimensional consensus ranking $\vrho_c$ is generated uniformly at random, i.e., $\vrho_c \sim \mathcal{U}(\mathcal{P}_{n^{*}})$.
Analogously, the true set of relevant items $\A^{*}_c$ is generated by selecting at random $n^{*}$ out of $n$ items; $\vrho_c$ defines the ranks of the items in $\A^{*}_c$.

When generating the observed full rankings $\mathbf{R}_j$ for $j=1,\ldots,N,$ the two DGPs proceed differently. In the top-rank DGP, for each $c=1,\ldots,C$, $N_c$ i.i.d.\ samples from the Mallows model are generated, i.e., $\forall j=1,\ldots,N:z_j=c,$ $\mathbf{R}_j|_{\A^{*}_c} \sim \text{Mallows}(\vrho_c, \alpha_c)$, given the set of relevant items $\A^{*}_c$, and the model parameters $\vrho_c$ and $\alpha_c$. The remaining $n-n^{*}$ items in $\A\backslash \A^{*}_{z_j}$ are assigned a higher rank uniformly at random: $\mathbf{R}_j|_{\A\backslash \A^{*}_{z_j}}\sim \mathcal{U}(\mathcal{P}_{\left\{n^{*}+1,\ldots,n\right\}})$\footnote{With a slight abuse of notation, which however we deem understandable from context, here $\mathcal{P}_A$ has been used to indicate the set of permutations of the objects included in the set $A$.}.
Thus, in the top-rank DGP, each $R_j$ is a rank vector of $n$ items, where the relevant items (those in $\A^{*}_{z_j}$) are ranked first and follow a $n^{*}\text{-dimensional}$ Mallows model, and the remaining items have larger (uniformly) random ranks.

The consistency DGP differs from the top-rank DGP in that the relevant items in $\A^{*}_{z_j}$ are not ranked top, but each rank vector $\mathbf{R}_j|_{\A^{*}_{z_j}}$ is randomly shifted upwards by at most $n-n^{*}$ positions.
That is, for each $j=1,\ldots,N$, $\mathbf{R}_j|_{\A^{*}_{z_j}}\sim \text{Mallows}(\vrho_{z_j}, \alpha_{z_j})$ is shifted of $s_j\sim \mathcal{U}(\{0,\ldots,n-n^{*}\})$, thus obtaining $\mathbf{\tilde{R}}_j|_{\A^{*}_{z_j}} = \mathbf{R}_j|_{\A^{*}_{z_j}} + s_j$.
The remaining items in $\A\backslash \A^{*}_{z_j}$ are assigned an available rank from $\{1,\ldots,n\}\backslash \mathbf{\tilde{R}}_j|_{\A^{*}_{z_j}}$, uniformly at random.
Note that the amount by which the rank sequence is shifted, $s_j$, is drawn independently for each assessor.
Thus, assessors belonging to the $c\text{-th}$ cluster share the set of relevant items, $\A^{*}_c$, which are ranked contiguously, and follow a Mallows model with consensus rank $\vrho_c$ in the restricted space.
However, their rank relative to other items is assessor-specific.
In Appendix \ref{app:consistencyDGP} in the Supplementary Materials, we present the analysis of a simulated scenario where data are generated from a modified version of the consistency DGP, where we relax the restriction of items in $\A^{*}_c$ being contiguously ranked for each assessor $j:z_j=c$.
This is achieved by swapping the ranks in $\mathbf{\tilde{R}}_j|_{\A^{*}_{z_j}}$ and $\mathbf{\tilde{R}}_j|_{\A\backslash \A^{*}_{z_j}}$ for a small assessor-specific percentage of the $n^{*}$ items.

To show that lowBM3 works under simulation scenarios in line with the model assumptions, we conducted two simulation experiments of differing dimensions, one smaller and one larger, using both DGPs, and specifying the true $C$, $n^{*}$, and $\{\alpha_c\}_{c=1,\ldots,C}$ in input to the lowBM3 algorithm.
Additionally, we performed an experiment in the larger data setup, where the data was generated using a mix of the two data-generating processes.
More specifically, for each cluster $c=1,\ldots,C$, half of the $n^{*}$ items in $\A^{*}_c$ were top-ranked, and the remaining half was shifted by an assessor-specific value $s_j\sim\mathcal{U}(\{0,\ldots,n-\lfloor n^{*}/2\rfloor\})$.
Therefore, in the restricted space determined by $\A^{*}_c$, the ranks of the relevant items still follow a Mallows model with consensus rank $\vrho_c$, but in the full space of $n$ items included in $\A$, top-ranked items and consistency-ranked items may show ranks that are randomly interspersed with those of non-relevant items.
This setup is also affected by noise: a random 10\% of the items in $\A^{*}_c$ had their rank swapped with random items from $\A\setminus{\A^{*}_c}$.

Table~\ref{tab:sim_setups}~(top-panel) contains an overview of the parameters that were used in each setup for generating the data sets.
We fix a unique $\alpha$ for all clusters: $\alpha_c=\alpha$, $c=1,\ldots,C,$ and run lowBM3 with a number of MCMC iterations equal to $M$, as reported in Table~\ref{tab:sim_setups}~(bottom-panel).
The $M$ iterations are then thinned taking every other ``thinning'' iterations, and then an initial percentage of iterations is discarded as burn-in after thinning.
The tuning parameters are set to $L=1$ and $l=\textnormal{round}(n^{*}/5)$ as suggested by the tuning parameter study in~\cite{eliseussen2022}.
We also set $\psi = N/C$ for the sampling of the cluster assignment probabilities as suggested in~\cite{vitelli2018}. The parameters $n^{*}$, $C$, and $\alpha$ in input to lowBM3 are set equal to the true one used for the DGPs (Table~\ref{tab:sim_setups}~top-panel).
The datasets used for Setup 1 and Setup 2 are shown in the supplementary material in figures~\ref{afig:dataplot-toprank-small} and \ref{afig:dataplot-consistency-small}, respectively.
Samples for Setup 3, Setup 4, and Setup 5 are depicted in figures~\ref{afig:dataplot-toprank-big}, \ref{afig:dataplot-consistency-big} and \ref{afig:dataplot-mixdata-big}, respectively, showing only the first out of five clusters.

In the supplementary material, we also include simulation results from two additional data scenarios where the DGPs were meant to test our methodology further, by adding more noise.
The DGPs are variations on the consistency design, and noise was added in the generation of the rankings.
We include small-scale (Setup 2.1 and Setup 2.2) and large-scale alternatives (Setup 4.1 and Setup 4.2); two of them have extreme noise (Setup 2.1 and Setup 4.1) and the remaining two have more moderate noise (Setup 2.2 and Setup 4.2).
All the details are deferred to the supplementary material.

\begin{table}[t]
  \centering
  \begin{tabular}{rlllll}
    & Setup 1  & Setup 2     & Setup 3  & Setup 4     & Setup 5 \\
    \toprule
    DGP      & top-rank & consistency & top-rank & consistency & mix     \\
    \midrule
    $n$      & 30       & 30          & 300      & 300         & 300     \\
    $N$      & 60       & 60          & 150      & 150         & 150     \\
    $n^{*}$  & 10       & 10          & 30       & 30          & 30      \\
    $C$      & 3        & 3           & 5        & 5           & 5       \\
    $\alpha$ & 3        & 3           & 3        & 3           & 3       \\
    \midrule
    $M$      & 25,000   & 25,000      & 100,000  & 100,000     & 100,000 \\
    thinning & 25       & 25          & 25       & 50          & 50      \\
    burnin   & 30\%     & 30\%        & 30\%     & 30\%        & 30\%    \\
    \bottomrule
  \end{tabular}
  \caption{Data generating processes (DGPs) and parameter values for all simulation scenarios.
    Top-panel: parameters used in the various data setups to generate samples.
    Bottom-panel: total number of MCMC iterations (M), thinning and burnin.
  }
  \label{tab:sim_setups}
\end{table}

\subsection{Results on the simulated data sets}\label{subsec:res-experimental-data}
We now present the results of the lowBM3 method on data sets generated according to the setups listed in Table~\ref{tab:sim_setups}.
In the postprocessing phase, described in Section~\ref{subsec:mh_postprocess}, we used different thresholds for determining the high probability set $\A'$ for the different DGPs.
Specifically, when data is generated according to the top-rank designs (Setup~1 and Setup~3), we set $k=n$ (or equivalently $p=0$), which means no selection.
In the consistency designs (Setup~2 and Setup~4) we use a frequency threshold $p=20\%$.
Finally, in the mixed design (Setup~5), we use $p=10\%$, averaging the two thresholds.
See subsection \ref{subsec:thresholds} for some specific considerations on the choice of these thresholds.

\begin{figure}[!htbp]
  \centering
  \begin{subfigure}{0.95\textwidth}
    \caption{Results from Setup 1 (top-rank), cluster 1.}
    \label{fig:selection-toprank-small-cluster1}
    \includegraphics[width=\textwidth]{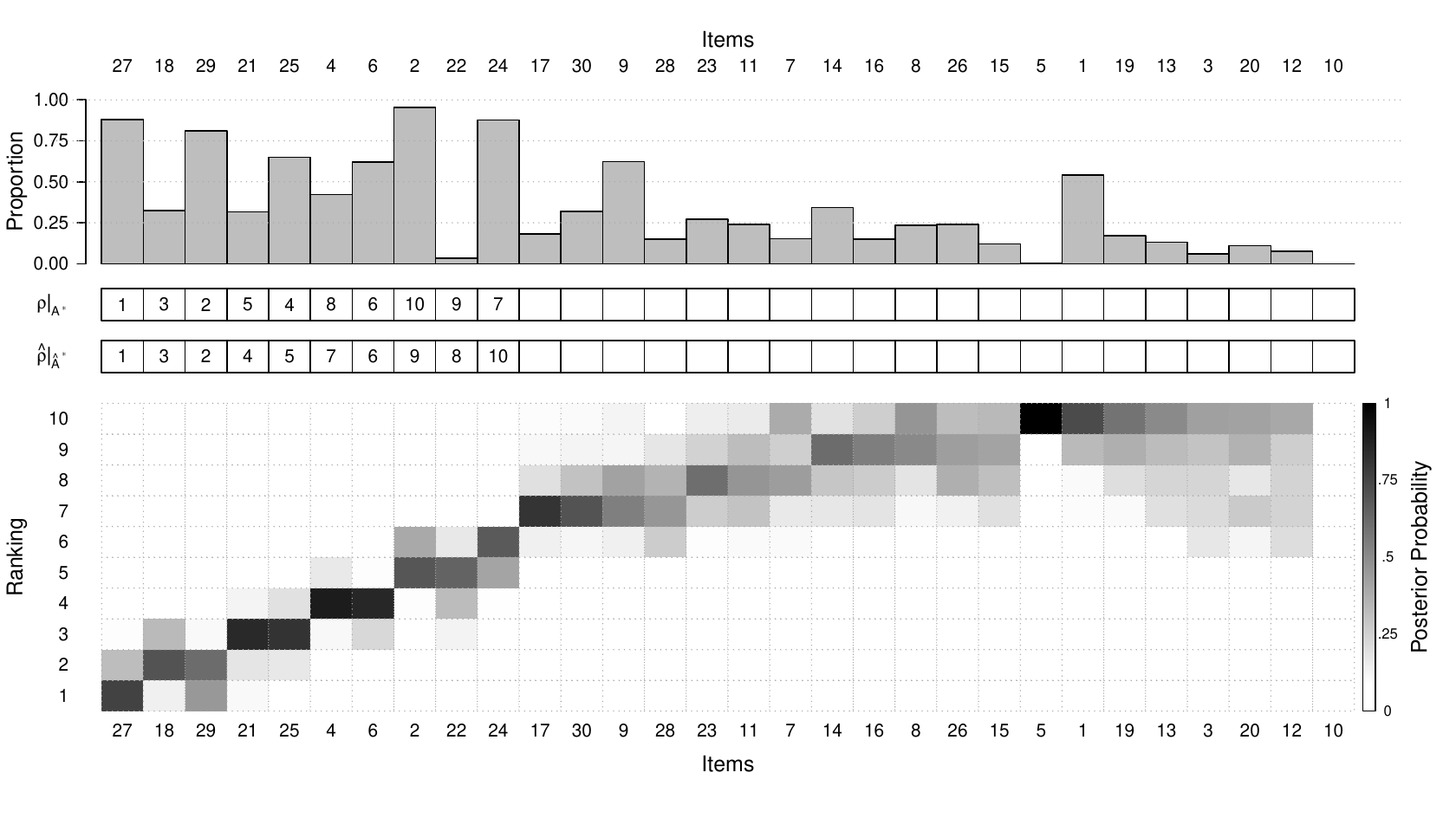}
  \end{subfigure}

  \vfill

  \begin{subfigure}{\textwidth}
    \caption{Results from Setup 2 (consistency), cluster 1.}
    \label{fig:selection-consistency-v2-small-cluster1}
    \includegraphics[width=\textwidth]{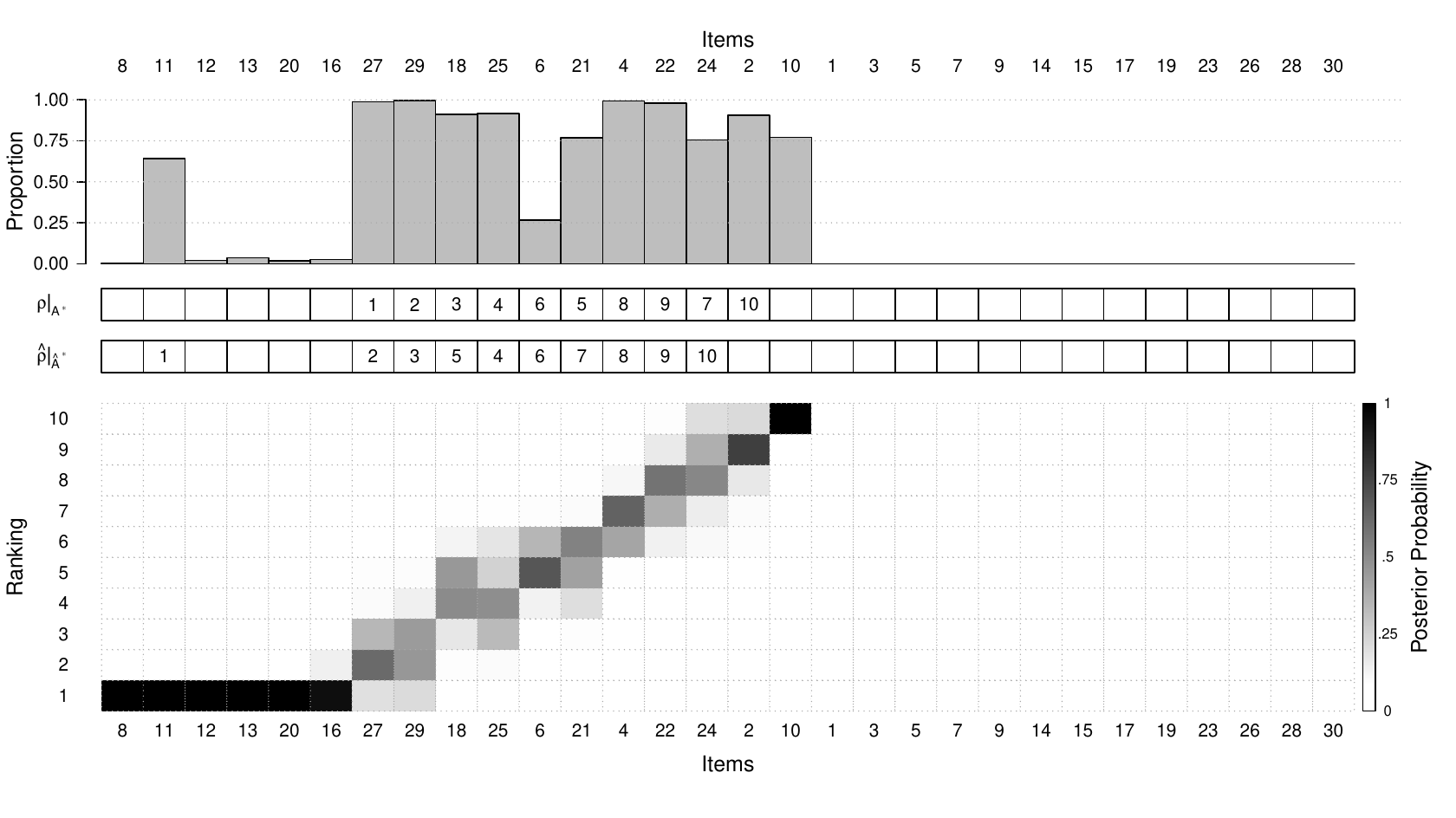}
  \end{subfigure}

  \caption{Results for \emph{cluster 1} from the smaller top-rank (panel a) and consistency (panel b) simulation designs.
  Each panel shows a composite graph with items listed on the x-axes; from top to bottom: bar-plot of the posterior probability with which an item is selected in the estimated set $\hat{\A}_c^{*}$; in the subsequent row, $\rho|_{A^{*}}$, true ranks of the items in $\A_c^{*}$; in the $\hat{\rho}|_{\hat{A}^{*}}$ row, estimated ranks of the items in $\hat{\A}_c^{*}$; in the heatmap, marginal posterior probability  of $\vrho_c$ (shades of gray) for each rank (y-axis) to be assigned to each item.
  On the x-axis, items are ordered by the latter.
  }
  \label{fig:selection-heatmap}
\end{figure}

Figure~\ref{fig:selection-toprank-small-cluster1} and Figure~\ref{fig:selection-consistency-v2-small-cluster1} display the marginal posterior distribution of $\hat{\bm{\rho}}_1$ (i.e., of the consensus ranking parameter associated to cluster 1) for one run of lowBM3, on data generated according to Setup 1 and Setup 2, respectively.
Both figures show the results for the first of three clusters; the results for the second and third cluster are qualitatively equivalent for both setups, and are reported in the supplementary material (Figure~\ref{afig:selection-heatmap-toprank} for Setup 1, and Figure~\ref{afig:selection-heatmap-consistency} for Setup 2).

The results indicate that the method effectively selects the relevant items for inclusion in the selection set.
In Setup 1, all relevant items are selected.
In Setup 2, most of the relevant items are consistently ranked \emph{and} selected.
From both figures, it is evident that while the method does extremely well in the estimation of $\A^{*}_c$, the estimated ranks, $\hat{\vrho}_c$, are less accurate for the true ranks $\vrho_c$, with items often obtaining a rank very close to, but not exactly, their true rank.

Analogous graphs for results obtained when running lowBM3 on data generated according to the larger simulation experiments, i.e, Setup 3 and Setup 4, are reported in Figure~\ref{afig:selection-heatmap-toprank-big} and Figure~\ref{afig:selection-heatmap-consistency-big}, respectively.
For Setup~3, the results in  Figure~\ref{afig:selection-heatmap-toprank-big} are qualitatively  equivalent to results from Setup~1: in almost all clusters the estimated sets $\hat{\A}_c^*$ contain mostly relevant items, although ranks not always precisely estimated.
In contrast, the performance deteriorates substantially in the larger consistency setup, Setup 4.
Here, the estimated sets $\hat{\A}_c^*$ include several irrelevant items.
However, it is worth noting that, even if this design is a very challenging one, still lowBM3 explores all relevant items in almost all cases, often assigning them a large probability of inclusion in the set of relevant items as estimated from MCMC samples $\A_{c,m}^*$.



In the mixed data setup, Setup 5, the performance of lowBM3 falls between that obtained for Setup 3 and Setup 4.
As illustrated in Figure~\ref{afig:selection-heatmap-mixdata-big}, the improvements relative to Setup~4 are likely due to the top-ranked items, which are more easily identified compared to consistency-ranked items.

\begin{table}[t]
  \centering
  \caption{Performance metrics results. The table shows performance on four metrics (rows) of the lowBM3 algorithm in the five setups (columns): top-rank DGP, small-scale design (Setup 1); consistency DGP, small-scale design (Setup 2); top-rank DGP, large-scale design (Setup 3); consistency DGP, large-scale design (Setup 4); mixed DGP, large-scale design (Setup 5). The four considered metrics are: proportion of correctly assigned assessors ($\hat{p}_z$); average posterior probability of assigning assessors to the correct cluster ($Z_{\text{post}}$); mean proportion of correctly selected items ($\hat{p}_{\A^*}$); normalized within-cluster mean footrule distance $d(\vrho, \hat{\vrho})$.}
  \label{tab:performance-selection}
  \begin{tabular}{@{}rlllll@{}}
    \toprule
    & Setup 1 & Setup 2 & Setup 3 & Setup 4 & Setup 5 \\
    \midrule
    $\hat{p}_z$             & 0.93    & 0.7     & 1       & 0.48    & 0.95    \\
    $Z_{post}$              & 0.78    & 0.7     & 1       & 0.48    & 0.93    \\
    $\hat{p}_{A*}$          & 1       & 0.87    & 0.91    & 0.46    & 0.52    \\
    $d(\vrho, \hat{\vrho})$ & 0.2     & 0.47    & 0.48    & 0.68    & 0.66    \\
    \bottomrule
  \end{tabular}
\end{table}

Table~\ref{tab:performance-selection} quantitatively summarizes the performance of lowBM3 in the five setups by reporting the measures defined in Section~\ref{sec:simulations} for each of them. Overall, lowBM3 performs extremely well with the top-rank DGP (Setup 1 and Setup 3): it achieves a high clustering accuracy ($\hat{p}_z$ and $Z_{\text{post}}$) and identifies almost perfectly all relevant items ($\hat{p}_{\A^*}$); estimated ranks are close to the true one in the smaller Setup 1, but less precise for the larger Setup 3 ($d(\vrho, \hat{\vrho})$ further from 0 in this scenario).
In Setup~5, where the top-rank DGP is contaminated with the consistency DGP, these results broadly hold: the clustering accuracy remains very high and 50\% of relevant items (mostly top-ranked items) are correctly identified on average.
The consistency-ranked items present a greater challenge for lowBM3, resulting in worse performance in Setups~2 and 4 relative to the same settings in Setups~1 and 3 in estimating $\A^*_c$ and $\vrho_c$.
This intuition justifies the very poor performance in Setup~4, which is the most challenging one: here all the measures of clustering accuracy, variable selection and rank estimation indicate degraded performance.
Nonetheless, despite the complexity of the simulation design, lowBM3 still identifies 46\% of the relevant items in $\A^*_c$, and achieves a clustering accuracy of 48\%.
In the simpler Setup~2, the smaller scale of the experimental design alleviates the difficulties encountered in Setup~4, and the results improve substantially.
Clustering accuracy reaches 70\%, and roughly 90\% of the relevant items are correctly identified.

It is worth mentioning that~\cite{cladagLowBM3} introduced the so-called \emph{stability postprocessing}, a novel postprocessing strategy for lowBM3 that specifically targets high-noise settings as Setup~2 and Setup~4.
This method is defined similarly to the approach described in Section~\ref{subsec:mh_postprocess}, with the only difference being in the estimation of the $\A^*_c$ set.
In particular, once the HPS set $\A'_c$ is defined, stability postprocessing identifies as relevant those items with the highest posterior probability of obtaining neighboring ranks, with a tuning parameter $\xi$ that can be adapted to define the most proper neighborhood.
This approach features a better recovery of relevant items whose rank is not in top positions, as in the consistency DGP: as shown in \cite{cladagLowBM3}, on Setup~4 the stability postprocessing improved the identification of relevant items to 58\%. However, this does not seem a strong enough gain to justify the introduction of the additional tuning parameter $\xi,$ which is particularly hard to tune in high-dimensional settings. For this reason, the stability postprocessing approach was not explored further, and we refer to~\cite{cladagLowBM3} for further details and considerations.

\subsubsection{\texorpdfstring{A note on the tuning of the threshold \(p\) for computing the HPS}{A note on the tuning of the threshold p for computing the HPS}}\label{subsec:thresholds}
The selection of the threshold $p$ for computing the HPS $\A'_c$ in the simulated scenarios described above is based on the following rationale.
In the top-rank DGP, as relevant items are all ranked at the top, selecting items with the lowest estimated rank on average is sufficient to produce good estimates of $\A^{*}_c$.
By design, items with a low rank in the data (lower than or equal to $n^{*}$) are all relevant, and it is highly unlikely that irrelevant items have a lower estimated rank.
In fact, discarding some objects based on their frequency of inclusion in the MCMC $\A^{*}_{c,m}$ samples can result in the rejection of relevant items that appear infrequently, while irrelevant items with higher estimated rank and higher frequency of inclusion could be preferred.
The latter could be due to the algorithm's inertia in estimating the $\A^{*}_{c,m}$ sets.
Our empirical observations indicate that excluding rarely selected items negatively impacts performance under the top-rank DGP.

The case for the consistency data generating process is different.
Under this scheme, irrelevant items may show lower ranks compared to relevant items.
This problem is partially mitigated by projecting the ranks onto a lower $n^{*}$-dimensional space during rank estimation.
Consider, for example, a simple single-cluster scenario in which the data is generated without dispersion around the consensus ranking, $\vrho$, according to the Mallows model.
In the case in which, by the $m$-th iteration, lowBM3 has selected only relevant items to be included in $\A^{*}_{c,m}$, the ranks of such items projected onto the $n^{*}$-dimensional space are equal across all sample data points.
In this situation, the consensus in the $m$-th iteration matches the true consensus rank, $\vrho_{c,m} = \vrho_c$, so that any other choice for $\A^{*}_{c,m+1}$ and $\vrho_{c,m+1}$ would result in lower acceptance probabilities~\eqref{eq:CLBMM 1} and~\eqref{eq:CLBMM 2}.
This is because irrelevant items do not feature consistent ranks among assessors, leading to varying ranks in the projected space.
Therefore, the consistency property of the ranks of relevant items aids in their selection once projected in the lower-dimensional space.
Nonetheless, there remains the possibility of selecting irrelevant items in $\A^{*}_{c,m}$ for some $m$, which may have lower estimated ranks compared to relevant objects.
However, based on the above intuition, irrelevant items are expected to be selected less frequently, hence the advantage of setting $p>0$: postprocessing the MCMC results by selecting items based solely on their estimated ranks is less effective compared to the top-rank case.
Performance is then enhanced by setting a non-zero threshold in constructing the HPS set.
In unreported experiments, using a threshold of $p=20\%$, i.e., selecting only items that appear in at least 20\% of the MCMC samples, improves results substantially\footnote{Using lower or higher thresholds resulted in selecting more irrelevant items: a lower threshold causes selecting irrelevant infrequent items based on their low estimated rank; an higher threshold leads to discard many relevant items due to their relatively lower frequency of inclusion in the MCMC samples.}.

Therefore, based on experimental evidence, it is preferable to use a void threshold in the construction of the high probability set, $\A'$,  when it can be reasonably assumed that relevant items are all top-ranked, while it is better to discard rarely selected objects otherwise.
Hence, for the mixed data generating process (Setup~5) and in the application to real data, we use a threshold $p=10\%$, achieving a compromise between the two mechanisms.

\subsection{\texorpdfstring{Misspecification of $\alpha$}{Misspecification of alpha}}\label{subsec:misspecific-alpha}

In the current implementation of lowBM3, the scale parameter $\alpha_c$ is fixed.
Consequently, it is crucial to examine whether misspecification of this parameter impacts performance.
To do so, we designed two experiments to assess the performance of the method under different scenarios.
In both experiments, we simulate data as in Setup 1 (see Table~\ref{tab:sim_setups}), with different choices for the $\alpha_c$  parameters.
In the first experiment, we used $\alpha_1 = \alpha_2 = \alpha_3 =  3$, simulating a case in which all the clusters feature the same dispersion around their true consensus rank, $\vrho_c$.
In the second experiment, we set $\alpha_1 = 1$, $\alpha_2 = 3.85$, and $\alpha_3 = 7$ to create a scenario where the dispersion around the true cluster consensus rank, $\vrho_c$, decreases from cluster 1 to cluster 3.
The value of $3.85$ for cluster 2 achieves a dispersion level approximately halfway between that of clusters 1 and 3.
A display of the top-ranked items for both data sets is available in Figure~\ref{afig:alphaexp-dataplot-toprank-small}, where the different dispersion around the true consensus ranking in the different clusters is rather evident.

To evaluate the performance of lowBM3 in the same situation as in Section \ref{subsec:res-experimental-data}, we set the value of $\alpha$ in input to the method to be equal for all clusters and search for its optimal value on a grid of values $\alpha_{\text{guess}}=\{1,3,5,7,10\}$.
We set the number of clusters in input to lowBM3  equal to three, matching the true number of clusters.
Results from the experiments are reported in Table~\ref{tab:alpha_sim}.

\begin{table}[t]
  \centering
  \begin{subtable}[t]{0.45\textwidth}
    \centering
    \caption{$[\alpha_1, \alpha_2, \alpha_3]=[3,3,3]$}
    \begin{tabular}{r|cccc}
      \hline
      $\alpha_\textnormal{guess}$ & $\hat{p}_z$   & $Z_\textnormal{post}$ & $\hat{p}_{\A^{*}}$ & $d(\rho,\rho_{\textnormal{est}})$ \\
      \hline
      1                           & 0.63          & 0.36                  & 0.73               & 0.61                              \\
      3                           & 0.93          & 0.78                  & \textbf{0.90}      & \textbf{0.33}                     \\
      5                           & 0.93          & 0.89                  & 0.83               & 0.36                              \\
      7                           & \textbf{0.95} & \textbf{0.93}         & 0.70               & 0.49                              \\
      10                          & 0.67          & 0.67                  & 0.63               & 0.56                              \\
      \hline
    \end{tabular}
  \end{subtable}
  \hfil
  \begin{subtable}[t]{0.45\textwidth}
    \centering
    \caption{$[\alpha_1, \alpha_2, \alpha_3]=[1,3.85,7]$}
    \begin{tabular}{r|cccc}
      \hline
      $\alpha_\textnormal{guess}$ & $\hat{p}_z$   & $Z_\textnormal{post}$ & $\hat{p}_{\A^{*}}$ & $d(\rho,\rho_{\textnormal{est}})$ \\
      \hline
      1                           & 0.57          & 0.35                  & 0.73               & 0.68                              \\
      3                           & \textbf{0.95} & 0.77                  & \textbf{0.90}      & 0.44                              \\
      5                           & 0.92          & \textbf{0.88}         & 0.83               & \textbf{0.43}                     \\
      7                           & 0.53          & 0.54                  & 0.53               & 0.52                              \\
      10                          & 0.68          & 0.67                  & 0.67               & 0.55                              \\
      \hline
    \end{tabular}
  \end{subtable}
  \caption{Performance measures for the simulation study testing the misspecification of $\alpha$, Setup 1, three balanced clusters. Results related to dispersion parameters $\alpha_c$ being set to $[\alpha_1, \alpha_2, \alpha_3]=[3,3,3]$ are reported in the left table (a), while those for $[\alpha_1, \alpha_2, \alpha_3]=[1,3.85,7]$ are in the right table (b). The first column in each table shows the value $a_{\text{guess}}$ used to run lowBM3 (same dispersion for all three clusters). Performance metrics are shown in the remaining four columns.}
  \label{tab:alpha_sim}
\end{table}

In the first experiment, as one may expect, setting $\alpha_{\text{guess}}=3$ in lowBM3 leads to the best performance in terms of rank estimation and identification of top-ranked items.
The clustering accuracy is only slightly better using a value larger than the true dispersion parameter, namely $\alpha_{\text{guess}}=7$.
In the second experiment, a value of $\alpha_{\text{guess}}=3$ seems to achieve the best performance overall.
With this value, lowBM3 obtains the highest clustering accuracy and identifies 90\% of relevant items on average.
Rank estimation is also good, achieving the second-best performance, extremely close to the first-best.

It is worth noting that the performance of the method is not extremely affected by a misspecified value of $\alpha$, as long as this stays close to the optimal one.
Indeed, in both experiments, the performance is excellent for $\alpha_\text{guess} \in \{3, 5\}$.
The method still delivers acceptable results for $\alpha_\textnormal{guess} = 1$ and, in the first experiment, for $\alpha_\textnormal{guess} = 7$.
Other values lead to poorer performance.
Finally, upon comparing the two sub-tables, it is evident that employing a unique value of $\alpha$ in lowBM3 for all clusters does not adversely impact overall performance, even with heterogenous within-cluster dispersions.
This is true as long as an appropriate $\alpha$ is chosen, balancing the clusters' heterogeneity.

\subsection{\texorpdfstring{Estimating $C$}{Estimating C}}\label{subsec:sim_clust_est}
As the lowBM3 method is based on a finite mixture model, it requires specifying the number of clusters in advance, which is typically unknown in practical applications.
In prior work, we proposed employing an elbow criterion for Bayesian model selection~\cite{vitelli2018,sorensen2019}, which is an adaptation of the frequentist approach of computing the within-cluster sum-of-squares for various values of $C$ to identify a plateau beyond which adding further clusters does not significantly improve performance.
To verify the efficacy of this criterion in the context of lowBM3, we examine the model's capacity to decide the correct number of groups through a simulation study.
In particular, we fit lowBM3 on the same data set varying the number of clusters in input to the method, while fixing other input parameters.
We use the posterior mean of the within-cluster footrule sum-of-distances (MWCD)  to evaluate the results:
\begin{equation}
  \mathrm{MWCD}(C) = \mathbb{E}_p\left[\sum\limits_{c=1}^C\sum\limits_{z_j=c}d_{A^{*}_{c}}(\textbf{R}_j, \vrho_{c})\right],
  \label{eq:mwcd}
\end{equation}
where $\mathbb{E}_p[\cdot]$ denotes the expected value under the posterior distribution
$P(\{\vrho_c, \A^{*}_c\}_{c=1}^C | \boldsymbol{R}_1,\ldots, \boldsymbol{R}_N)$.
The expected outcome is a decreasing trend in the MWCD as the number of clusters increases until the correct number of clusters is reached, at which point the MWCD should stabilize.
The ``elbow-plot'' provides a visual indication of the optimal number of clusters.

\begin{figure}[!htbp]
  \centering

  \begin{subfigure}{\textwidth}
    \centering
    \caption{Results from Setup 1, 3 true clusters}
    \includegraphics[width=\textwidth]{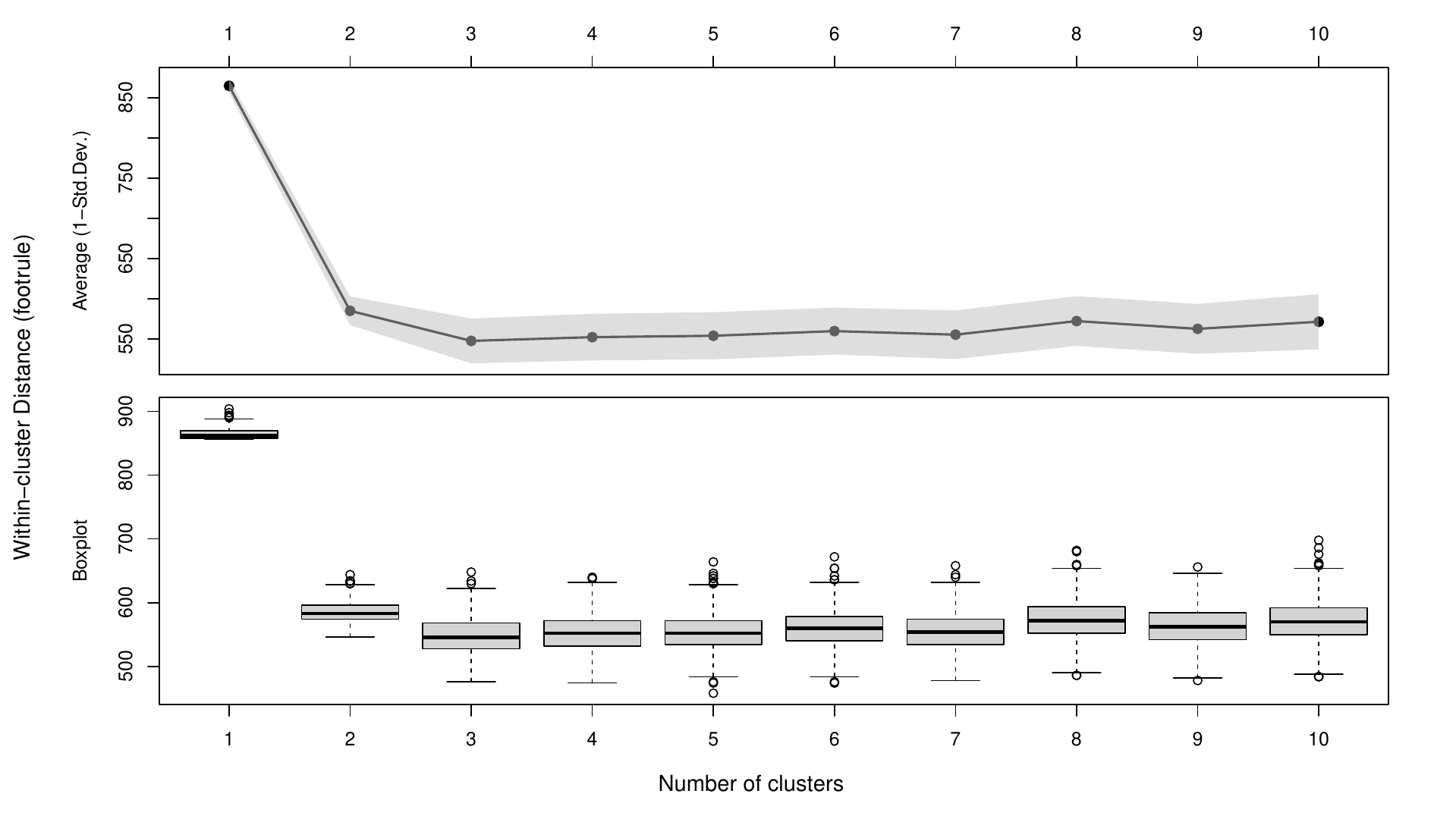}
  \end{subfigure}

  \vfill

  \begin{subfigure}{\textwidth}
    \centering
    \caption{Results from Setup 5, 5 true clusters}
    \includegraphics[width=\textwidth]{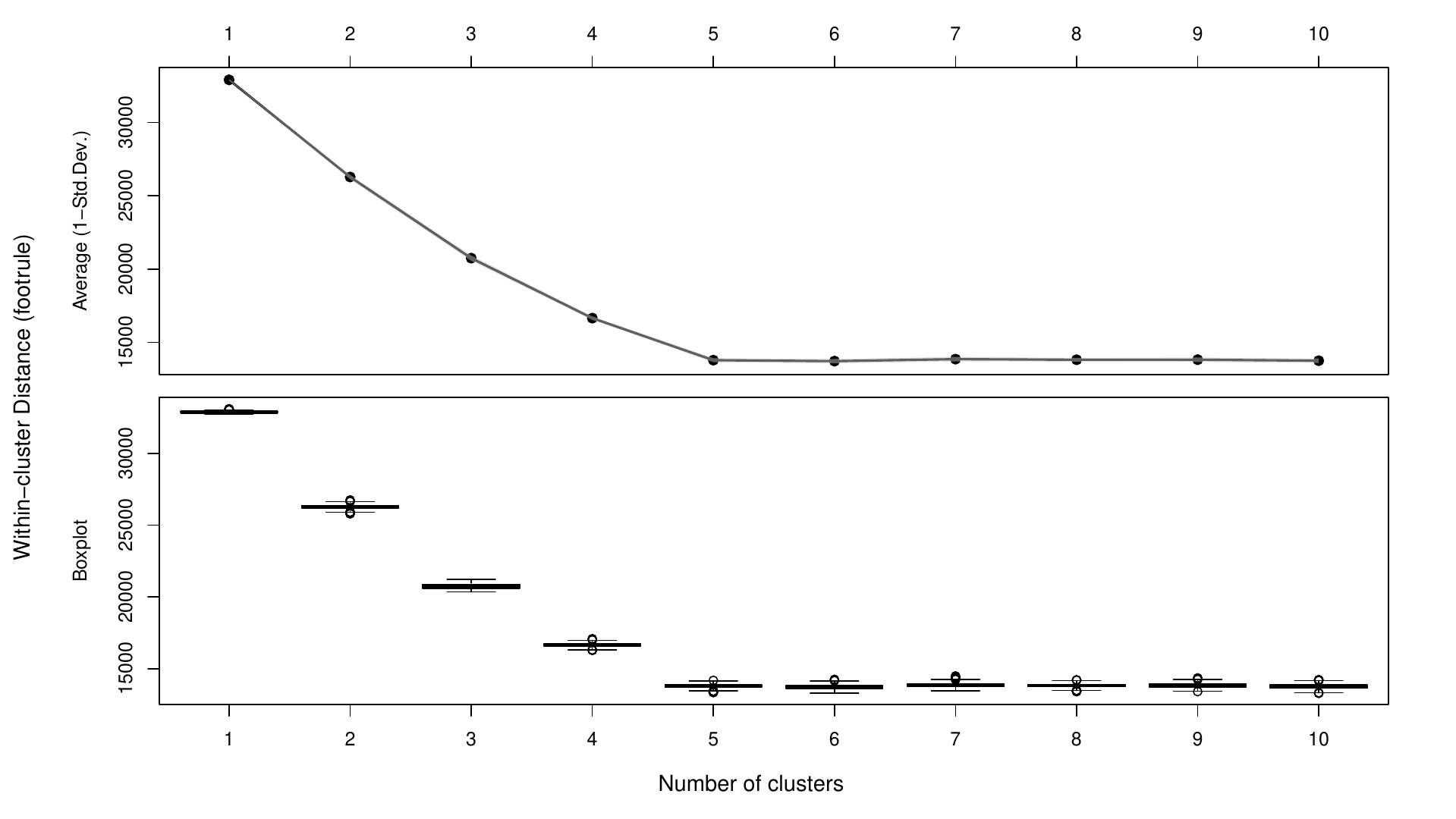}
  \end{subfigure}

  \caption{Elbow-plots of the within-cluster sum-of-distances for the different number of clusters set for lowBM3. The top and bottom panels show results for the top-rank small-scale and large-scale designs, respectively. In each panel, the top graph shows the MWCD$\pm$ the standard deviation of the same quantity (shaded area), while the bottom graph shows a boxplot of the within-cluster sum-of-distances posterior distribution.}
  \label{fig:wcd-toprank}
\end{figure}

In this simulation study, we run lowBM3 with different numbers of clusters, $C \in \{1,2,\ldots,10\}$, on data generated from both Setup 1 and Setup 3.
The true number of clusters is 3 and 5, respectively.
Figure~\ref{fig:wcd-toprank} displays the results for both data sets, with Setup 1 depicted in the top panel and Setup 3 in the bottom panel.
Each panel is composed of two graphs: the upper graph shows the MWCD at different values of $C$, while the bottom graph shows a boxplot of the posterior distribution of this same quantity before taking the expectation.

In both setups, there is a clear ``elbow'' in correspondence of the true number of clusters, 3 for Setup~1, and 5 for Setup~3.
The elbow-plot shows that the variability of the within-cluster distance is not greatly affected by the number of clusters in input to lowBM3 being larger than necessary, and that the results are very similar to one another once the true number of clusters is reached\footnote{The variability for the results of Setup 5 is less visible due to scaling issues.}.
This is likely due to the fact that lowBM3 can identify empty clusters.
When an overly complex clustering structure is fit to the data, some clusters may remain empty with no assessors assigned to them.
Consequently, overly complex solutions might become nearly identical if they share the same number of non-empty clusters.
In summary, the elbow criterion seems to be a suitable approach in our context to determine the optimal number of clusters.

\section{Case study: Breast Cancer RNA-seq data}\label{sec:application}

We obtained RNA-seq data from breast cancer (BRCA) patients from the TCGA Data Portal\footnote{TCGA Data Portal: \href{https://tcga-data.nci.nih.gov/tcga/}{https://tcga-data.nci.nih.gov/tcga/}. The BAM files were downloaded from CGHub (www.cghub.ucsc.edu), and converted to FASTQs, where all samples were processed similarly aligning reads to the hg19 genome assembly using MapSplice~\cite{mapsplice2010}. Gene expression was quantified for the transcript models corresponding to the TCGA GAF 2.13, using RSEM4 and normalized within-sample to a fixed upper quartile~\cite{rsem2011}.}.
As lowBM3 assumes complete data, we included in the analyses only the genes showing less than 50$\%$ missing values over the samples, imputing missing values using k-nearest neighbor averaging~\cite{troyanska2001}.
The result was a final dataset of ultra-high-dimension with $n=15348$ genes and $N=845$ patients.
Finally, we transformed the RNA-seq dataset to ranks, by ordering each RNA-seq patient measurement according to their continuous value (i.e.,\ the highest gene expression value gets 1, and the lowest gets $n$).
In addition, we also obtained clinical information connected to the patients, among which the PAM50 subtype~\cite{sorlie2001, parker2009} was further used in results interpretation.

\begin{figure}[!htbp]
  \centering
  \includegraphics[width=\textwidth]{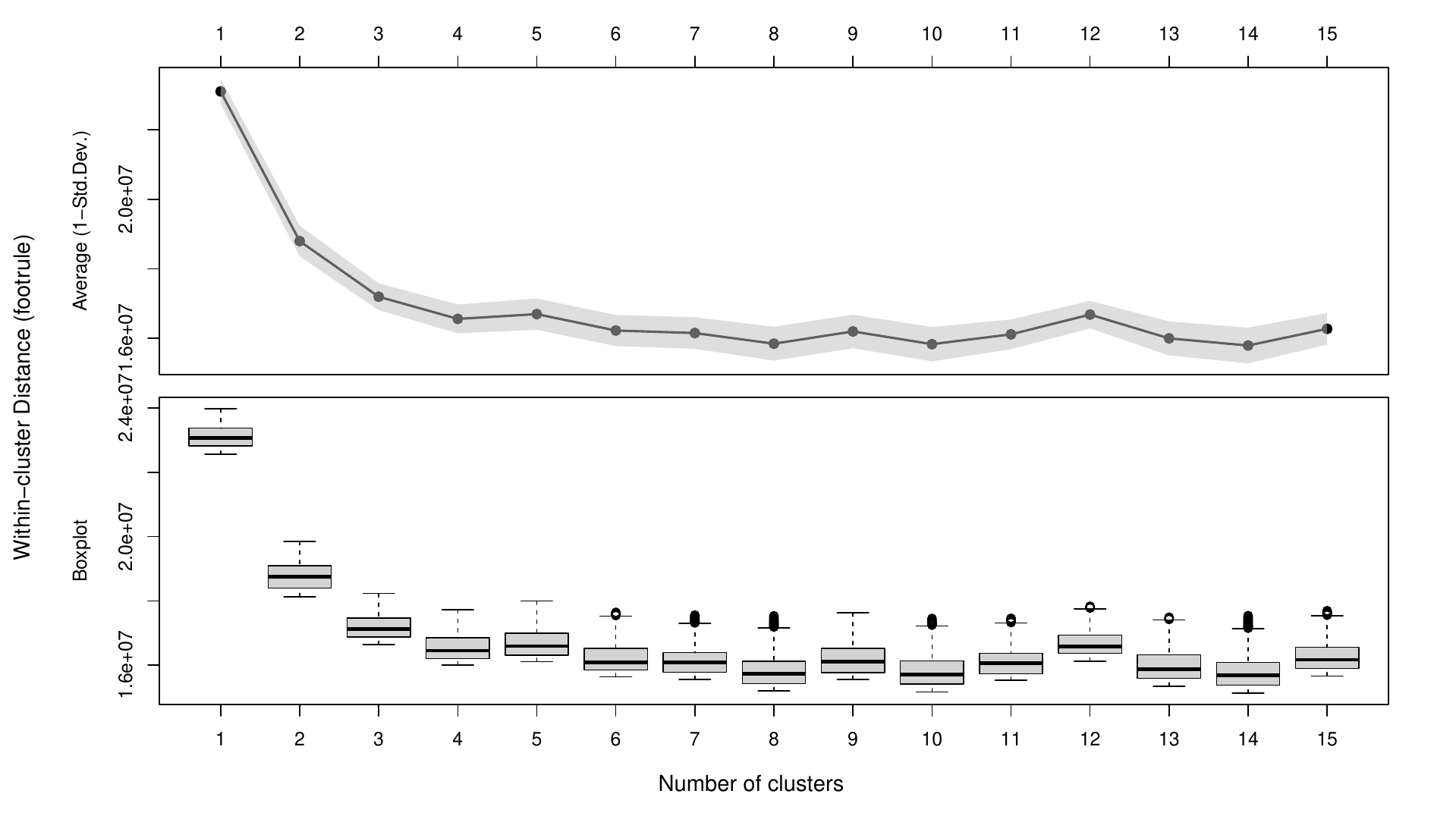}
  \caption{Results of the BRCA case study.
    Posterior distribution of the within-cluster footrule distances (WCD) of the gene expression individual patient ranks from the corresponding cluster consensus (y-axis), for different choices of $C$ (x-axis) when using lowBM3. Top panel: mean WCD (MWCD, dots), $\pm$ 1-standard-deviation (shaded area); bottom panel: boxplots of the WCD posterior distribution for the different values of $C$.}
  \label{fig:WCD-brca}
\end{figure}

We set the lowBM3 parameters to the following values: $n^{*}=500$, $l=\textnormal{round}(n^{*}/5)=100$, $L_c=1$ for all $c$, $\psi=N/C,$ and set $\alpha=1$ for the burn-in period, to then increase it to $\alpha=5$ for the remaining MCMC iterations.
This was done to explore the permutation space sufficiently well at the beginning of the Markov chain iterations, as a smaller $\alpha$ induces more variability in the sampling, and to facilitate convergence after the burn-in by reducing the variance (i.e., by increasing $\alpha$).
The method was run for a range of possible values for the number of clusters $C \in \{1, \ldots, 15\}$, with $M=2\cdot10^5$ MCMC iterations.
The first 50$\%$ of the MCMC samples were discarded as burn-in, and the chain was thinned by retaining every 100th sample for posterior analysis.
To select the optimal $C$, we inspected the elbow-plot of the within-cluster distances (WCD) reported in Figure~\ref{fig:WCD-brca}, using the approach described in Section~\ref{subsec:sim_clust_est}.

Figure~\ref{fig:WCD-brca} shows a smooth elbow plot, where it is possible to identify only one abrupt decrease corresponding to $C=4$: here the mean WCD drops to visibly smaller values than those obtained for $C<4$, while the decrease is less pronounced for $C>4$.
Other significant yet less pronounced decreases in WCD occur at $C=8$, $C=10$ and $C=14$.
In fact, these three cluster configurations should be preferred to $C=4$ as they achieve smaller WCD values, both in terms of mean ($\sim 1.579\times 10^7$ and $\sim 1.655\times 10^7$, respectively) and median ($\sim 1.568\times 10^7$ and $\sim 1.645\times 10^7$, respectively).

We note that the mean WCD does not monotonically decrease with $C$. This has to be expected in Bayesian clustering, as the posterior samples from sub-optimal large-$C$ solutions might not easily reach convergence, so that a poorly estimated cluster consensus after burn-in might lead to the WCD being larger on average than for solutions with smaller $C$s.

It is also worth noting that the final solution of lowBM3 does not force all $C$ clusters to be non-empty.
Indeed, the maximum number of non-empty clusters that are discovered in our experiments is 6, which is obtained in all optimal configurations with $C=8$, $C=10$, and  $C=14$.
Figure~\ref{afig:alluvial} shows alluvial plots for the patient-to-cluster assignments, both along all the clustering solutions, and focusing on the 6 non-empty cluster solutions, showing an excellent adherence of non-empty-cluster assignments.
As all 6 non-empty clusters solutions obtained with $C=8$, $C=10$, and $C=14$ are similar to one another both in terms of patients-to-cluster assignments and in terms of discovered genes, and since all such solutions result in the lowest WCD, we decided to select and further analyze the lowBM3 solution obtained with $C=10$, as it achieves a slightly lower WCD value than the other two.

\begin{figure}[!htbp]
  \centering
  \includegraphics[width=\linewidth]{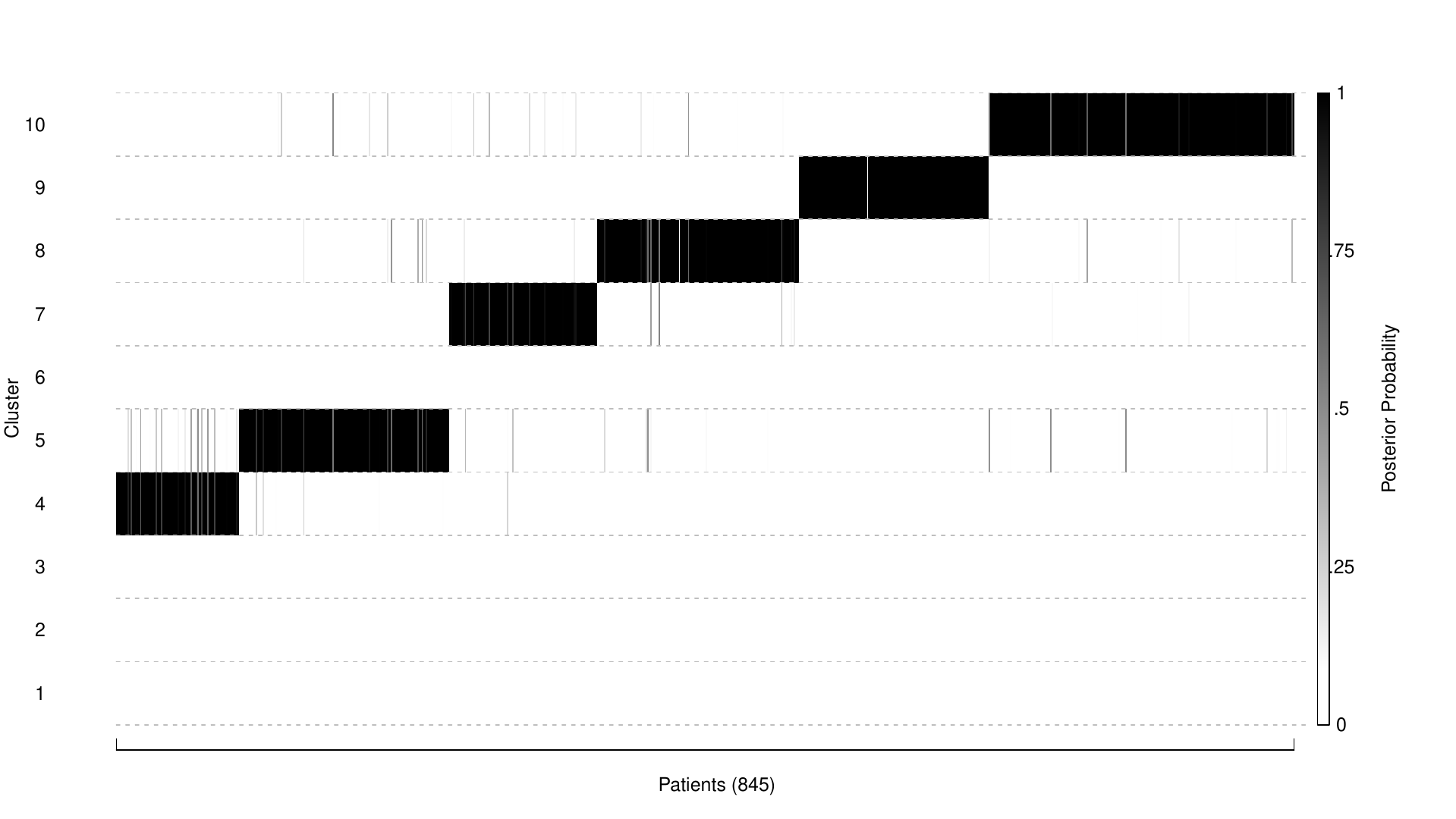}
  \caption{Results of the BRCA case study.
    Heatplot of posterior probabilities for all $N=845$ patients (x-axis) of being assigned to each of 10 clusters (y-axis). Posterior probabilities are represented in shades of gray, with darker colors corresponding to higher posterior probability of being assigned to the cluster.
  }
  \label{fig:assessprob-BRCA-10}
\end{figure}

The stability of the cluster assignments for the $C=10$ solution is visually presented in Figure~\ref{fig:assessprob-BRCA-10}, showing a heatplot of the posterior probabilities of all 845 patients (on the x-axis) of being assigned to each of the $C=10$ clusters (on the y-axis).
The vast majority of the individuals' posterior probabilities of assignment are polarized around a single cluster, suggesting a persistent and stable pattern in the cluster assignments.
This confirms the model ability to capture the inherent structures in the data with low uncertainty.
The average acceptance probabilities over the clusters for $\bm{\rho}_c$ and $\A_c^{*}$ were respectively 29.3\% and 5.7\% during the burn-in phase, and dropped to 9.9\% and 0.4\% post burn-in, indicating sufficient mixing in the MCMC, especially in terms of $\bm{\rho}_c$, which has previously shown consistently low acceptance rates in the BMM mixture model.

\begin{figure}[!htbp]
  \centering
  \includegraphics[width=\textwidth]{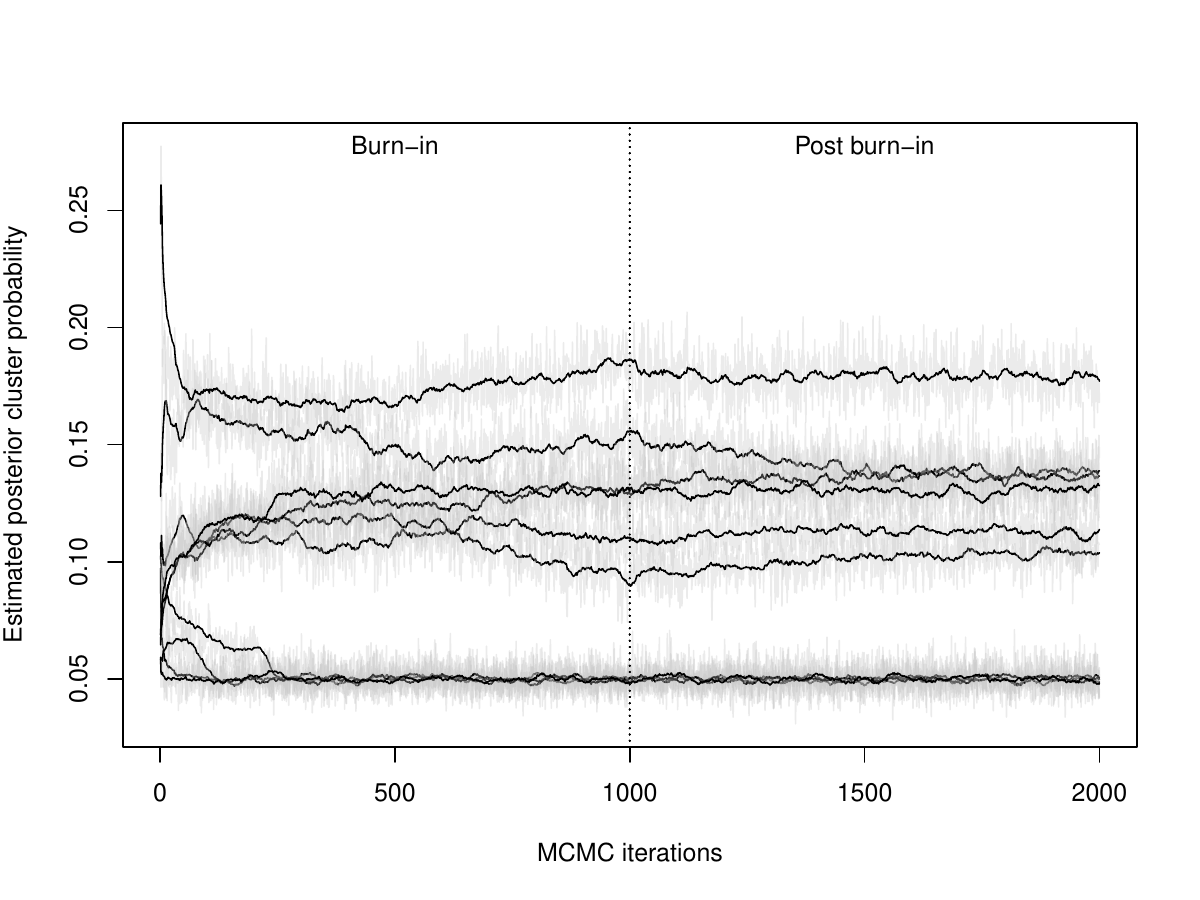}
  \caption{Results of the BRCA case study. Trace plots of the mixture parameters $\tau_c$, for $c=1,\ldots,C$, obtained via lowBM3 with $C=10$ on the y-axis, with MCMC iterations on the x-axis.. The graph shows one trace plot for each cluster (gray), together with a 25-points rolling average (black solid lines). Burn-in and post-burn-in observations are separated by a vertical line. The entire MCMC chain (burn-in and post-burn-in) is thinned, taking every other 100th observation.
  }
  \label{fig:tautrace-BRCA}
\end{figure}

The number of MCMC iterations $M$ is deemed sufficient to achieve convergence, as confirmed by the convergence plot for $\{\tau_c\}_{c=1,\ldots, C}$, shown in Figure~\ref{fig:tautrace-BRCA}.
The figure shows the trace plot for the estimated posterior cluster probabilities, for all the clusters, along the MCMC chain, including both burn-in and post burn-in observations.
The chain is thinned taking every other 100th iteration, and the traces are drawn together with a 25-points rolling average, to ease visualization.
From the figure, it is clear that the mixture parameters $\tau_c$ undergo initial adjustments during the burn-in phase, quickly stabilizing around their final level.
It is also possible to notice that for 4 out of 10 clusters, $\tau_c$ drops soon to a value close to 0, motivating the empty clusters in the final solution.
Convergence is also demonstrated in Figure~\ref{afig:rhoAstarTrace-BRCA}, which shows the MCMC trace plots for the average gene rank in the estimated $\hat{\A}^*_c$ and the average fraction of selected genes that are also identified as top-ranked genes (i.e., $\A_{m,c}^* \cap \hat{A}_c^*$).

There are four widely clinically accepted and implemented intrinsic subtypes of breast cancer, named PAM50 subtypes~\cite{parker2009}: Luminal A (LumA), Luminal B (LumB), Her2-enriched (Her2) and basal-like (Basal).
The originally defined normal-like (Normal) breast cancer subtype is now less frequently used~\cite{weigelt2010, elloumi2011, bastien2012}.
Thus, we can use the four subtypes to interpret the clustering results.
In Table~\ref{tab:BRCA_contingency} a comparison of MAP cluster assignments from lowBM3 with $C=10$ and the associated four PAM50 subtypes is reported.

\begin{table}[htbp!]
  \centering
  \renewcommand{\arraystretch}{1.5}
  \begin{tabular}{r|rrrrrr|r}
    \hline
    PAM50 Subtype & Cl. 4 & Cl. 5 & Cl. 7 & Cl. 8 & Cl. 9 & Cl. 10 & Row Total \\
    \hline
    Basal         & 2     & 0     & 1     & 3     & 134   & 0      & 140       \\
    Her2          & 0     & 0     & 15    & 45    & 1     & 6      & 67        \\
    LumA          & 65    & 142   & 58    & 42    & 0     & 113    & 420       \\
    LumB          & 2     & 9     & 29    & 54    & 0     & 100    & 194       \\
    Normal        & 19    & 0     & 3     & 1     & 1     & 0      & 24        \\
    \hline
    Column Total  & 88    & 151   & 106   & 145   & 136   & 219    & 845       \\
    \hline
  \end{tabular}
  \caption{Results of the BRCA case study. Contingency table of PAM50 Subtype and MAP cluster assignments estimated via lowBM3. The PAM50 Subtypes are shown on the rows, while the 6 non-empty clusters from lowBM3 with $C=10$ are reported on the columns. Each cell shows the total number of patients falling within a given PAM50-cluster combination. Rows and columns marginals are also shown.}
  \label{tab:BRCA_contingency}
\end{table}

The table indicates that a clear correspondence between clusters and subtypes exists only for the Basal subtype and Cluster~9, while other clusters display more substantial overlap between subtypes.
The Her2 and Normal subtypes are primarily associated with Cluster~8 and Cluster~4, respectively; however, both clusters also include patients from other subtypes.
The LumA and LumB subtypes are distributed across three to four clusters, potentially reflecting relevant subgroups within these subtypes. Cluster~4 predominantly comprises LumA patients, but it also includes individuals from the Normal subtype, as mentioned earlier. Cluster~5, in contrast, is almost exclusively composed of LumA patients.
Cluster~7 contains a mixture of Her2, LumB, and LumA subtypes, with the latter being the most prominent. Cluster~8 has the highest proportion of Her2 patients but also includes significant representation from LumA and LumB subtypes, resulting in a balanced composition across the three. Similarly, Cluster~10 exhibits a balanced distribution of LumA and LumB patients, which have been previously observed to be rather similar~\cite{aure_vitelli2017}.
In general, the PAM50 subtype classification is still under scrutiny and other stratifications of breast cancer patients have been proposed (see e.g.~\cite{curtis2012genomic,prat2015clinical}), which may partially explain the observed mixing of subtypes across different clusters.

It is generally of great interest for biological interpretation, in addition to the mere patients subtyping, the possibility to also recover biomarkers capable of further characterising the subtypes at the molecular level~\cite{goel2026uncovering}. This is one of the crucial advantages of using lowBM3 for clustering transcriptomics data as compared to non-rank-based methods, for instance. In Table~\ref{tab:genes-postprob-BRCA-C10}, we investigate the cluster-specific variable selection results by inspecting the top-ranked genes in the  posterior distribution of $\vrho_c,$ when ordering them according to their posterior probability of being assigned a top-10 rank (i.e., rank lower or equal to 10) in the consensus of each cluster, for all the 6 non-empty clusters in the selected clustering solution with $C=10$.

\begin{table}[htbp!]
  \centering
  \renewcommand{\arraystretch}{1.2}
  \begin{tabular}{r|llllll}
    \hline
    & Cl. 4 & Cl. 5    & Cl. 7    & Cl. 8    & Cl. 9   & Cl. 10   \\
    \hline
    1  & PLIN1 & GATA3    & GATA3    & GATA3    & GABRP   & ESR1     \\
    2  & GFRA1 & CILP     & AZGP1    & CILP     & SFRP1   & NKAIN1   \\
    3  & APOD  & AZGP1    & ESR1     & COL11A1  & MIA     & MAPT     \\
    4  & GATA3 & PGR      & MAPT     & LRRC15   & CHI3L2  & SCUBE2   \\
    5  & CILP  & PRLR     & GFRA1    & PRLR     & CRABP1  & RET      \\
    6  & AZGP1 & GREB1    & CILP     & MAPT     & TTYH1   & GATA3    \\
    7  & PI16  & FSIP1    & CHAD     & KIAA1324 & S100A1  & GREB1    \\
    8  & THBS4 & AFF3     & SERPINA3 & APOD     & KRT16   & SERPINA3 \\
    9  & F2RL2 & CLSTN2   & EFHD1    & COL10A1  & IL12RB2 & SLC1A2   \\
    10 & MGP   & SERPINA3 & TFF3     & SCUBE2   & SOX11   & TRPS1    \\
    \hline
  \end{tabular}
  \caption{Results of the BRCA case study: genes ranked according to posterior probability of being assigned a rank lower or equal to 10 in the cluster consensus $\vrho_c$, for each of the 6 non-empty clusters in the selected clustering solution ($C=10$). Within each cluster, the genes are ordered by their posterior probability of receiving a rank lower or equal to 10 in decreasing order, and the top-10 genes in this ordering are reported in the table.}
  \label{tab:genes-postprob-BRCA-C10}
\end{table}

Here we notice that GATA3 appears on top of the ranking in 5 out of 6 clusters (absent in Cluster~9), and in top position for Cluster~5, Cluster~7 and Cluster~8, which is reasonable as this gene is a known driver gene in hormone receptor positive (luminal) breast cancer.
Furthermore ESR1, encoding the estrogen receptor (ER), also shows up in top position in Cluster~10, and among the first three positions in Cluster~7.
These two clusters are composed of both LumA and LumB subtypes.
ESR1 is indeed known to be a key driver of luminal breast tumors, and is believed to be epigenetically regulated in breast cancer~\cite{fleischer2017dna}.

To gain insight into the results provided by the method and add to the biological interpretation, we selected the top-10 genes in $\mathcal{\hat{A}}_c$ for each of the 6 non-empty clusters in the selected clustering solution ($C=10$), and analyzed them with a Gene Set Enrichment Analysis (GSEA)~\cite{ mootha2003, subramanian2005} using the Molecular Signatures Database v2023.2.Hs (MSigDB) H and C5 collections. The results from such GSEA are reported in Table~\ref{tab:GSEA}.
By inspecting Table~\ref{tab:GSEA}, we see that the hallmark gene sets Estrogen Response Early and Estrogen Response Late are both enriched in Clusters 7 and 10, and the former is also enriched in Cluster 5, confirming that these key pathways are activated in tumors in these clusters.
In Cluster 5, pathways associated to adhesion and extracellular matrix are enriched, potentially reflecting that Luminal A tumors (which nearly entirely compose Cluster 5) have a more rigorous luminal structure and grow more orderly.
Pathways enriched in Cluster 9 include regulation of cell cycle and proliferation, reflecting the high aggressiveness of basal-like tumors~\cite{perou2000}.
For Cluster 8, the gene set Epithelial to Mesenchymal Transition is enriched, possibly reflecting the increased invasive potential of these tumors.
Taken together, the GSEA results illustrate that our method identifies genes that are important for breast cancer development and progression.

To assess the biological similarity of the clusters, we computed the overlap of top-ranked genes between the identified clusters. Here, we used the genes that had an 80\% probability of being ranked top 100 (Figure~\ref{fig:genes_upset_inclusive}). Between 40 and 20 genes were shared between pairs of clusters with luminal tumors, and between 19 and 10 genes were shared between three of the luminal clusters. Conversely, genes in cluster 9 (basal-like tumors) shared almost no genes with other clusters. Together, this illustrates that our method captures similar but non-identical biology in the five clusters associated to hormone receptor positive breast cancer.

\begin{figure}[!htbp]
  \centering
  \includegraphics[width=\linewidth, height=0.75\textheight, keepaspectratio]{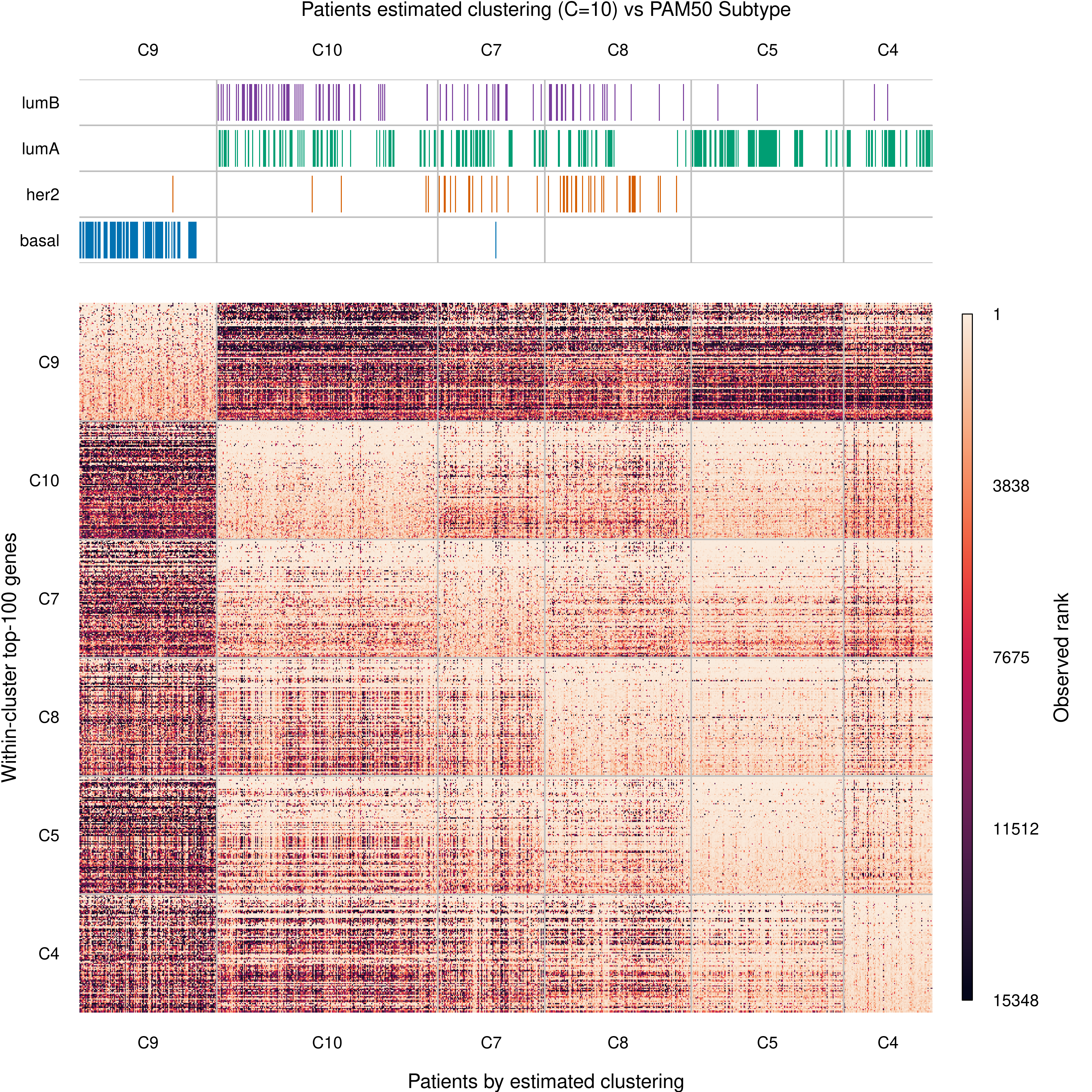}
  \caption{Results of the BRCA case study.
    Bottom panel: heatmap of the ranked RNA-seq data for all patients (columns), displaying the top-100 genes (rows) selected based on posterior probabilities of assignment to a rank of 100 or lower in $\vrho_c$. Both columns and rows are ordered and grouped according to the six non-empty clusters identified in the optimal clustering solution ($C=10$). Top panel: estimated MAP cluster assignments $\hat{z}_1,\ldots,\hat{z}_N$ stratified by PAM50 subtype, with patient ordering consistent with the heatmap below.
  }
  \label{fig:BRCA_annotated}
\end{figure}

The ranked RNA-seq values associated to the union of the top-100 genes in terms of posterior probability of having a top-100 rank in $\vrho_c$ for each of the 6 non-empty clusters in the selected clustering solution ($C=10$) are also visualized in an annotated heatmap in Figure~\ref{fig:BRCA_annotated}. It is clear from the heatmap that, even with a much broader gene selection, the non-empty groups can be characterized in terms of block structure in the associated RNA-seq values, which adds to the interpretation of each group in terms of a more flexible overlapping variable selection than what is allowed by classical bi-clustering methods \cite{nicholls2021comparison}.

In conclusion, the results from the BRCA analysis demonstrate how lowBM3 can successfully identify essential genes influencing cancer development within each cluster in a completely unsupervised framework and from a genome-wide ultra-high-dimensional dataset.
This proves that this method can be used for signature discovery in cancer genomics, as the proposed variable selection strategy acts genome-wide, and therefore important unknown drivers of the disease could potentially be discovered.

\begin{table}[htbp!]
  \scriptsize
  \centering
  \caption{Results of the BRCA case study: top-10 results from the GSEA analysis on the top-50 genes in $\mathcal{\hat{A}}_c$ for the selected clustering solution ($C=10$). Results are displayed for the six non-empty clusters.}
  \label{tab:GSEA}
  \begin{tabular}{llrr}
    \toprule
    Cluster &Gene Set Name                                            & \makecell{\# Genes\\in overlap} & p-value \\
    \midrule

    Cluster 4 &External Encapsulating Structure (GOCC)              & 30                  & 5.5e-30 \\
    &Collagen Containing Extracellular Matrix (GOCC) & 25 & 3.7e-26 \\
    &Extracellular Matrix Structural Constituent (GOMF) & 16 & 3e-20 \\
    &Biological Adhesion (GOBP) & 32 & 4.4e-20 \\
    &Regulation Of Multicellular Organismal Development (GOBP) & 30 & 9.8e-19 \\
    &Circulatory System Development (GOBP) & 26 & 7.1e-17 \\
    &Enzyme Linked Receptor Protein Signaling Pathway (GOBP) & 25 & 1.9e-16 \\
    &Signaling Receptor Binding (GOMF) & 29 & 6.2e-16 \\
    &Animal Organ Morphogenesis (GOBP) & 24 & 7.4e-16 \\
    &External Encapsulating Structure Organization (GOBP) & 17 & 1.1e-15 \\

    \\

    Cluster 5 &External Encapsulating Structure (GOCC)              & 31                  & 7.4e-32 \\
    &Epithelial Mesenchymal Transition (HALLMARK) & 23 & 3.7e-31 \\
    &Collagen Containing Extracellular Matrix (GOCC) & 27 & 1.7e-29 \\
    &Extracellular Matrix Structural Constituent (GOMF) & 19 & 1.4e-25 \\
    &External Encapsulating Structure Organization (GOBP) & 22 & 9.2e-23 \\
    &Biological Adhesion (GOBP) & 28 & 2e-16 \\
    &Circulatory System Development (GOBP) & 25 & 3.8e-16 \\
    &Estrogen Response Early (HALLMARK) & 13 & 1.1e-14 \\
    &Vasculature Development (GOBP) & 20 & 3.2e-14 \\
    &Structural Molecule Activity (GOMF) & 19 & 3.3e-14 \\
    \\

    Cluster 7 &Estrogen Response Late (HALLMARK)               & 14                  & 6e-17   \\
    &Estrogen Response Early (HALLMARK) & 13 & 2.2e-15 \\
    &External Encapsulating Structure (GOCC) & 14 & 8.6e-11 \\
    &Collagen Containing Extracellular Matrix (GOCC) & 11 & 6.1e-09 \\
    &Regulation Of Cell Differentiation (GOBP) & 19 & 8.2e-09 \\
    &Somatodendritic Compartment (GOCC) & 14 & 1.1e-08 \\
    &Embryonic Skeletal System Development (GOBP) & 7 & 1.4e-08 \\
    &Cell Surface (GOCC) & 14 & 2.1e-08 \\
    &Neuron Differentiation (GOBP) & 17 & 2.2e-08 \\
    &Neurogenesis (GOBP) & 18 & 4.6e-08 \\
    \\

    Cluster 8 &External Encapsulating Structure (GOCC)              & 39                  & 6.5e-46 \\
    &Collagen Containing Extracellular Matrix (GOCC) & 34 & 4.5e-42 \\
    &Epithelial Mesenchymal Transition (HALLMARK) & 27 & 1.5e-39 \\
    &Extracellular Matrix Structural Constituent (GOMF) & 25 & 1.3e-37 \\
    &External Encapsulating Structure Organization (GOBP) & 28 & 6.5e-33 \\
    &Biological Adhesion (GOBP) & 35 & 4.7e-25 \\
    &Structural Molecule Activity (GOMF) & 25 & 3e-22 \\
    &Collagen Binding (GOMF) & 11 & 1.8e-17 \\
    &Collagen Fibril Organization (GOBP) & 10 & 1.3e-16 \\
    &Glycosaminoglycan Binding (GOMF) & 14 & 6.9e-16 \\
    \\

    Cluster 9 &Cell Cycle (GOBP)                    & 28                  & 4.5e-15 \\
    &Cell Population Proliferation (GOBP) & 27 & 1e-13 \\
    &Cell Cycle Process (GOBP) & 23 & 3.7e-13 \\
    &Epithelium Development (GOBP) & 22 & 4e-13 \\
    &Supramolecular Complex (GOCC) & 21 & 4.9e-12 \\
    &Mitotic Cell Cycle (GOBP) & 19 & 6.6e-12 \\
    &G2M Checkpoint (HALLMARK) & 10 & 7.1e-11 \\
    &Cytoskeleton Organization (GOBP) & 20 & 1.2e-10 \\
    &Mitotic Nuclear Division (GOBP) & 11 & 1.8e-10 \\
    &Chromosome (GOCC) & 22 & 2.6e-10 \\
    \\

    Cluster 10 &Estrogen Response Late (HALLMARK)                & 19                  & 5.5e-25 \\
    &Estrogen Response Early (HALLMARK) & 18 & 2.8e-23 \\
    &Gland Morphogenesis (GOBP)                   & 9 & 9.5e-12 \\
    &Neurogenesis (GOBP) & 21 & 3.3e-10 \\
    &Mammary Gland Development (GOBP) & 8 & 1.5e-09 \\
    &Somatodendritic Compartment (GOCC) & 15 & 1.9e-09 \\
    &Secretion (GOBP) & 19 & 2.8e-09 \\
    &Neuron Differentiation (GOBP) & 18 & 5.5e-09 \\
    &Gland Development (GOBP) & 11 & 6.7e-09 \\
    &Positive Regulation Of Molecular Function (GOBP) & 20 & 8.2e-09 \\
    \bottomrule
  \end{tabular}
\end{table}

\section{Discussion and conclusions}\label{sec:discussion}

In this paper, we have expanded the scope of the lowBMM method introduced in~\cite{eliseussen2022} to accommodate the existence of multiple clusters within a data set.
To achieve this, we combined the clustering technique utilized in BMM with the framework of lowBMM.
The resulting model, called lowBM3, serves as a joint Bayesian clustering and variable selection method for high-dimensional heterogeneous ranked data.

We conducted several simulation studies to evaluate lowBM3,  to assess its performance and effectiveness in identifying clusters and estimating the relevant variables.
Results from the simulation experiments demonstrated the model capability to accurately determine the number of clusters and effectively differentiate between them.
Moreover, lowBM3 exhibited robustness in separating clusters based on their distinct data-generating processes.
Notably, the accuracy in the clustering had minimal impact on the model capability to estimate the relevant variables accurately.
Finally, a case study on RNA-seq data from breast cancer patients demonstrated how the method can be successfully used in the context of clustering transcriptomic data genome-wide, to estimate the patients' molecular heterogeneity and to detect the relevant genes characterizing each cluster.
It is clear that lowBM3 successfully identifies essential genes influencing cancer development in a completely unsupervised way, such that the method proves to be useful for signature discovery and subtyping in cancer genomics.

The current implementation of lowBM3 faces few limitations, including relying on complete data, and assuming a fixed scale parameter $\alpha$ and a fixed set dimension $n^{*}$.
Concerning the former limitation, even if the method does not currently handle missing data, a data augmentation step as presented in~\cite{vitelli2018} could be easily included inside the inferential framework to solve this issue.
For what concerns $\alpha$ and $n^{*}$, instead, including their estimation in the inferential procedure would heavily influence computational choices, and approaches to deal with these aspects are left for future speculation.

In terms of the tuning parameters $l$ and $L_c$, it is worth exploring alternative methods for incorporating their tuning into the lowBM3 algorithm.
The current implementation selects the parameter $L_c$ uniformly, with an upper limit of $L$.
The choice to use a uniform selection for the parameter $L_c$ was made due to its implementation simplicity and cost-effectiveness.
This approach enables better exploration of the parameter space without significantly compromising the convergence speed of the algorithm.
While tuning the parameter $L_c$ does not directly impact the acceptance probability in Equation~\eqref{eq:CLBMM 2}, it can have a direct effect on the convergence speed and exploration of the parameter space.
Therefore, selecting an appropriate value for $L_c$ is important to ensure efficient and effective sampling during the MCMC process.
Future versions of lowBM3 could investigate different techniques for tuning $l$ and $L_c$, such as adaptive sampling methods \cite{andrieu2008tutorial,zhu2019sample,bell2024adaptive}.

The clustering scheme implemented in this paper is a finite mixture, so the true number of clusters $C$ is not estimated by lowBM3.
Instead, we ran the method for several values of $C$ and evaluated the results in postprocessing, which gave reliable estimates of the true number of clusters in simulation experiments as well as in the presented case study, when essentially the same clustering structure was detected with different values of $C$.
However, estimating $C$ would be greatly beneficial, and could be implemented with an infinite mixture using Bayesian nonparametrics techniques, or via Mixture of Finite Mixtures approaches.


In conclusion, lowBM3 is a novel rank-based clustering model that handles variable selection, providing a scalable Bayesian framework for the unsupervised analysis of ultra-high-dimensional ranking data.
To the best of our knowledge, no rank-based model is currently available that can simultaneously handle clustering and unsupervised item selection in a scalable manner for the typical data dimension observed in -omics applications. In this sense, the proposal contributes a tremendously impactful contribution to the scientific literature.

\bibliographystyle{plain}
\bibliography{bib-v2}

\appendix
\setcounter{table}{0}
\setcounter{figure}{0}
\setcounter{page}{0}
\renewcommand{\thetable}{A\arabic{table}}
\renewcommand{\thefigure}{A\arabic{figure}}

\clearpage
\thispagestyle{empty}
\begin{minipage}[c][0.5\textheight][c]{0.9\textwidth}
  \begin{center}
    \bfseries
    \huge Supplementary Material \\[1em]
    \Large Bayesian genome-wide clustering and variable selection of transcriptomic data via rank-based mixtures\\
  \end{center}
\end{minipage}

\clearpage
\footnotesize

\section{Noisy data generating processes}\label{app:consistencyDGP}
\begin{figure}[t]
  \centering
  \includegraphics[width=\textwidth]{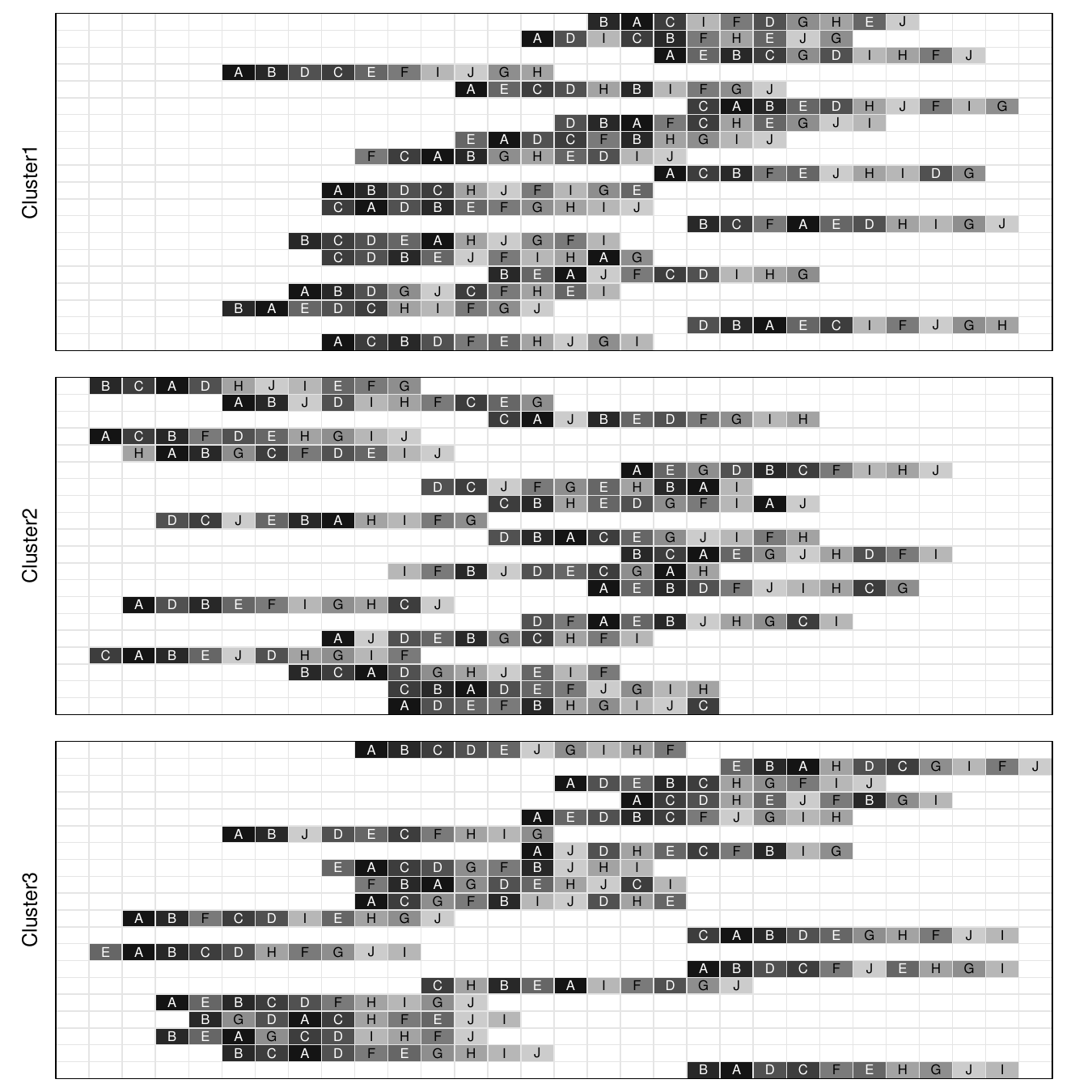}
  \caption{Data set generated according to the smaller consistency design (Setup 2).
  Each matrix shows one of three clusters from the data set.
  Assessors are stored on the rows (20 per cluster) and ranks on the columns in increasing order (30 ranks in total).
  Only the 10 items in $\mathcal{A}^*_c$ are highlighted, associating each of them with a unique shade of gray and a capital letter.
  The association is independent across clusters, meaning that the same color and letter does not denote the same item across different clusters.
  For the consistency design, relevant items have contiguous ranks, but the whole sequence is shifted by a random amount for each assessor.
  \label{afig:dataplot-consistency-small}}
\end{figure}

We devise two additional data generating processes (DGPs) that add noise to the consistency DGP.
The new designs are meant to further challenge the proposed methodology.
Both DGPs are based on the \textit{consistency} design illustrated in the main paper.
We recall that under this design, for each cluster, the ranks for each assessor in the cluster are sampled from a Mallows model around the consensus rank, and then the entire rank sequence for each assessor is shifted by a random quantity in the higher-dimensional space (an example of consistency-generated data is shown in Figure \ref{afig:dataplot-consistency-small}).

Once the data has been generated according to the consistency DGP, the \textit{noise} design adds a fixed amount of noise for each assessor by exchanging the ranks of the relevant items with that of randomly-chosen irrelevant ones, in a fixed proportion.
The \textit{random-noise} design is less noisy in that each assessor has only a 50\% probability of having added noise.
Specifically, for those assessors who have added noise, the ranks of items in $\mathcal{A}^*_c$ are exchanged with that of random items in $\mathcal{A}\setminus{\mathcal{A}^*_c}$, up to a fixed proportion (the exact amount of swapped ranks is chosen at random, independently for each assessor).
While the first DGP adds a lot of noise in the ranking sequences, the random-noise design is more moderate and seems to produce more realistic data sets.
Still, both noisy DGPs are built in such a way that they present very challenging cases for lowBM3.
We note that, for both designs, the ranks of the relevant items in the sampled rankings of the assessors do not follow a Mallows model once projected in the lower-dimensional space of dimension $n^*$.

\begin{figure}[t]
  \begin{subfigure}{.90\textwidth}
    \centering
    \caption{Setup 2.1: for each assessor, 20\% of the items in $\mathcal{A}^*_c$ get their rank swapped with random items in $\mathcal{A}\setminus{\mathcal{A}^*_c}$.}
    \includegraphics[width=0.65\textwidth]{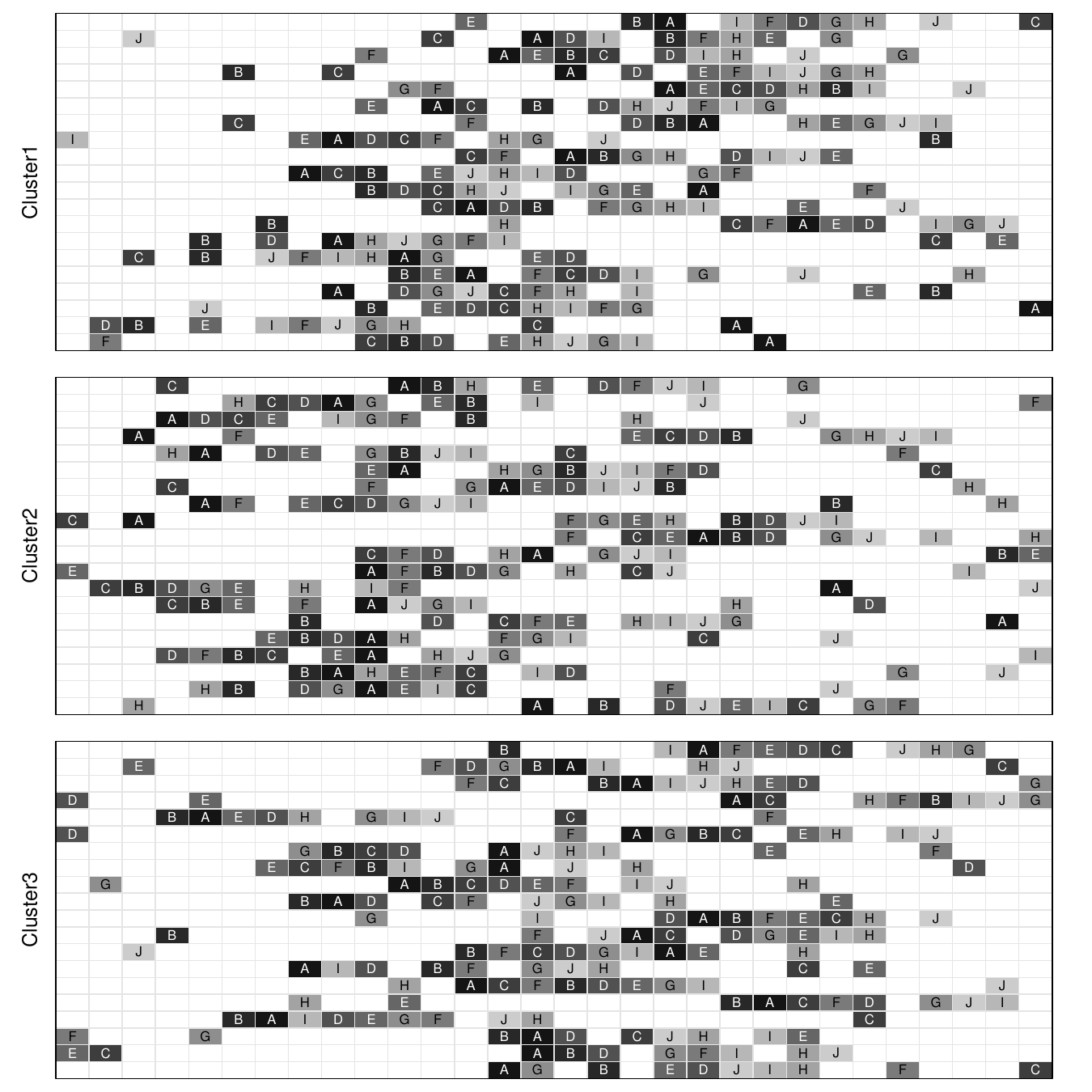}
  \end{subfigure}

  \begin{subfigure}{0.90\textwidth}
    \centering
    \caption{Setup 2.2: each assessor has a 50\% probability of getting up to 20\% of the ranks for items in $\mathcal{A}^*_c$ swapped with ranks of random items in $\mathcal{A}\setminus{\mathcal{A}^*_c}$.}
    \includegraphics[width=0.65\textwidth]{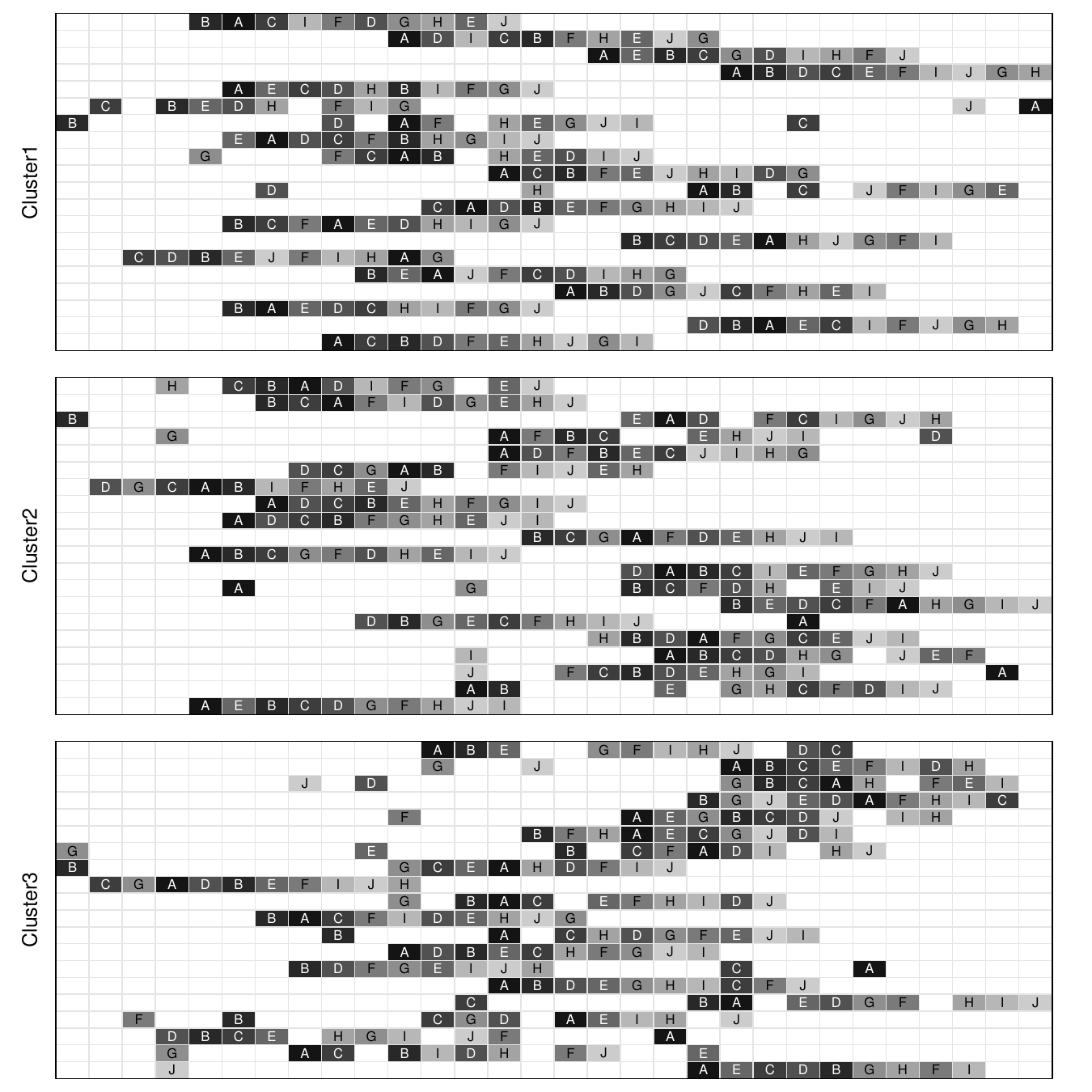}
  \end{subfigure}
  \caption{Data set generated according to the smaller consistency design with noise (additional setups).
  Panels are constructed as in Figure~\ref{afig:dataplot-consistency-small}, adding noise in ranked items for each assessor.
  \label{afig:dataplot-consistency-noise-small}}
\end{figure}

\begin{figure}[t]
  \centering
  \begin{subfigure}{0.9\textwidth}
    \centering
    \caption{Setup 5: mixed DGPs.
    Half of the items in $A^*_c$ are top-ranked, and the other half are randomly shifted.
    10\% of the items in $\mathcal{A}^*_c$ have their rank swapped with items from $\mathcal{A}\setminus{\mathcal{A}^*_c}$, at random.\label{afig:dataplot-mixdata-big}}
    \includegraphics[width=\textwidth]{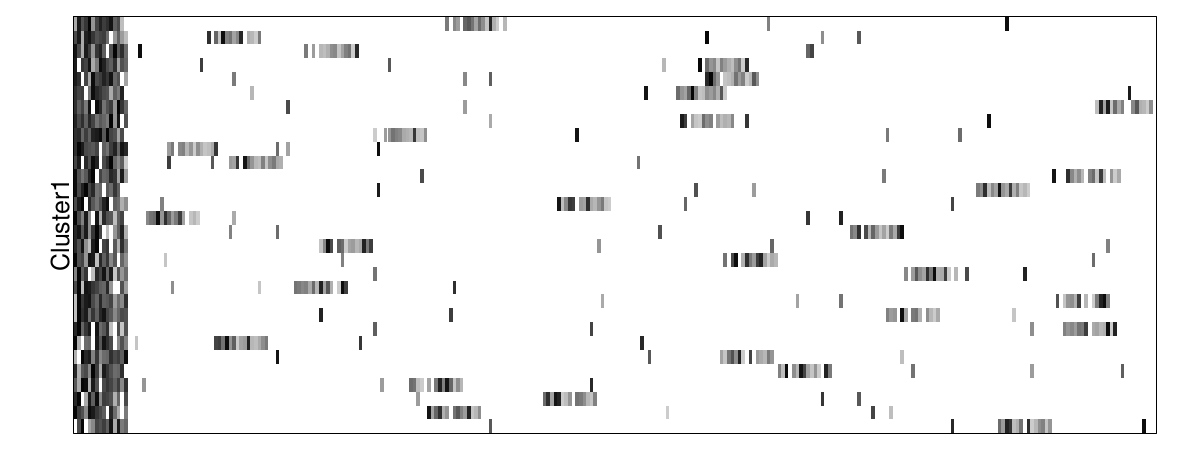}
  \end{subfigure}

  \bigskip
  \begin{subfigure}{0.4\textwidth}
    \centering
    \caption{Setup 3: larger top-rank DGP.\label{afig:dataplot-toprank-big}}
    \includegraphics[width=\textwidth]{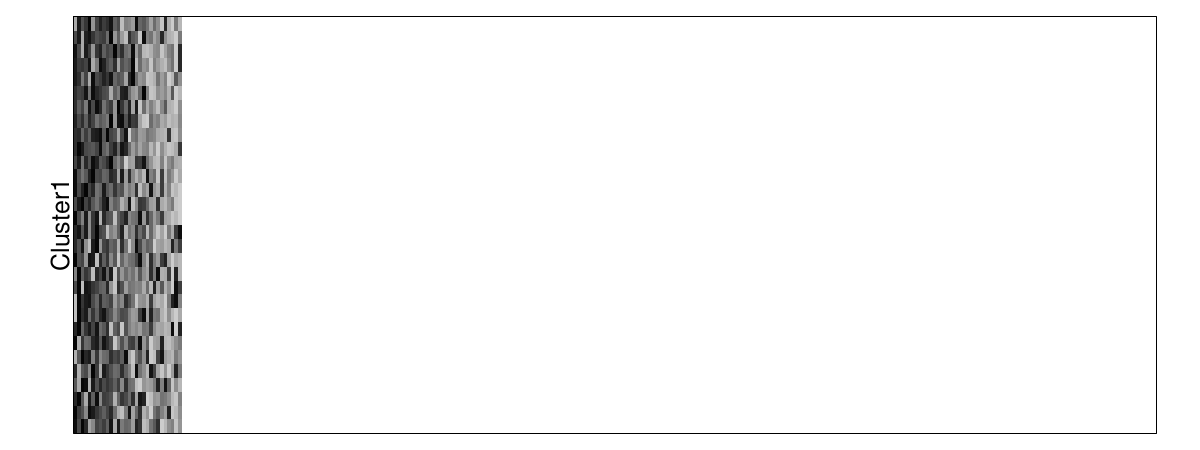}
  \end{subfigure}
  \qquad
  \begin{subfigure}{0.4\textwidth}
    \centering
    \caption{Setup 4: larger consistency DGP.\label{afig:dataplot-consistency-big}}
    \includegraphics[width=\textwidth]{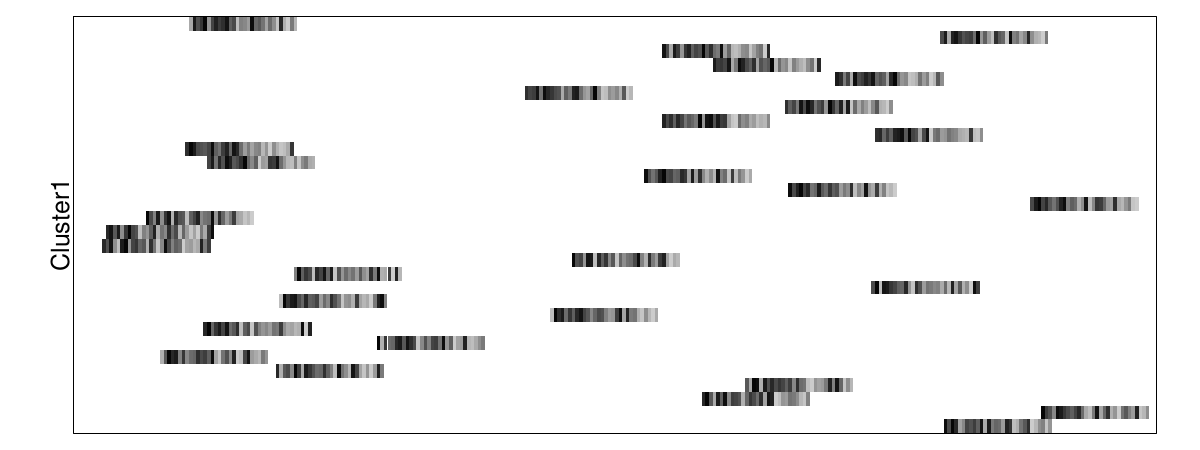}
  \end{subfigure}

  \bigskip
  \begin{subfigure}{0.4\textwidth}
    \centering
    \caption{Setup 4.1: larger consistency DGP, with 20\% of the items affected by noise.\label{afig:dataplot-consistency-noise-big}}
    \includegraphics[width=\textwidth]{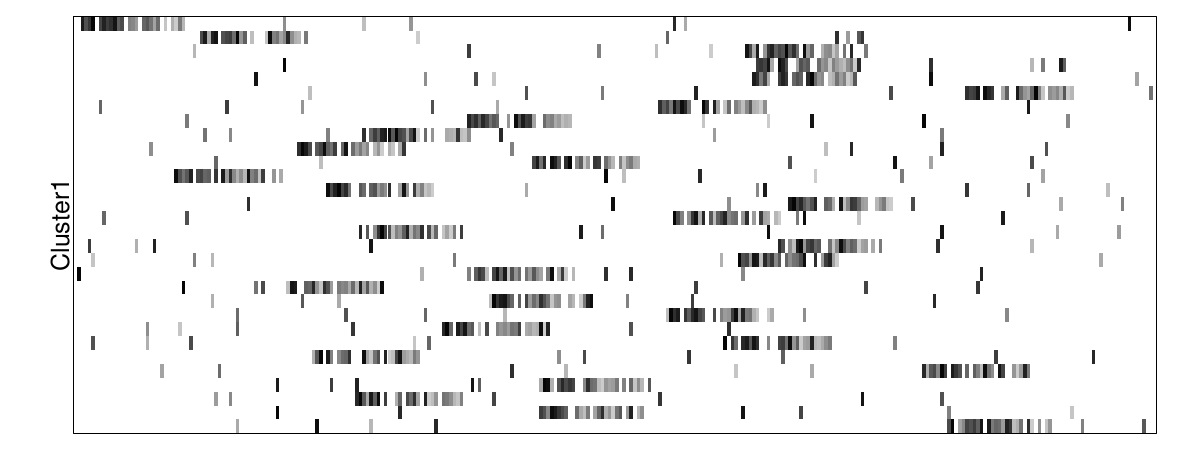}
  \end{subfigure}
  \qquad
  \begin{subfigure}{0.4\textwidth}
    \centering
    \caption{Setup 4.2: larger consistency DGP, with a 50\% probability of up to 20\% of the items being affected by noise.\label{afig:dataplot-consistency-randomnoise-big}}
    \includegraphics[width=\textwidth]{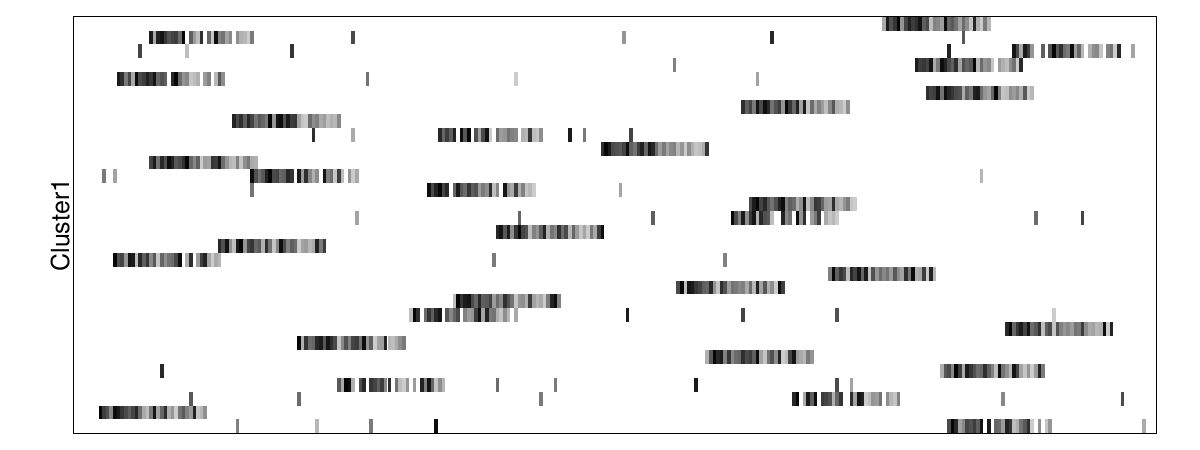}
  \end{subfigure}
  \caption{Each panel shows data generated for one of the larger experimental DGPs, for only one of five clusters (the plots for the other clusters are extremely similar).
  Assessors are showed on the rows (30 in total) and ranks on the columns in increasing order (300 items in total).
  Only the 30 items in $\mathcal{A}^*_c$ are highlighted, associating each of them with a unique shade of gray.
  Setup 5 (panel a) generates data that appear as a mix of Setup 3 (panel b) and Setup 4 (panel c).
  Setup 4.1 (panel d) and Setup 4.2 (panel e) are respectively the larger versions of Figure~\ref{afig:dataplot-consistency-noise-small}, panel a and b.
  Setup 4.2 results in overall less noise than Setup 4.1.
  \label{afig:dataplot-big}}
\end{figure}

Table~\ref{atab:sim_setups} shows the parameters used with the noise and random-noise DGPs, to produce four additional data sets: Setup 2.1 and Setup 4.1 are generated with the noise DGP, with the latter being a higher-dimensional version of the former; analogously, Setup 2.2 and Setup 4.2 use the random-noise design.
Data sets for the four setups are shown in Figure~\ref{afig:dataplot-consistency-noise-small}, Figure~\ref{afig:dataplot-consistency-noise-big}, and Figure~\ref{afig:dataplot-consistency-randomnoise-big}.

\begin{table}[!
htb]
  \centering
  \begin{tabular}{rlllll}
    & Setup 2.1 & Setup 2.2    & Setup 4.1 & Setup 4.2   \\
    \toprule
    DGP      & noise     & random-noise & noise     & random-noise \\
    \midrule
    $n$      & 30        & 30           & 300       & 300          \\
    $N$      & 60        & 60           & 150       & 150          \\
    $n^*$    & 10        & 10           & 30        & 30           \\
    $C$      & 3         & 3            & 5         & 5            \\
    $\alpha$ & 3         & 3            & 3         & 3            \\
    \midrule
    $M$      & 25,000    & 25,000       & 100,000   & 100,000      \\
    thinning & 25        & 25           & 25        & 50           \\
    burnin   & 30\%      & 30\%         & 30\%      & 30\%         \\
    noise    & 20\%      & 20\%         & 20\%      & 20\%         \\
    \bottomrule
  \end{tabular}
  \caption{Data generating processes (DGPs) and parameter values used in the noise and random-noise DPGs, to prouduce setups 2.1, 2.2, 4.1, and 4.2. The ``noise'' row indicates the maximum amount of noise added in each Setup: for the noise DGP, all assessors get that\% of the ranks of relevant items swapped with ranks of irrelevant items; for the random-noise DGP, the assessors that get the noise perturbation, have at most that \% of the ranks of relevant items swapped with ranks of irrelevant items.}
  \label{atab:sim_setups}
\end{table}

\begin{figure}[t]
  \centering
  \includegraphics[width=\textwidth]{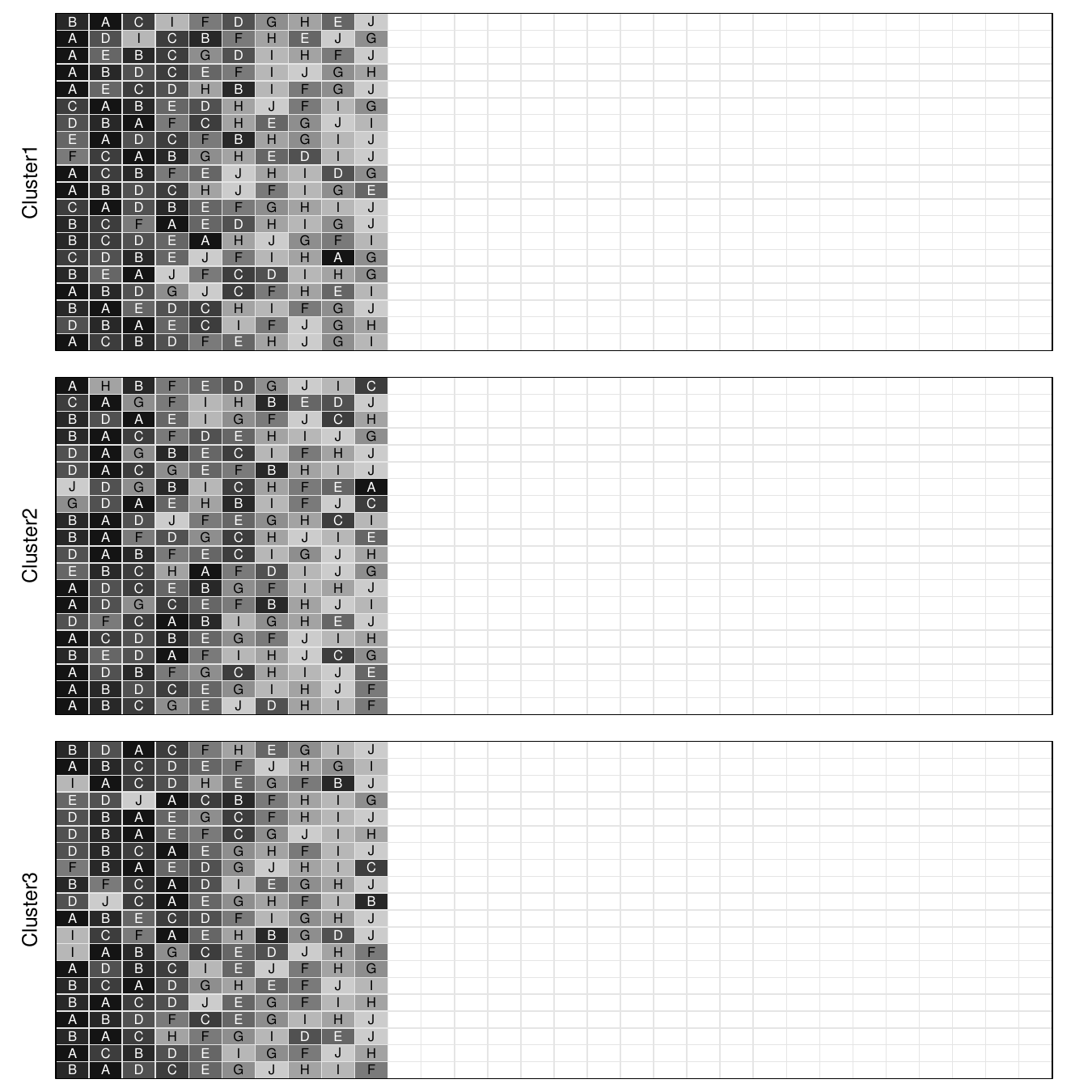}
  \caption{Data set generated according to the smaller top-rank design (Setup 1).
  Each matrix shows one of three clusters from the data set.
  Assessors are stored on the rows (20 per cluster) and ranks on the columns in increasing order (30 ranks in total).
  Only the 10 items in $\mathcal{A}^*_c$ are highlighted, associating each of them with unique shade of gray and a capital letter.
  The association is independent across clusters, meaning that the same color and letter does not denote the same item across different clusters.
  The figure makes it evident that all relevant items are top-ranked: they are shifted to the left of the matrices, having lower (better) rank.
  \label{afig:dataplot-toprank-small}}
\end{figure}

\begin{figure}[!t]
  \centering
  \begin{subfigure}{\textwidth}
    \caption{Data from Setup 1 with $[\alpha_1, \alpha_2, \alpha_3] = [3, 3, 3]$}
    \label{afig:alphaexp-dataplot-toprank-small-333}
    \includegraphics[width=\textwidth]{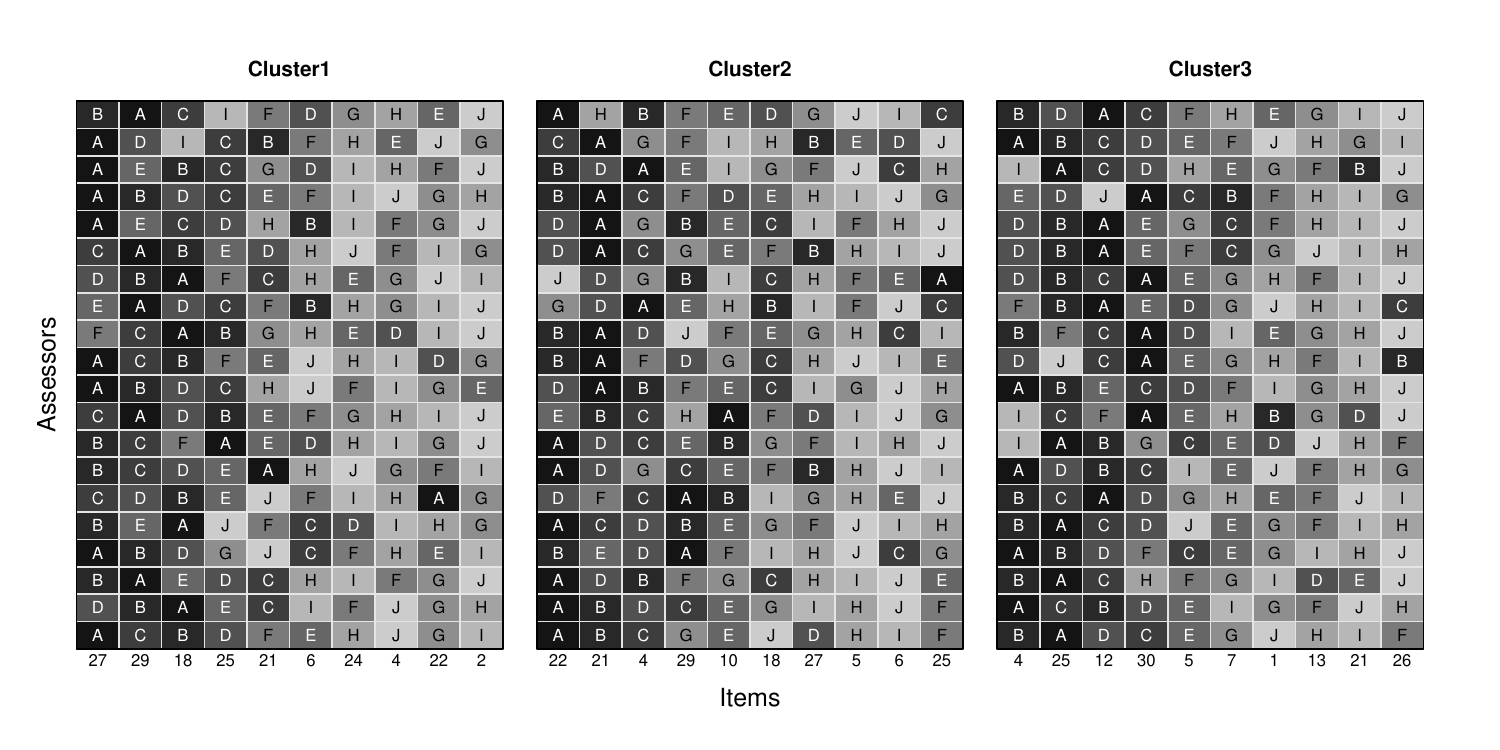}
  \end{subfigure}

  \vfill

  \begin{subfigure}{\textwidth}
    \caption{Data from Setup 1 with $[\alpha_1, \alpha_2, \alpha_3] = [1, 3.85, 7]$}
    \label{afig:alphaexp-dataplot-toprank-small-1m7}
    \includegraphics[width=\textwidth]{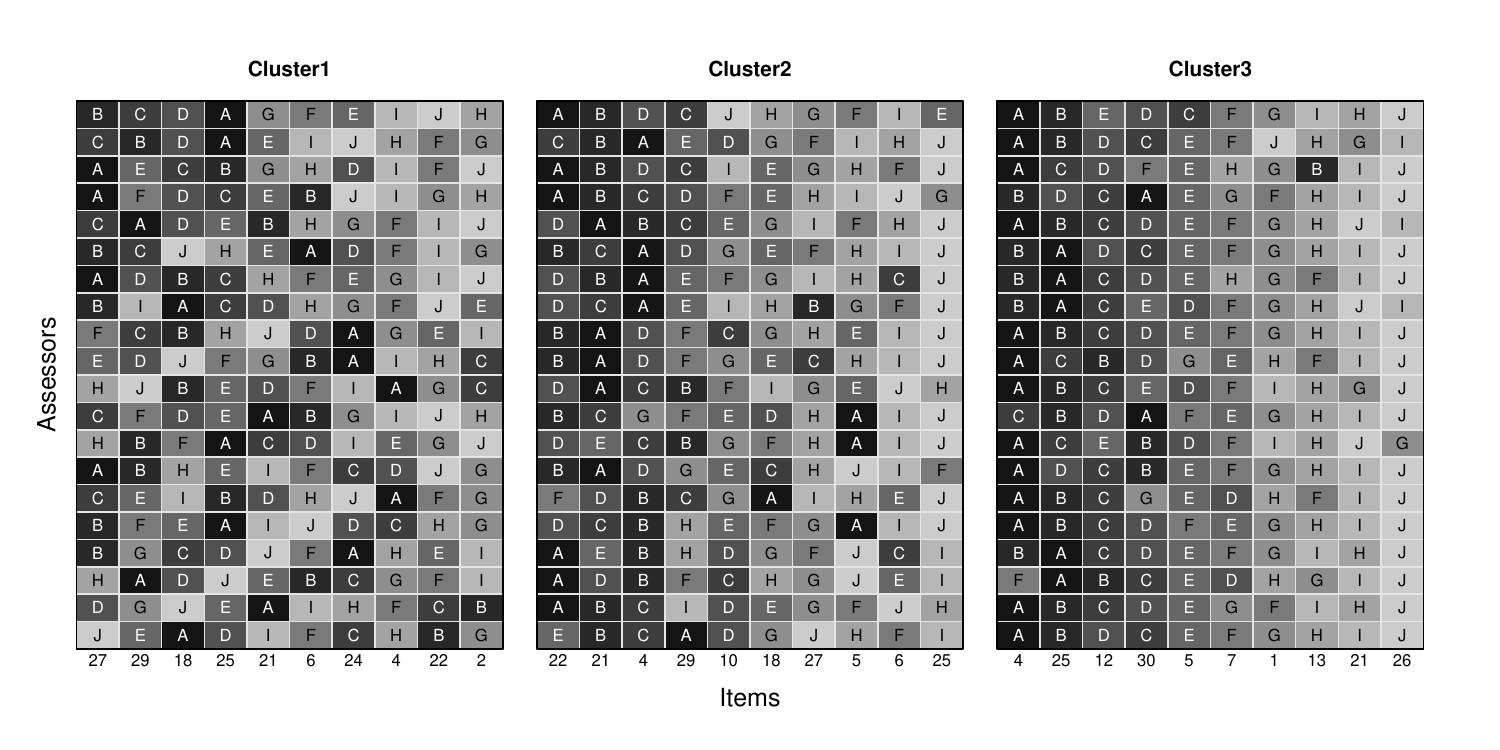}
  \end{subfigure}

  \caption{Display of the data sets used to test the misspecification of $\alpha$. Each of the two subplots shows the sample ranks by assessors (y-axis) of top-ranked items (x-axis), in each of the three clusters characterizing the data set. Both data sets were generated from Setup 1, setting the Mallows' model dispersion parameter to either $[\alpha_1, \alpha_2, \alpha_3] = [3, 3, 3]$ (top panel) or $[\alpha_1, \alpha_2, \alpha_3] = [1, 3.85, 7]$ (bottom panel).}
  \label{afig:alphaexp-dataplot-toprank-small}
\end{figure}

\clearpage

\section{Additional results on simulation experiments}
\begin{figure}[t]
  \centering
  \begin{subfigure}{.9\textwidth}
    \caption{Results from Setup 1 (top-rank), cluster 2.}
    \label{afig:selection-toprank-small-cluster2}
    \includegraphics[width=\textwidth]{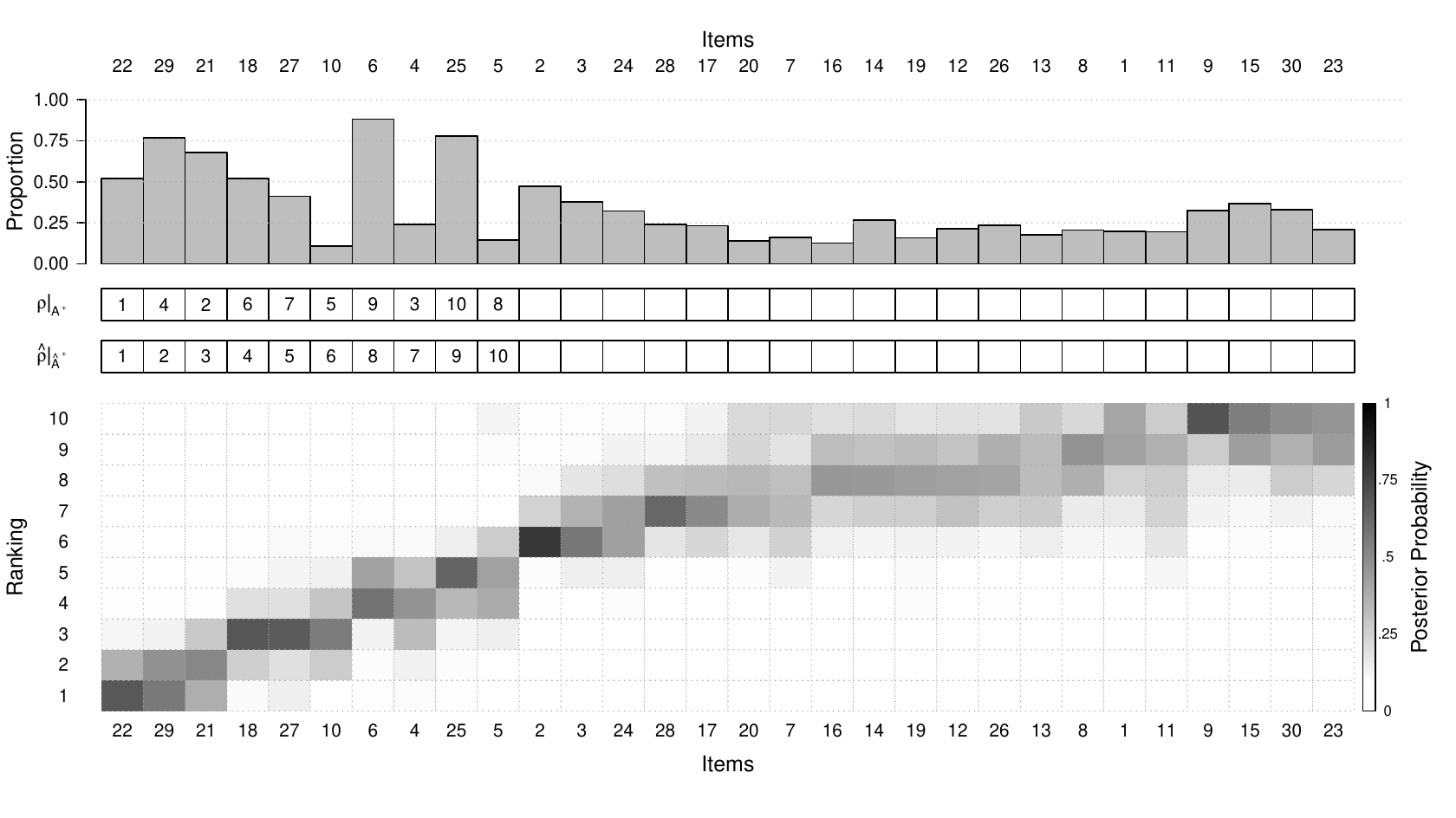}
  \end{subfigure}

  \vfill

  \begin{subfigure}{.9\textwidth}
    \caption{Results from Setup 1 (top-rank), cluster 3.}
    \label{afig:selection-toprank-small-cluster3}
    \includegraphics[width=\textwidth]{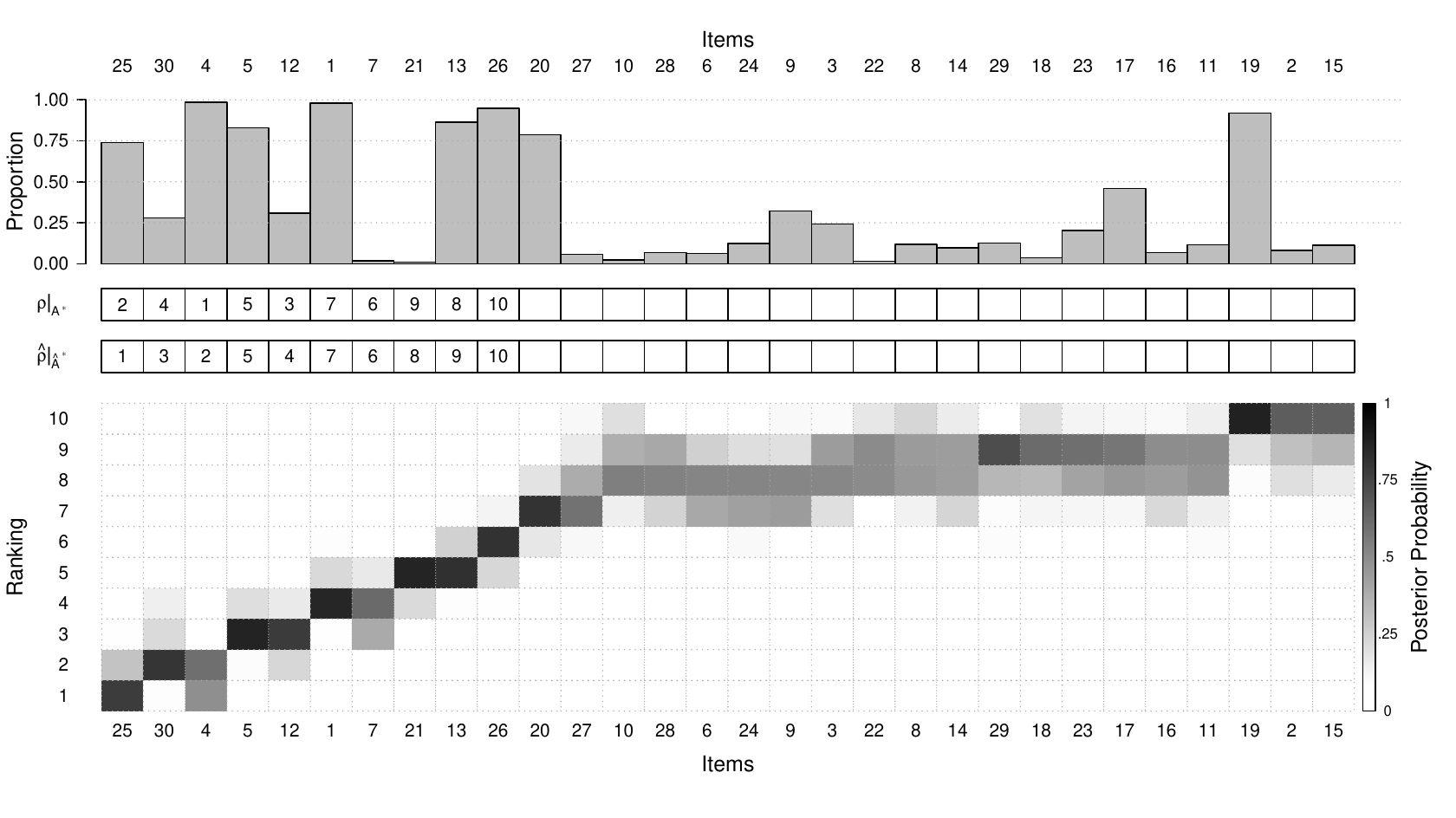}
  \end{subfigure}

  \caption{Results for \emph{cluster 2} (panel a) and \emph{cluster 3} (panel b) from the smaller top-rank simulation design (Setup 1).
  Each panel shows a composite graph with items listed on the x-axes; from top to bottom: bar-plot of the posterior probability with which an item is selected in the estimated set $\hat{\A}_c^{*}$; in the subsequent row, $\rho|_{A^{*}}$, true ranks of the items in $\A_c^{*}$; in the $\hat{\rho}|_{\hat{A}^{*}}$ row, estimated ranks of the items in $\hat{\A}_c^{*}$; in the heatmap, marginal posterior probability  of $\vrho_c$ (shades of gray) for each rank (y-axis) to be assigned to each item.
  On the x-axis, items are ordered by the latter.
  The selection of items in $\hat{\mathcal{A}}_c^*$ uses no threshold to determine the High Probability Set, $\mathcal{A}'_c$, (see Definition~\ref{def:hps1}).
  \label{afig:selection-heatmap-toprank}
  }
\end{figure}

\begin{figure}[t]
  \centering
  \begin{subfigure}{.9\textwidth}
    \caption{Results from Setup 2 (consistency), cluster 2.}
    \label{afig:selection-consistency-small-cluster2}
    \includegraphics[width=\textwidth]{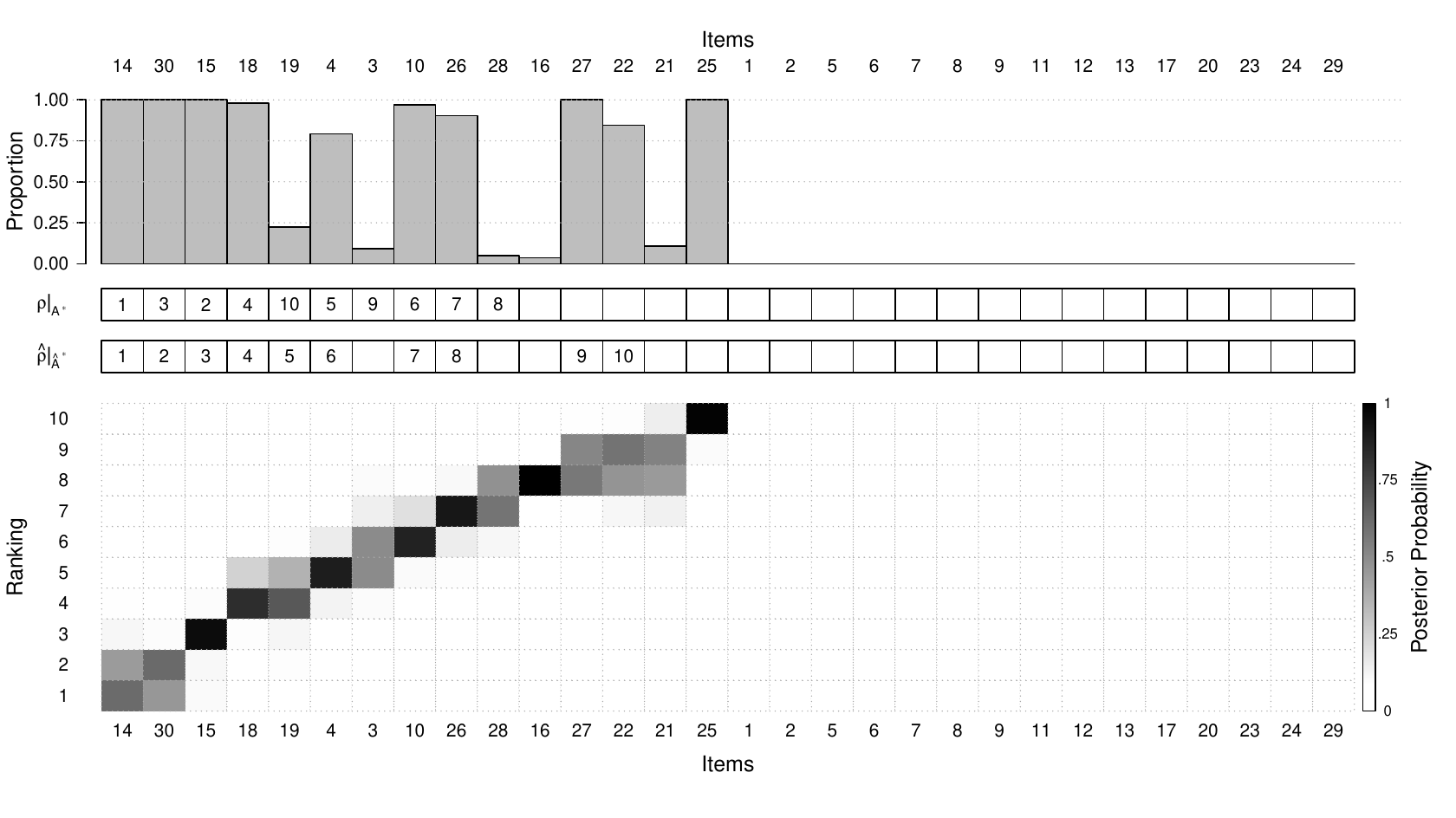}
  \end{subfigure}

  \vfill

  \begin{subfigure}{.9\textwidth}
    \caption{Results from Setup 2 (consistency), cluster 3.}
    \label{afig:selection-consistency-small-cluster3}
    \includegraphics[width=\textwidth]{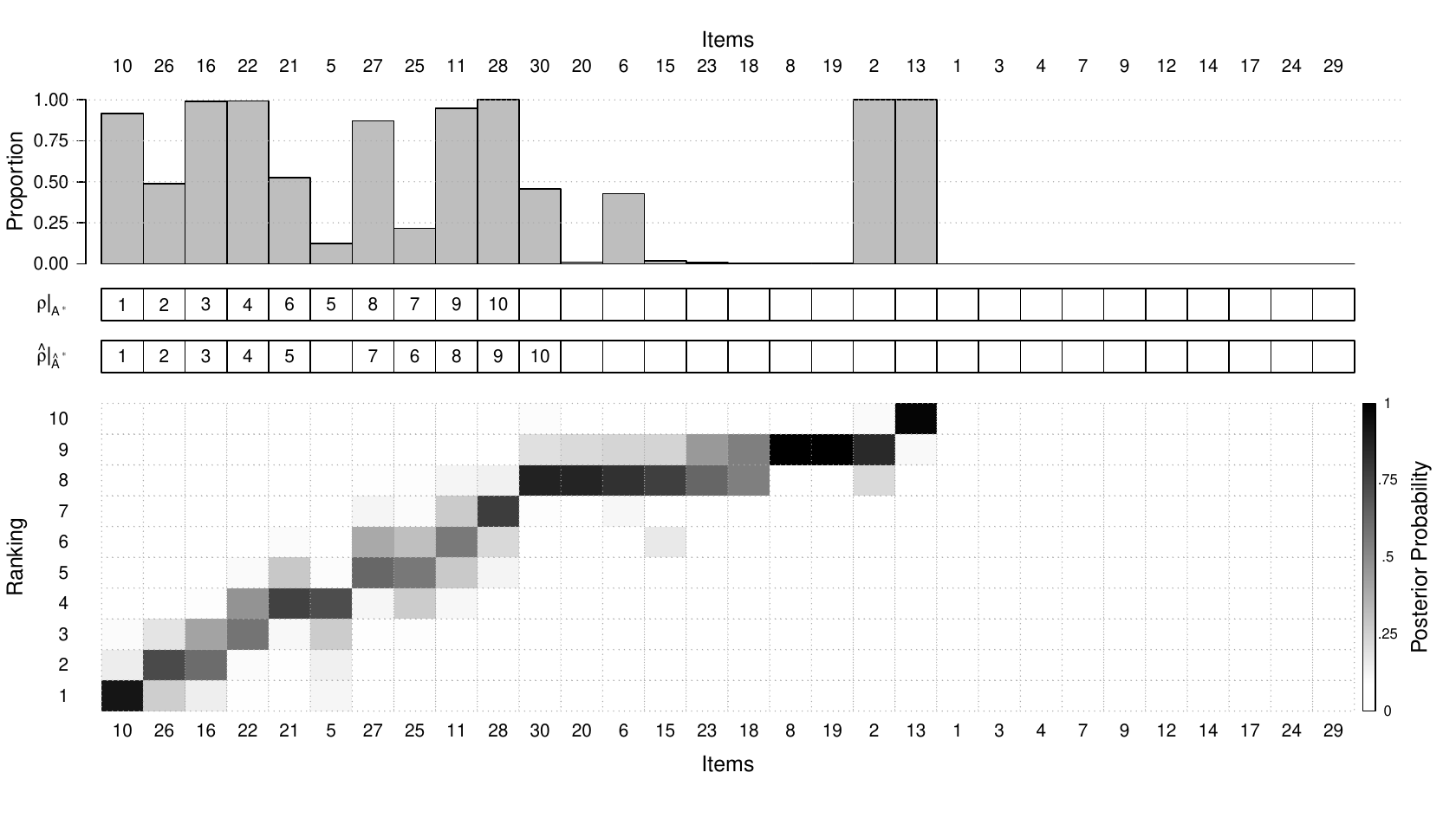}
  \end{subfigure}

  \caption{Results for \emph{cluster 2} (panel a) and \emph{cluster 3} (panel b) from the smaller consistency simulation design (Setup 2).
  Each panel shows a composite graph with items listed on the x-axes; from top to bottom: bar-plot of the posterior probability with which an item is selected in the estimated set $\hat{\A}_c^{*}$; in the subsequent row, $\rho|_{A^{*}}$, true ranks of the items in $\A_c^{*}$; in the $\hat{\rho}|_{\hat{A}^{*}}$ row, estimated ranks of the items in $\hat{\A}_c^{*}$; in the heatmap, marginal posterior probability  of $\vrho_c$ (shades of gray) for each rank (y-axis) to be assigned to each item.
  On the x-axis, items are ordered by the latter.  The selection of items in $\hat{\mathcal{A}}_c^*$ uses a threshold $p=20\%$ to determine the High Probability Set, $\mathcal{A}'_c$, (see Definition~\ref{def:hps1}).
  }
  \label{afig:selection-heatmap-consistency}
\end{figure}

\begin{figure}[!t]
  \centering
  \caption{Results for \emph{clusters 1 to 5} (panel \textit{a} to \textit{e}) from the larger top-rank simulation design (Setup 3).
  Each panel shows a composite graph with items listed on the x-axes; from top to bottom: bar-plot of the posterior probability with which an item is selected in the estimated set $\hat{\A}_c^{*}$; in the subsequent row, $\rho|_{A^{*}}$, true ranks of the items in $\A_c^{*}$; in the $\hat{\rho}|_{\hat{A}^{*}}$ row, estimated ranks of the items in $\hat{\A}_c^{*}$; in the heatmap, marginal posterior probability  of $\vrho_c$ (shades of gray) for each rank (y-axis) to be assigned to each item.
  On the x-axis, items are ordered by the latter.  The selection of items in $\hat{\mathcal{A}}_c^*$ uses no threshold to determine the High Probability Set, $\mathcal{A}'_c$, (see Definition~\ref{def:hps1}).
  }
  \label{afig:selection-heatmap-toprank-big}
  \begin{subfigure}{\textwidth}
    \caption{Results from Setup 3 (top-rank), cluster 1.}
    \label{afig:selection-toprank-big-cluster1}
    \includegraphics[width=\textwidth]{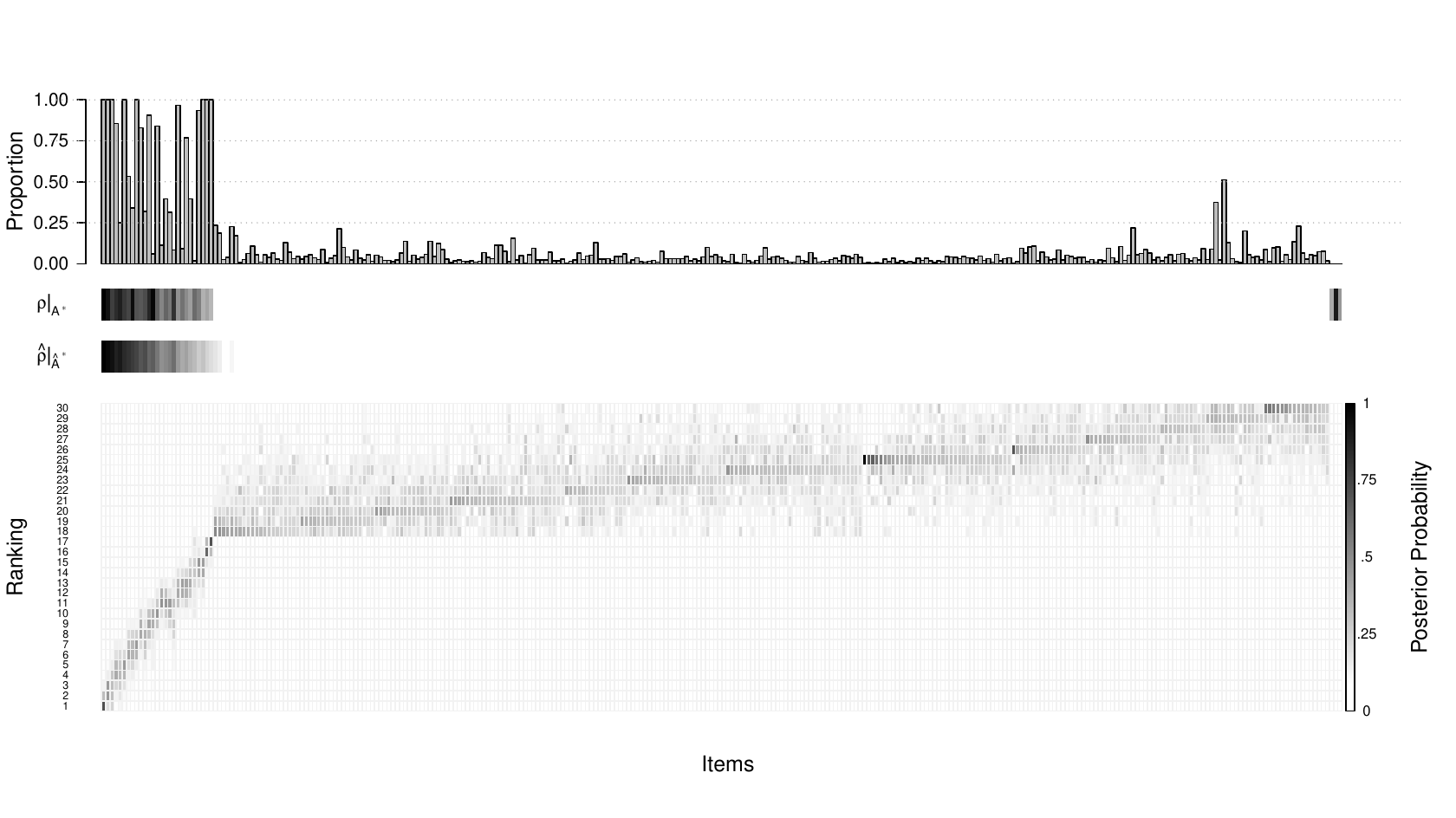}
  \end{subfigure}
\end{figure}

\begin{figure}[!t]
  \centering
  \ContinuedFloat

  \begin{subfigure}{\textwidth}
    \caption{Results from Setup 3 (top-rank), cluster 2.}
    \label{afig:selection-toprank-big-cluster2}
    \includegraphics[width=\textwidth]{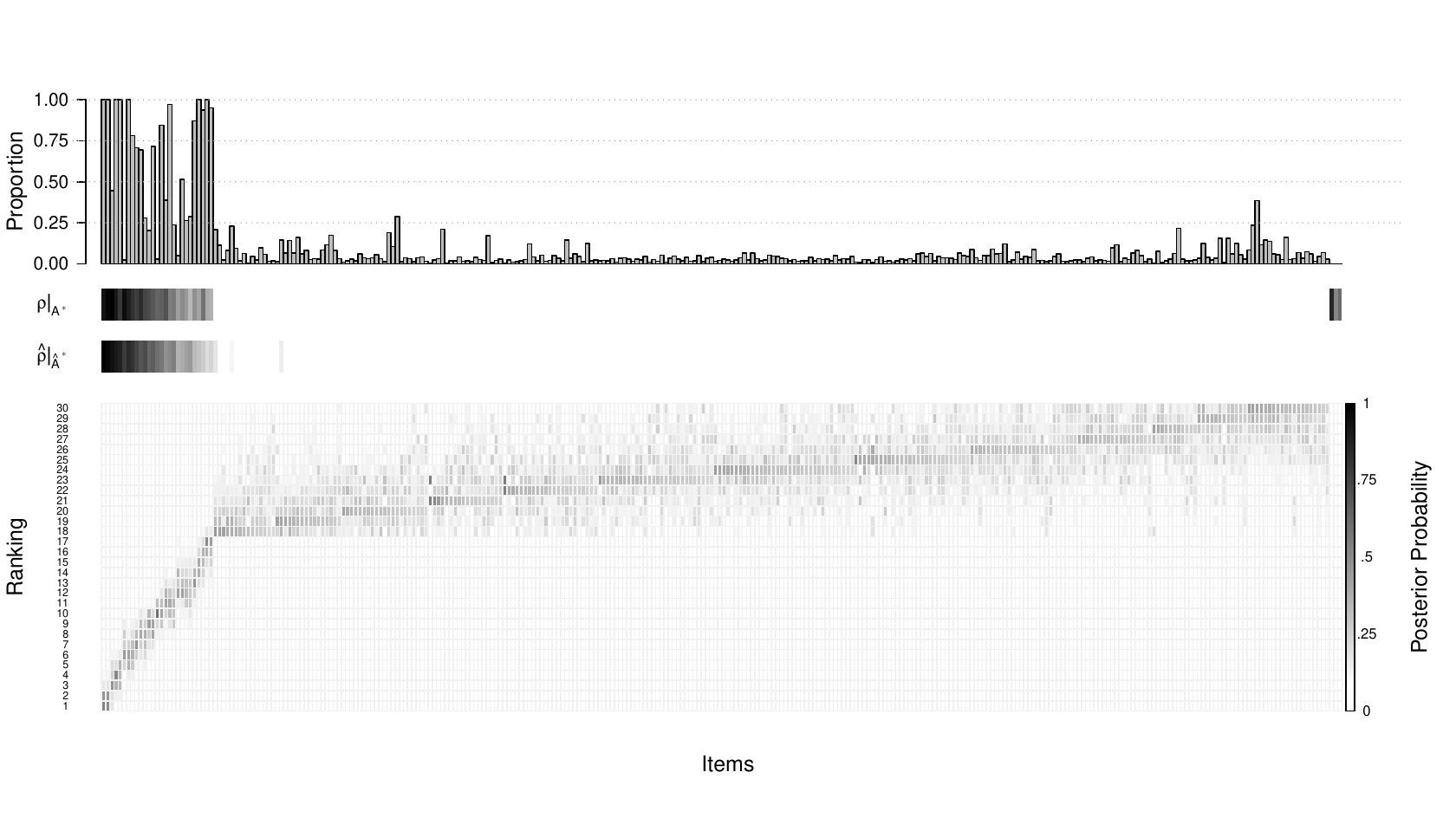}
  \end{subfigure}

  \vfill

  \begin{subfigure}{\textwidth}
    \caption{Results from Setup 3 (top-rank), cluster 3.}
    \label{afig:selection-toprank-big-cluster3}
    \includegraphics[width=\textwidth]{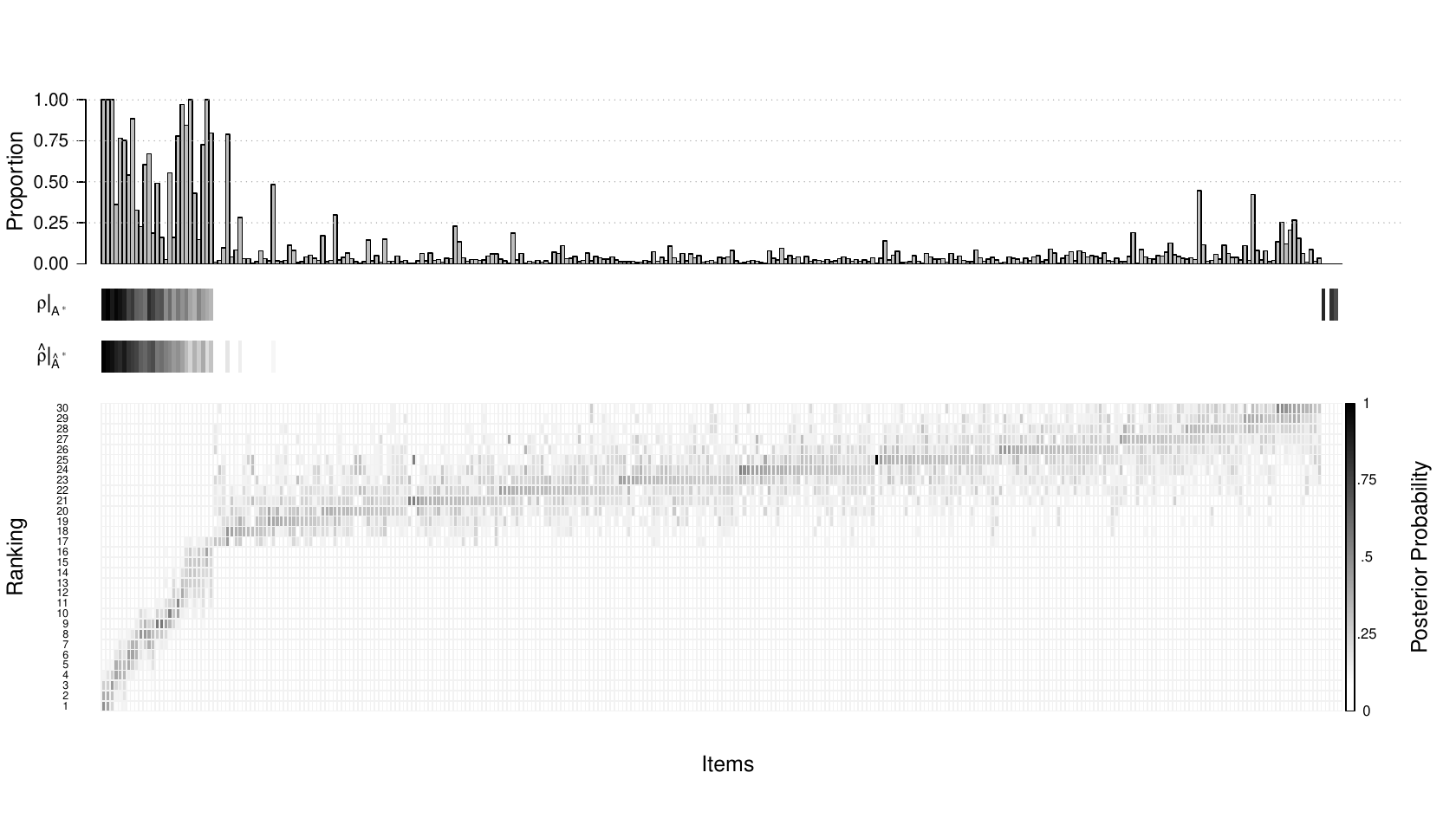}
  \end{subfigure}
\end{figure}

\begin{figure}[!t]
  \centering
  \ContinuedFloat
  \begin{subfigure}{\textwidth}
    \caption{Results from Setup 3 (top-rank), cluster 4.}
    \label{afig:selection-toprank-big-cluster4}
    \includegraphics[width=\textwidth]{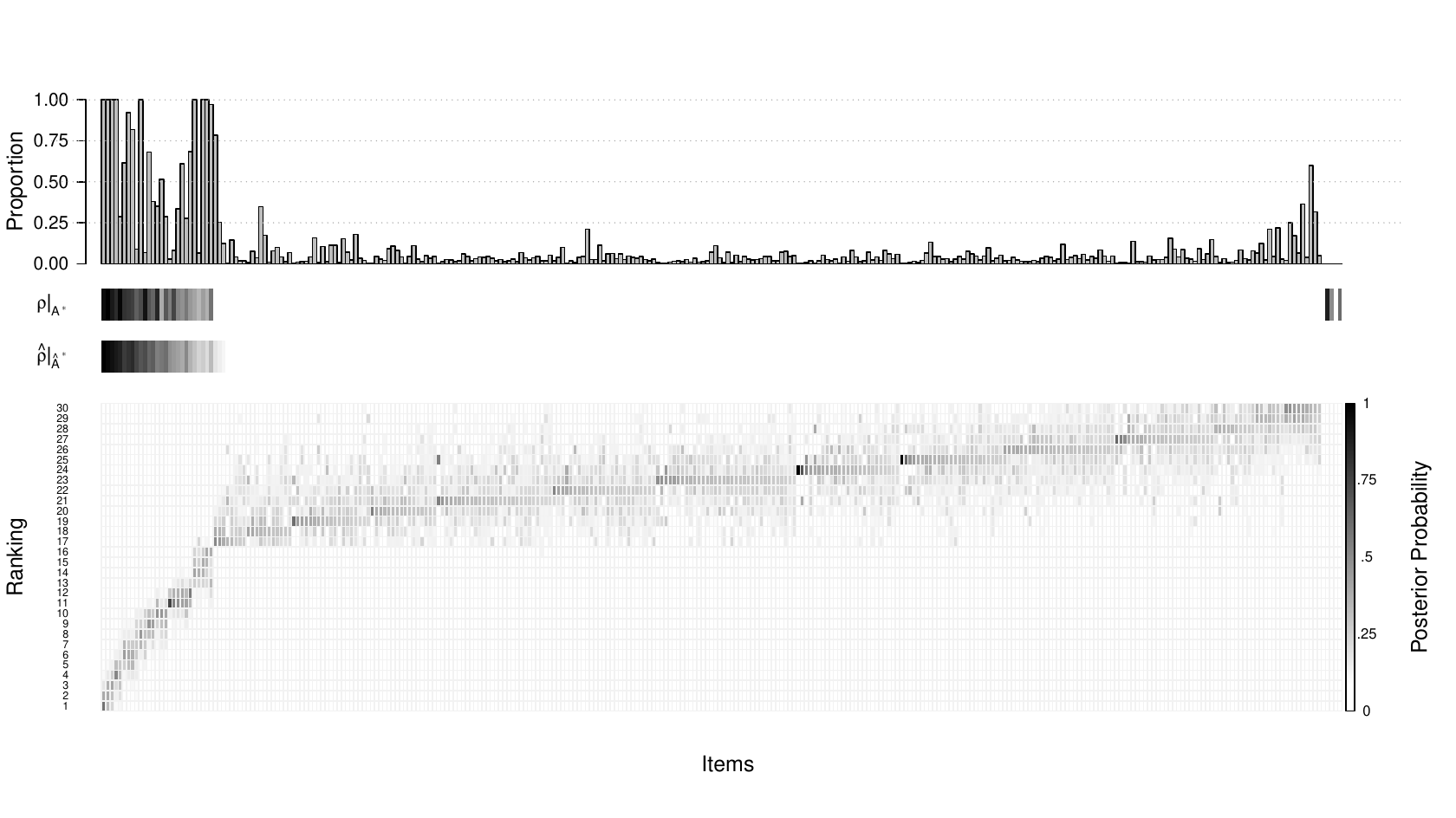}
  \end{subfigure}

  \vfill

  \begin{subfigure}{\textwidth}
    \caption{Results from Setup 3 (top-rank), cluster 5.}
    \label{afig:selection-toprank-big-cluster5}
    \includegraphics[width=\textwidth]{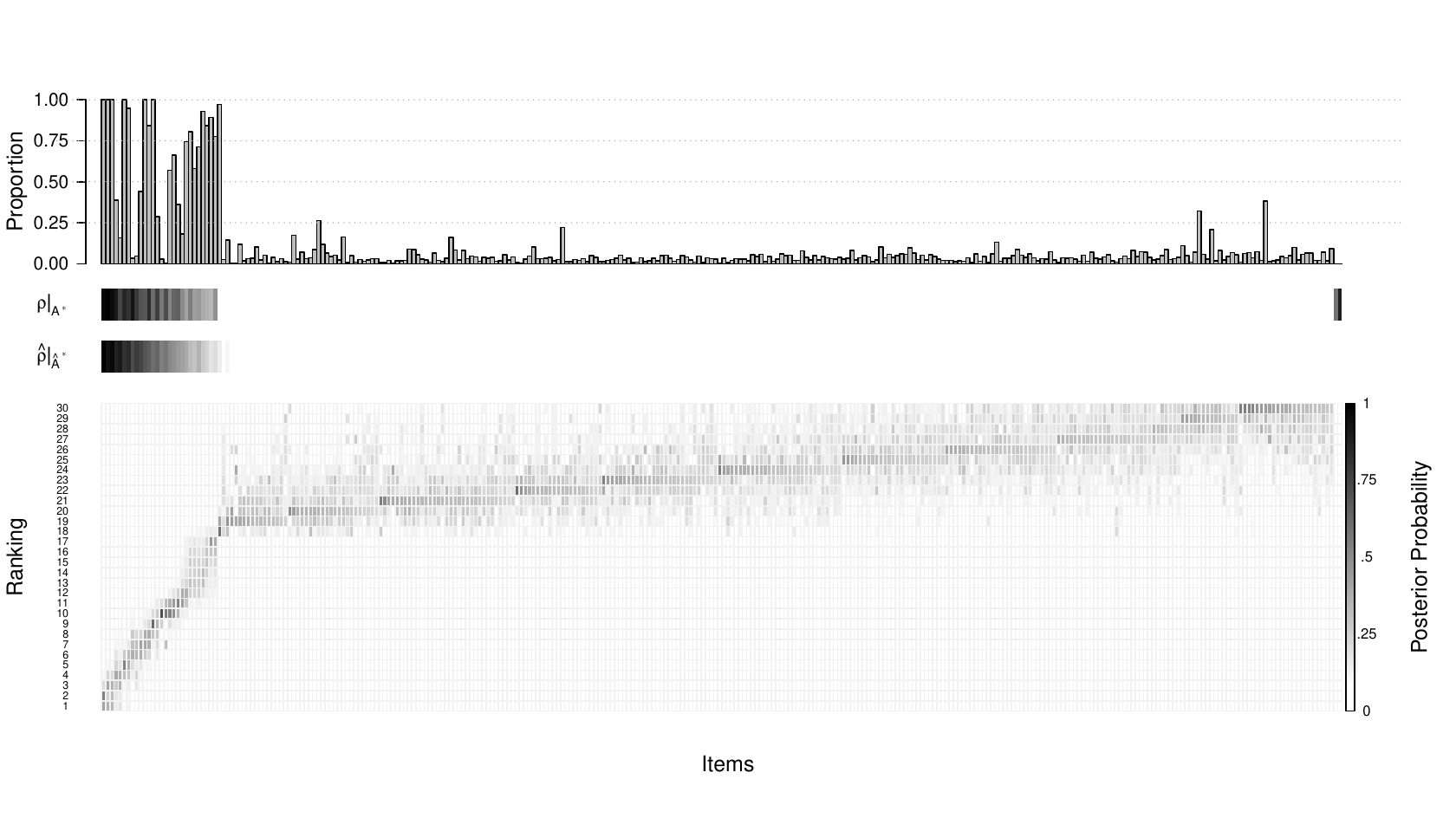}
  \end{subfigure}
\end{figure}

\begin{figure}[!t]
  \centering
  \caption{Results for \emph{clusters 1 to 5} (panels \textit{a} to \textit{e}) from the larger consistency simulation design (Setup 4).
  Each panel shows a composite graph with items listed on the x-axes; from top to bottom: bar-plot of the posterior probability with which an item is selected in the estimated set $\hat{\A}_c^{*}$; in the subsequent row, $\rho|_{A^{*}}$, true ranks of the items in $\A_c^{*}$; in the $\hat{\rho}|_{\hat{A}^{*}}$ row, estimated ranks of the items in $\hat{\A}_c^{*}$; in the heatmap, marginal posterior probability  of $\vrho_c$ (shades of gray) for each rank (y-axis) to be assigned to each item.
  On the x-axis, items are ordered by the latter. The selection of items in $\hat{\mathcal{A}}_c^*$ uses a threshold $p=20\%$ to determine the High Probability Set, $\mathcal{A}'_c$, (see Definition~\ref{def:hps1}).
  }
  \label{afig:selection-heatmap-consistency-big}
  \begin{subfigure}{\textwidth}
    \caption{Results from Setup 4 (consistency), cluster 1.}
    \label{afig:selection-consistency-big-cluster1}
    \includegraphics[width=\textwidth]{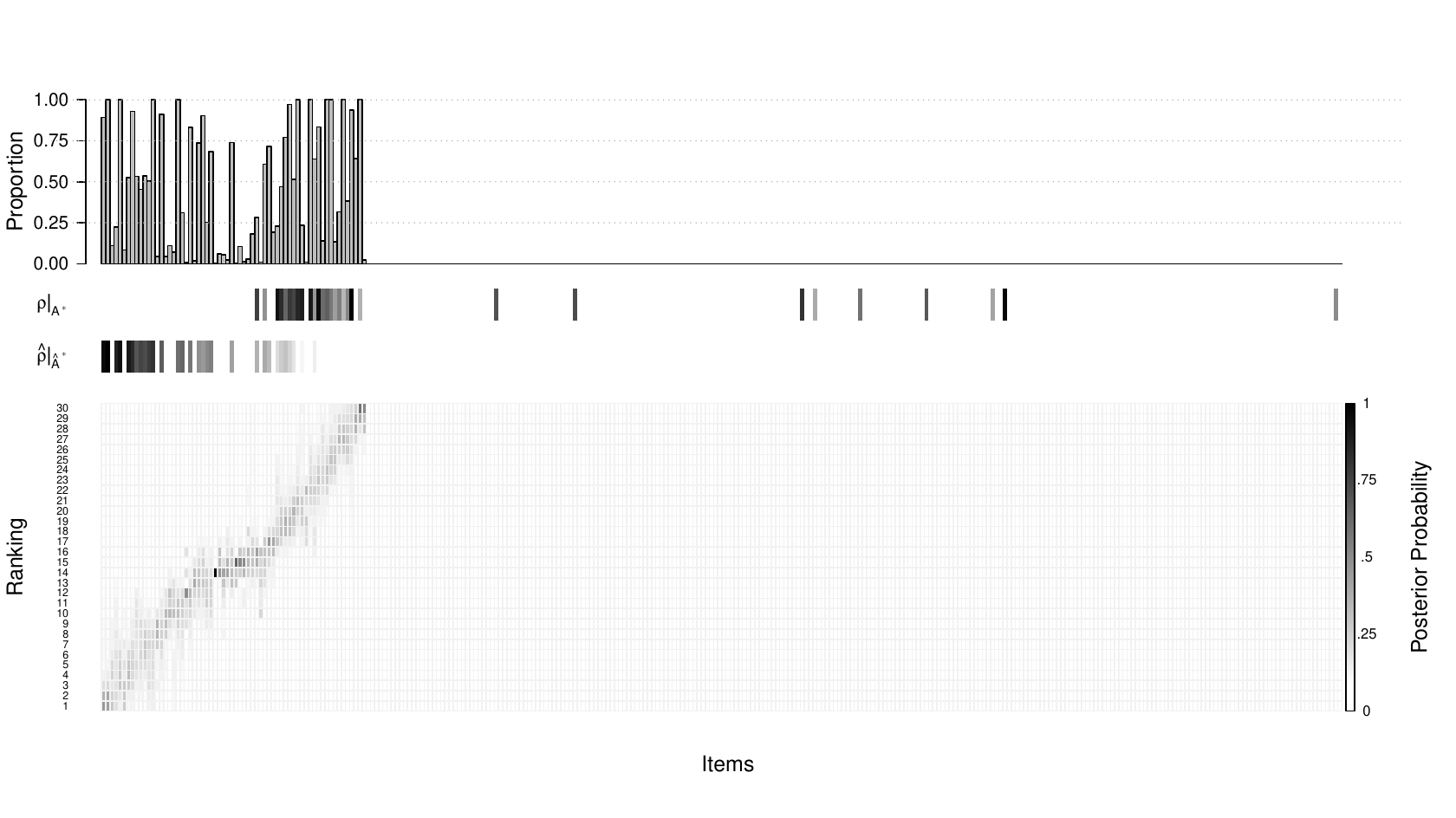}
  \end{subfigure}
\end{figure}

\begin{figure}[!t]
  \centering
  \ContinuedFloat

  \begin{subfigure}{\textwidth}
    \caption{Results from Setup 4 (consistency), cluster 2.}
    \label{afig:selection-consistency-big-cluster2}
    \includegraphics[width=\textwidth]{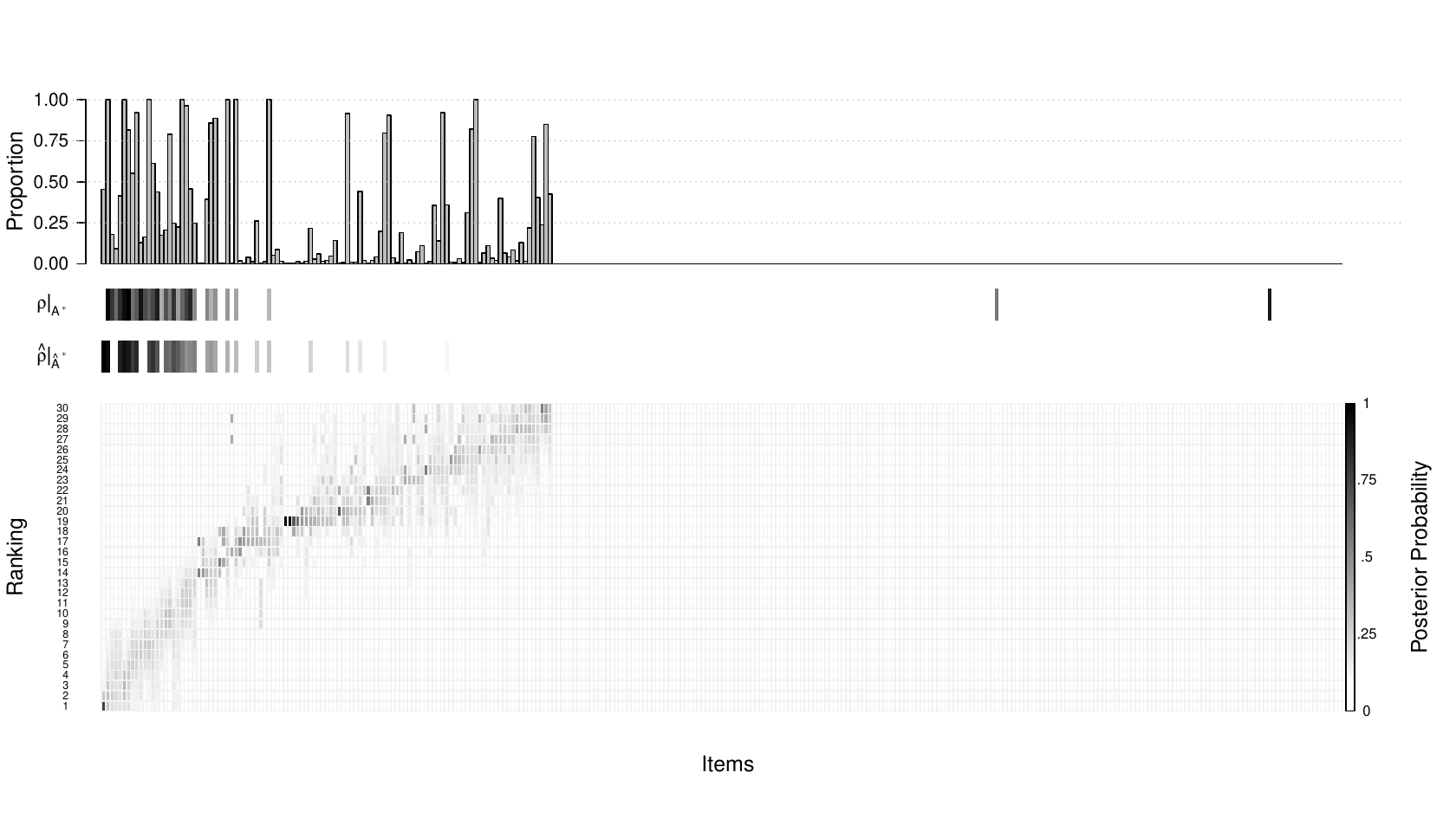}
  \end{subfigure}

  \vfill

  \begin{subfigure}{\textwidth}
    \caption{Results from Setup 4 (consistency), cluster 3.}
    \label{afig:selection-consistency-big-cluster3}
    \includegraphics[width=\textwidth]{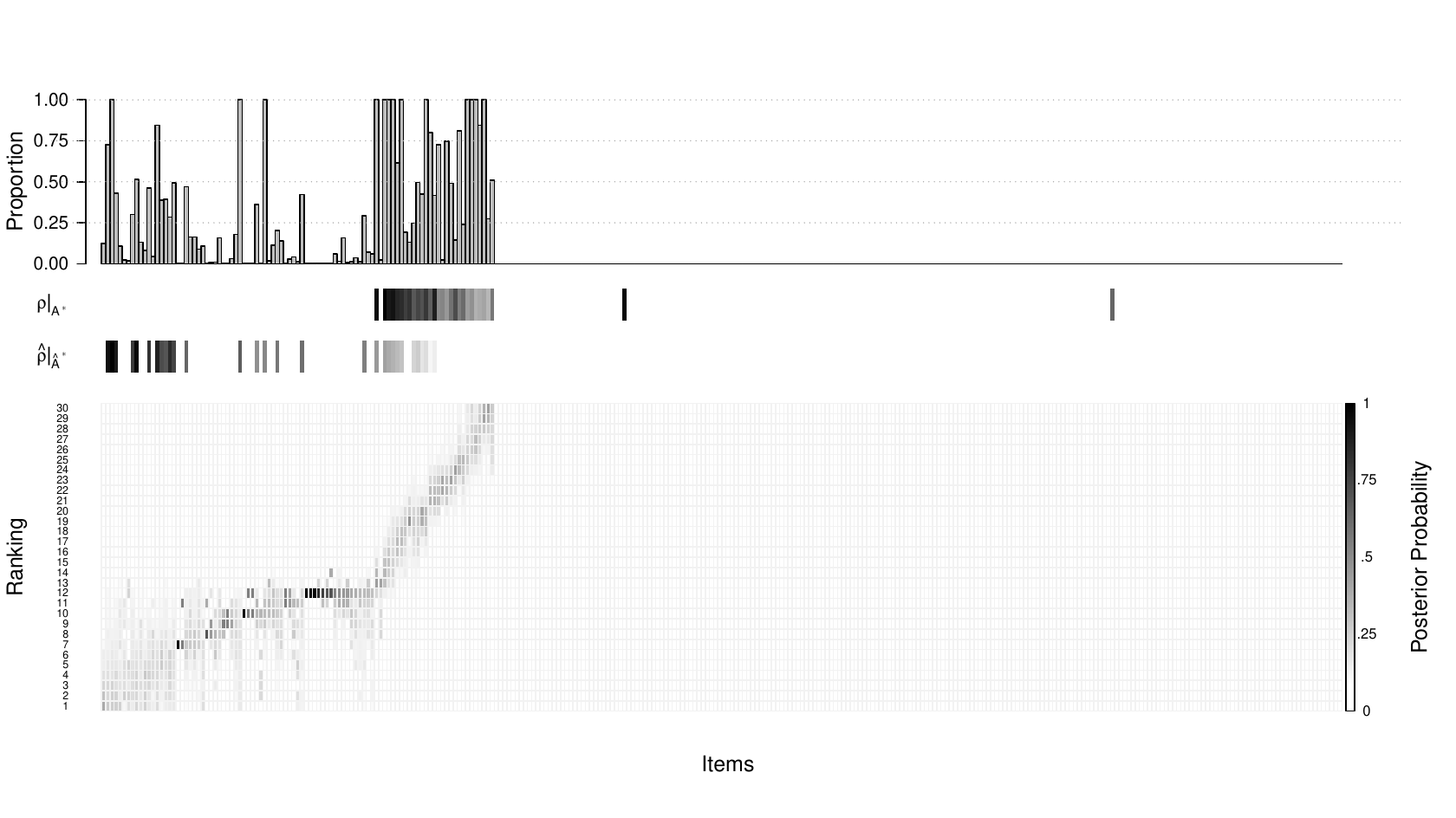}
  \end{subfigure}
\end{figure}

\begin{figure}[!t]
  \centering
  \ContinuedFloat
  \begin{subfigure}{\textwidth}
    \caption{Results from Setup 4 (consistency), cluster 4.}
    \label{afig:selection-consistency-big-cluster4}
    \includegraphics[width=\textwidth]{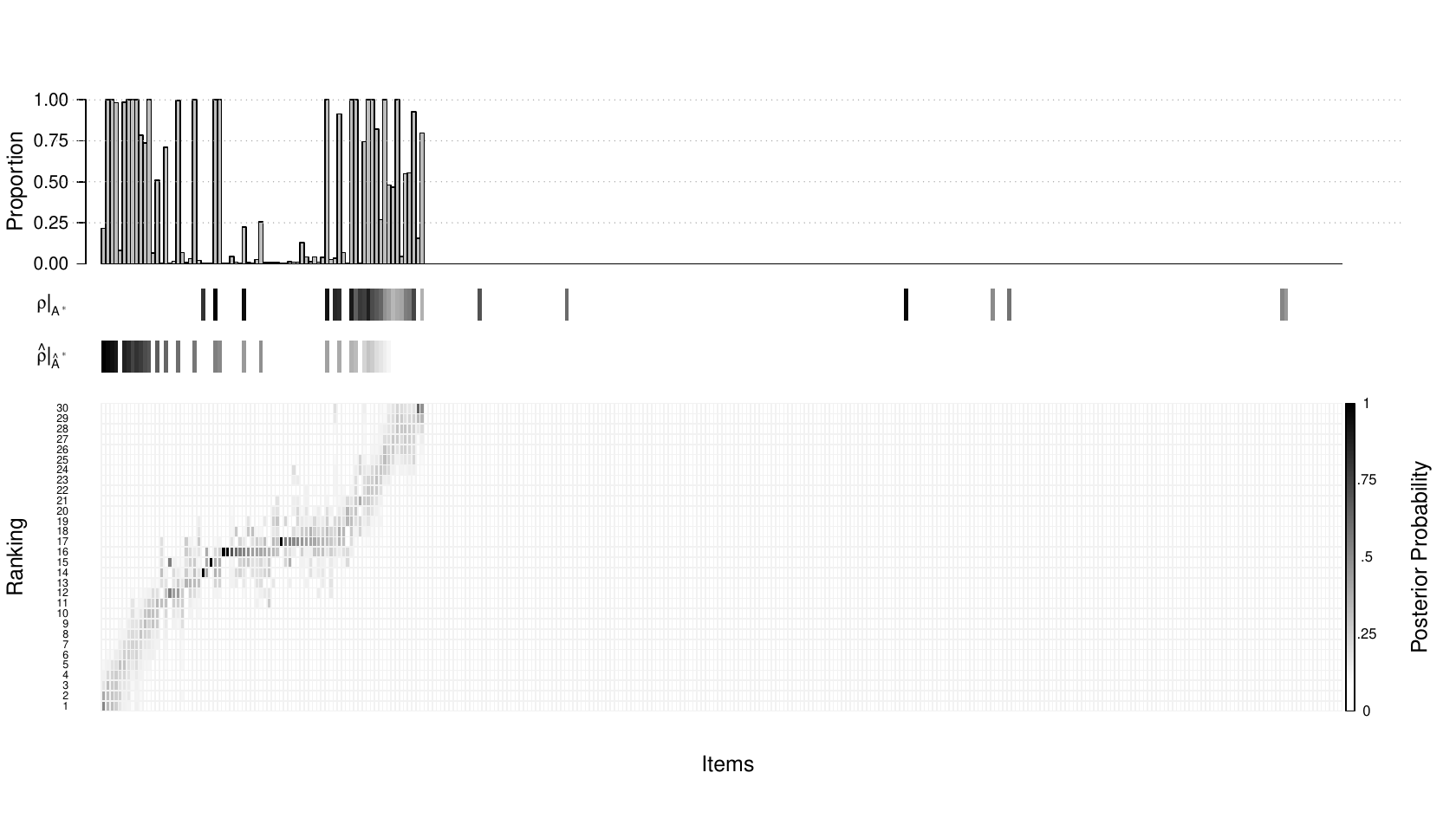}
  \end{subfigure}

  \vfill

  \begin{subfigure}{\textwidth}
    \caption{Results from Setup 4 (consistency), cluster 5.}
    \label{afig:selection-consistency-big-cluster5}
    \includegraphics[width=\textwidth]{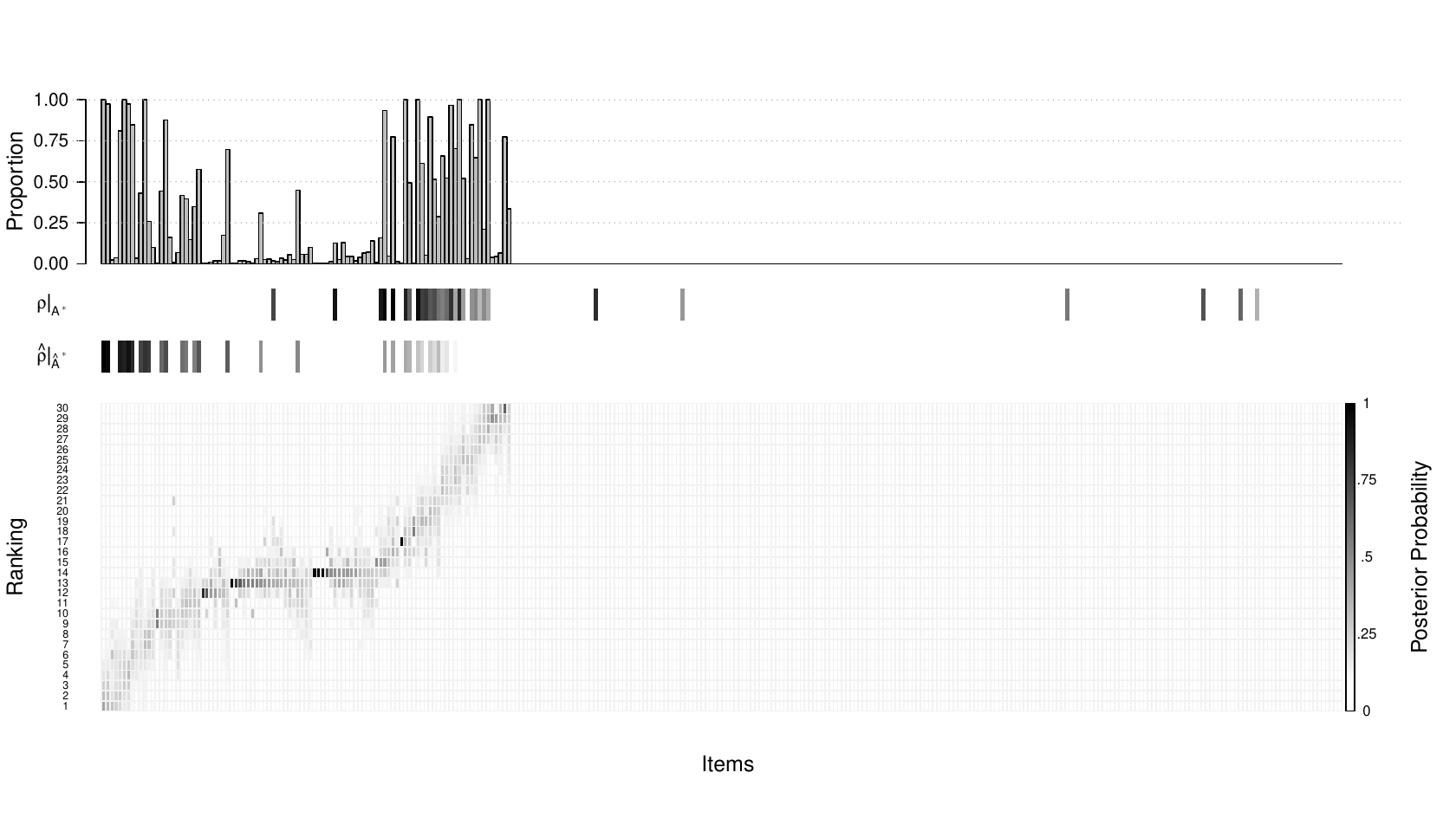}
  \end{subfigure}
\end{figure}

\begin{figure}[!t]
  \centering
  \caption{Results for \emph{clusters 1 to 5} (panels \textit{a} to \textit{e}) from the mixed simulation design (Setup 5).
Each panel shows a composite graph with items listed on the x-axes; from top to bottom: bar-plot of the posterior probability with which an item is selected in the estimated set $\hat{\A}_c^{*}$; in the subsequent row, $\rho|_{A^{*}}$, true ranks of the items in $\A_c^{*}$; in the $\hat{\rho}|_{\hat{A}^{*}}$ row, estimated ranks of the items in $\hat{\A}_c^{*}$; in the heatmap, marginal posterior probability  of $\vrho_c$ (shades of gray) for each rank (y-axis) to be assigned to each item.
  On the x-axis, items are ordered by the latter. The selection of items in $\hat{\mathcal{A}}_c^*$ uses a threshold $p=10\%$ to determine the High Probability Set, $\mathcal{A}'_c$, (see Definition~\ref{def:hps1}).
  }
  \label{afig:selection-heatmap-mixdata-big}
  \begin{subfigure}{\textwidth}
    \caption{Results from Setup 5 (mixed), cluster 1.}
    \label{afig:selection-mixdata-big-cluster1}
    \includegraphics[width=\textwidth]{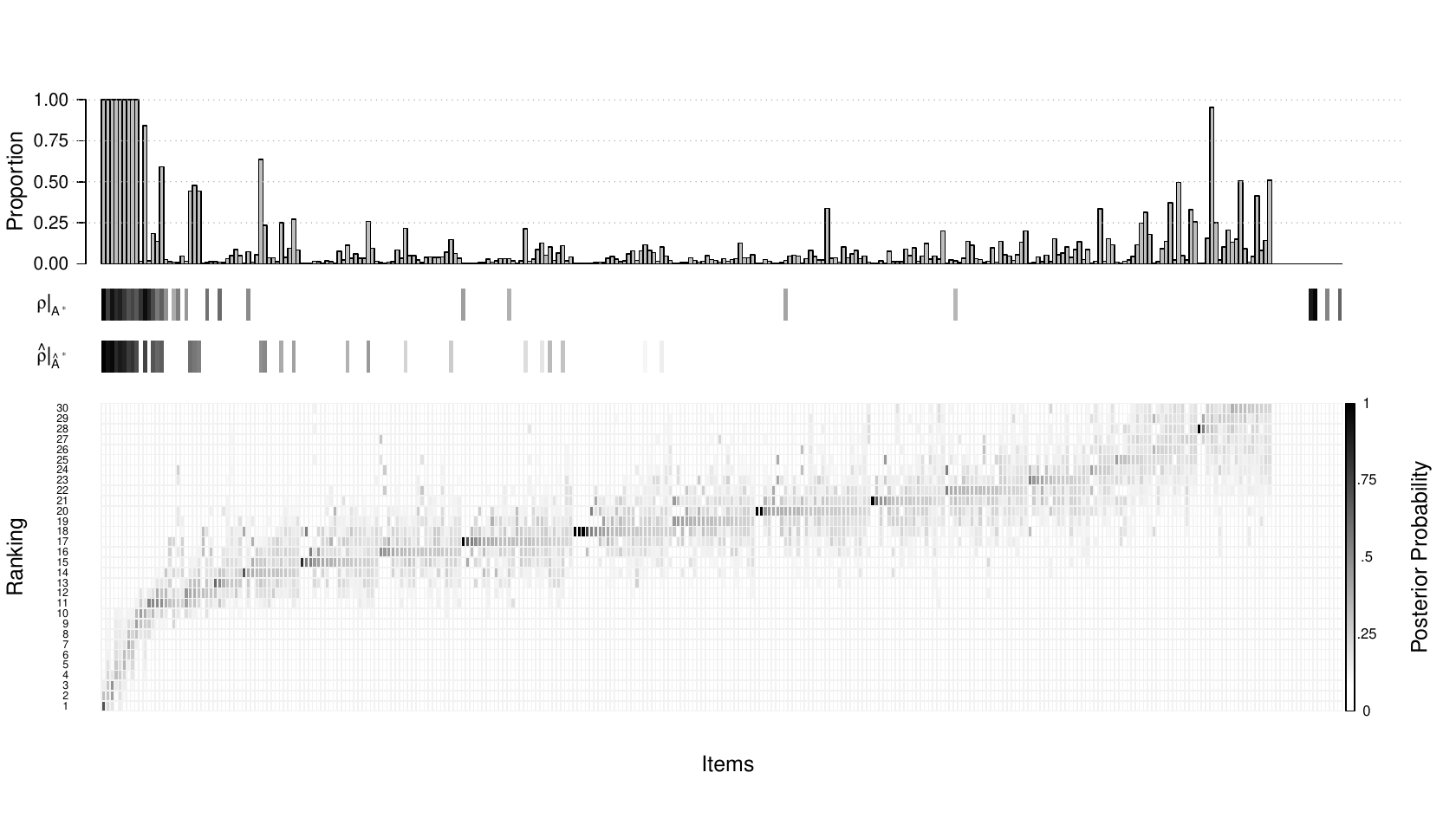}
  \end{subfigure}
\end{figure}

\begin{figure}[!t]
  \centering
  \ContinuedFloat

  \begin{subfigure}{\textwidth}
    \caption{Results from Setup 5 (mixed), cluster 2.}
    \label{afig:selection-mixdata-big-cluster2}
    \includegraphics[width=\textwidth]{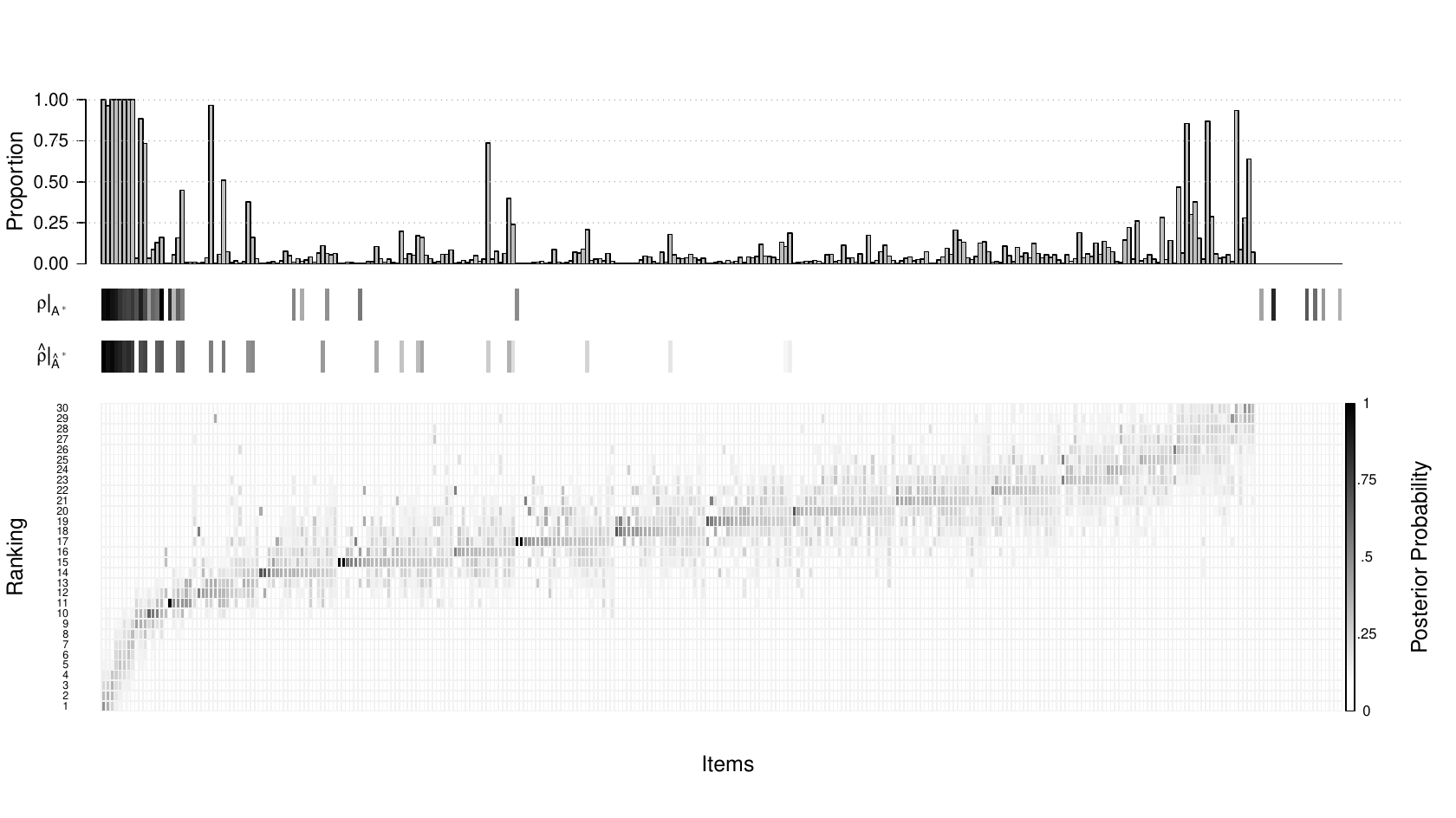}
  \end{subfigure}

  \vfill

  \begin{subfigure}{\textwidth}
    \caption{Results from Setup 5 (mixed), cluster 3.}
    \label{afig:selection-mixdata-big-cluster3}
    \includegraphics[width=\textwidth]{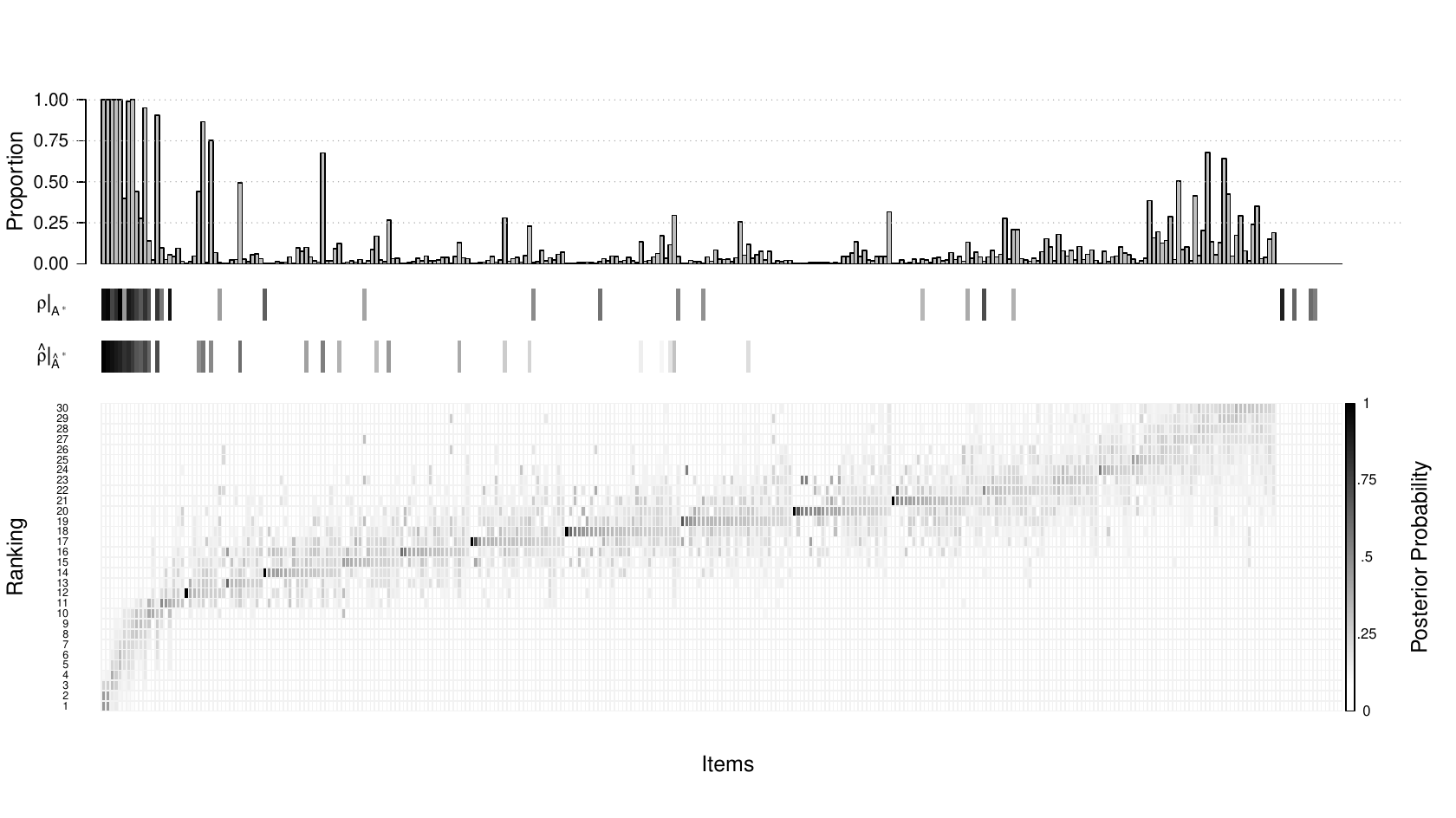}
  \end{subfigure}
\end{figure}

\begin{figure}[!t]
  \centering
  \ContinuedFloat
  \begin{subfigure}{\textwidth}
    \caption{Results from Setup 5 (mixed), cluster 4.}
    \label{afig:selection-mixdata-big-cluster4}
    \includegraphics[width=\textwidth]{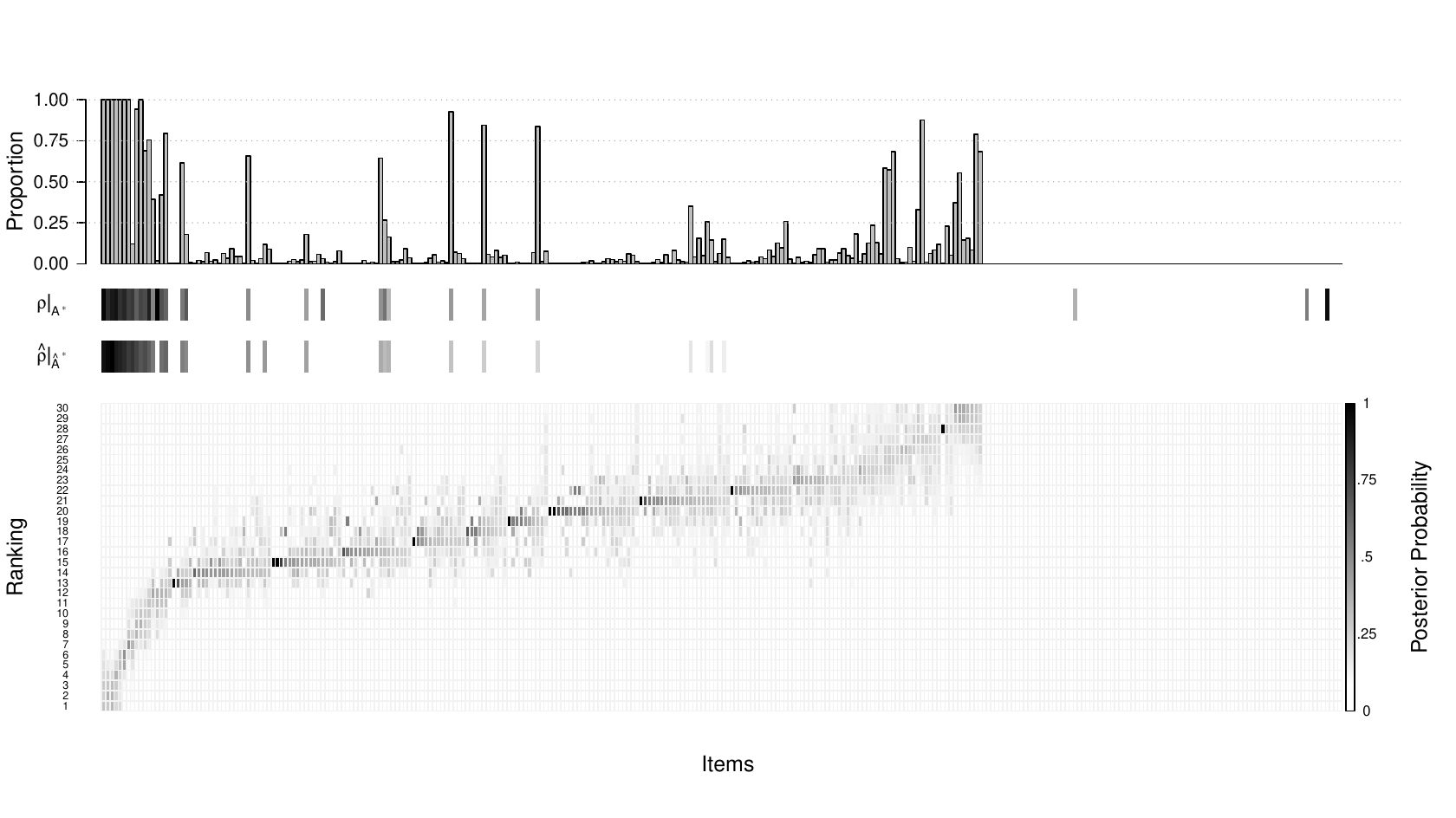}
  \end{subfigure}

  \vfill

  \begin{subfigure}{\textwidth}
    \caption{Results from Setup 5 (mixed), cluster 5.}
    \label{afig:selection-mixdata-big-cluster5}
    \includegraphics[width=\textwidth]{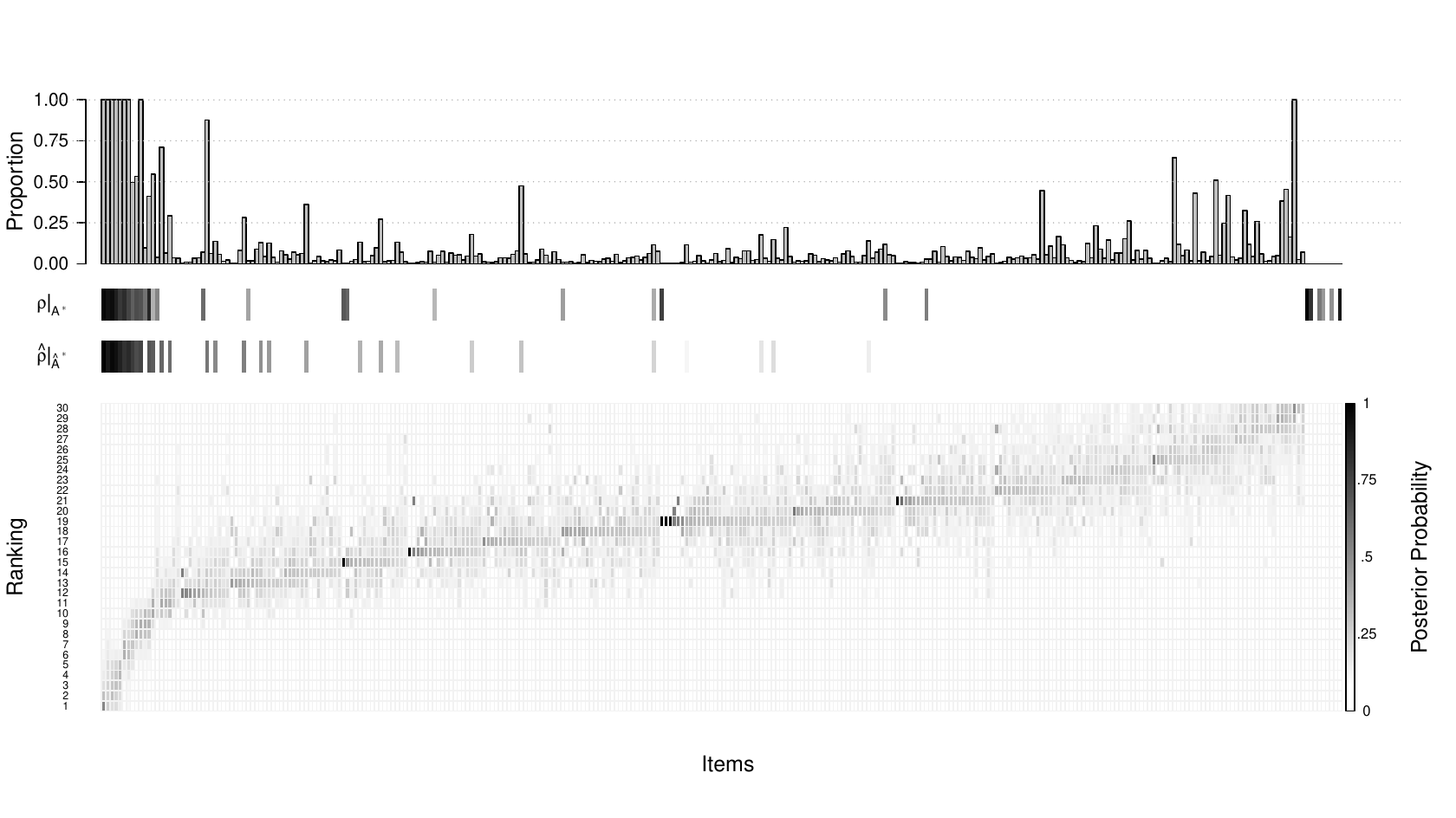}
  \end{subfigure}
\end{figure}

\clearpage

\section{Additional results on BRCA data}
\begin{figure}[!th]
  \centering
  \includegraphics[width=\textwidth]{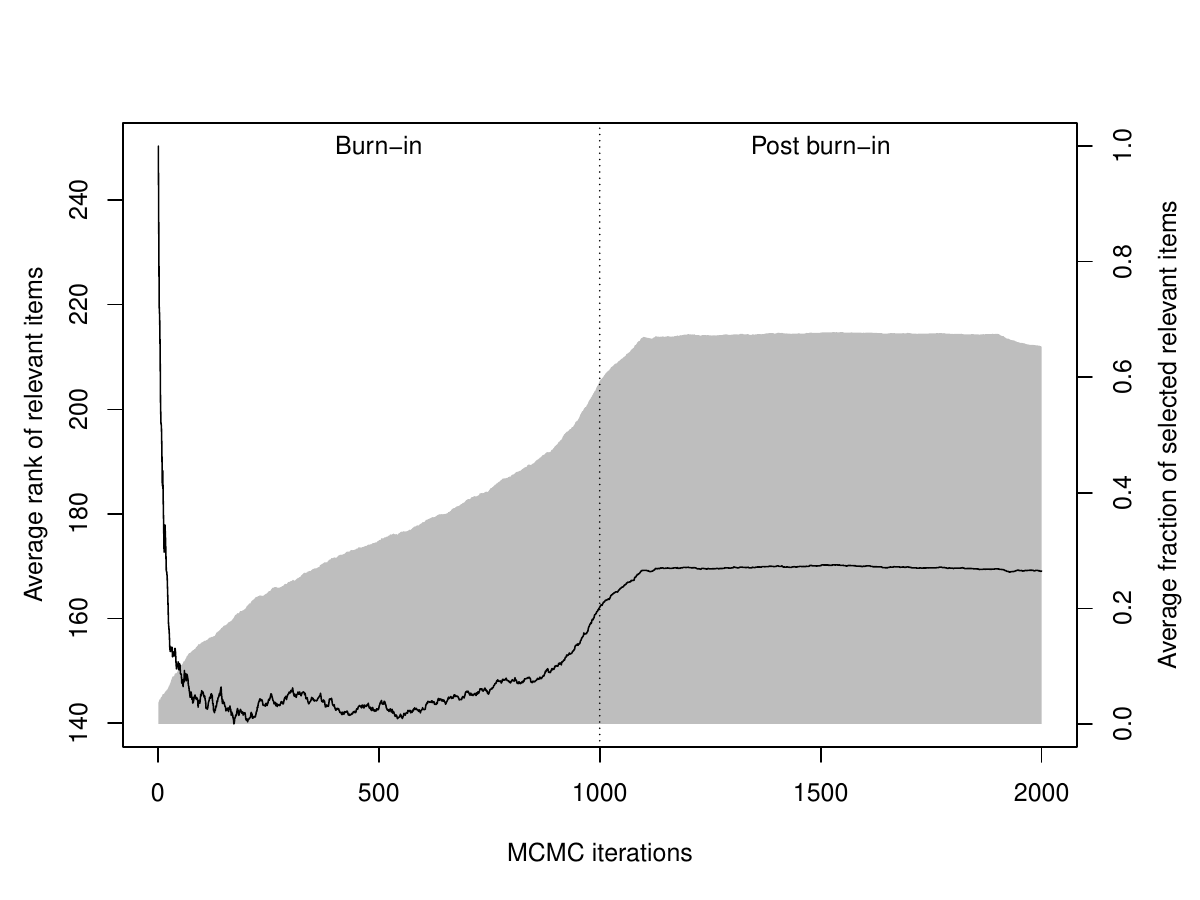}
  \caption{MCMC trace plots for average rank and average inclusion fracion of top-genes. The x-axis plots Monte Carlo iterations for both the burn-in period and after burn-in (separated by the dotted vertical line), thinned by considering only every other 100th observation. For a given MCMC iteration, the left y-axis (solid line) plots the average rank assigned to genes in the estimated top-genes sets, $\hat{\A}^*_c$, for $c\in\{1,2, \ldots, 10\}$, where the average is taken with respect to all genes and all clusters. Writing $\bar{A}_{m,c} = \{A_i \in \A_{m,c}^* \cap \hat{\A}^*_c\}$, this is formally $(\sum_{c=1}^{10}\sum_{A_i \in \bar{A}_{m,c}} \rho_i^{c,m}) / (\sum_{c=1}^{10} | \bar{A}_{m,c} |).$
    The right y-axis (shaded area) plots the average fraction of genes in the estimated top-genes sets, $\hat{\A}^*_c$, for $c\in\{1,2, \ldots, 10\}$, selected at each MCMC iteration, averaging across cluster, that is $(\sum_{c=1}^{10} | \bar{A}_{m,c} |) / (\sum_{c=1}^{10} \hat{\A}^*_c) $.}
  \label{afig:rhoAstarTrace-BRCA}
\end{figure}

\begin{figure}[!th]
  \centering
  \begin{subfigure}{\textwidth}
    \caption{Alluvial plot for all clustering solutions}
    \includegraphics[width=\textwidth]{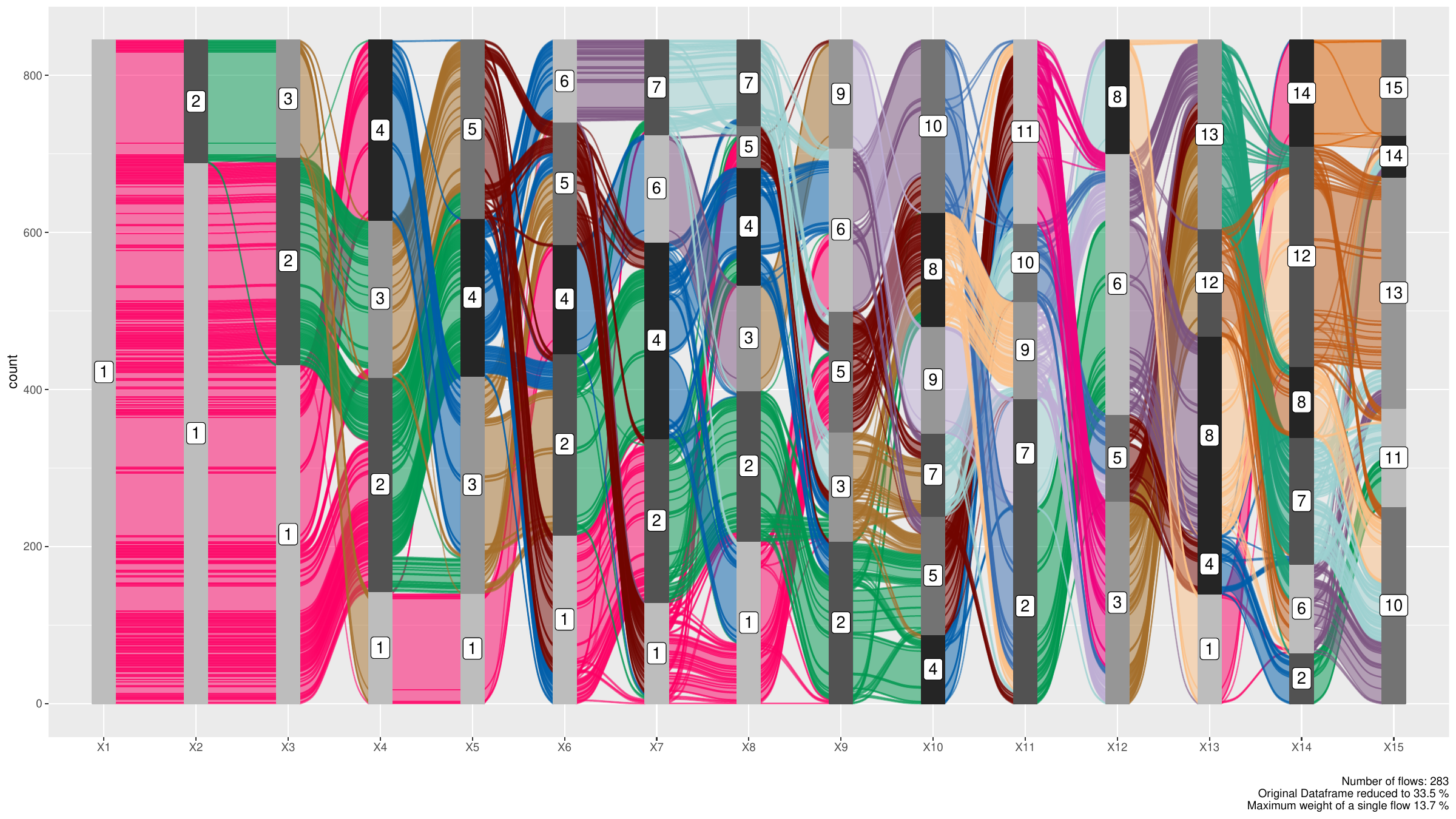}
  \end{subfigure}

  \begin{subfigure}{\textwidth}
    \caption{Alluvial plot for solutions $C=10$, $C=14$, and $C=8$.}
    \includegraphics[width=\textwidth]{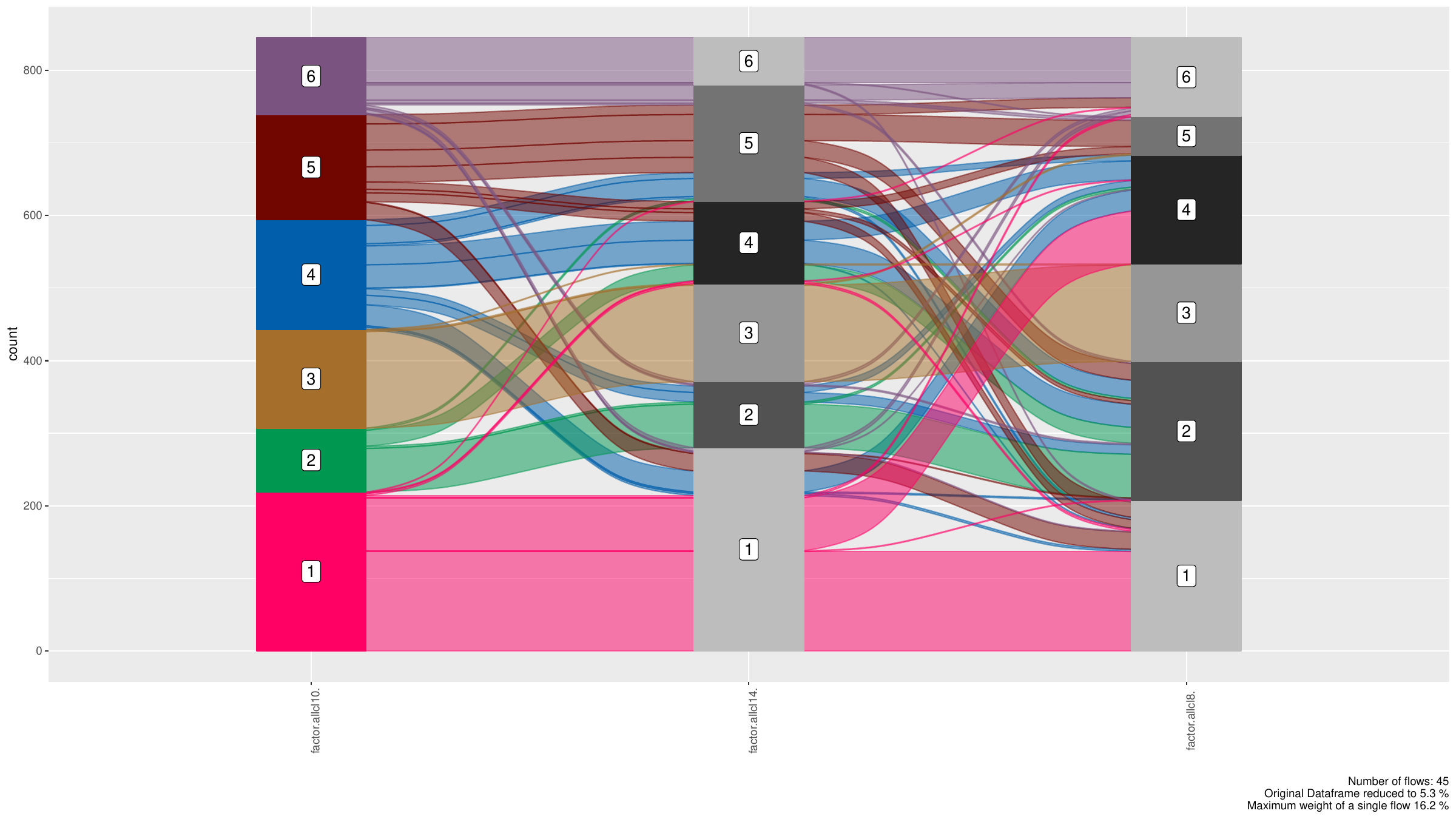}
  \end{subfigure}
  \caption{Alluvial plot for clustering solutions. The plot shows the flows of patients (y-axis) across clusters, for the lowBM3 estimated clustering solutions (x-axis). The top graph shows transitions for all the clustering solutions ($C\in\{1, 2, \ldots 15\}$), while the bottom graph focuses on the solutions with 6 non-void clusters, starting from the selected solution with $C=10$.}
  \label{afig:alluvial}
\end{figure}






\clearpage


\begin{figure}
    \centering
    \includegraphics[width=\textwidth]{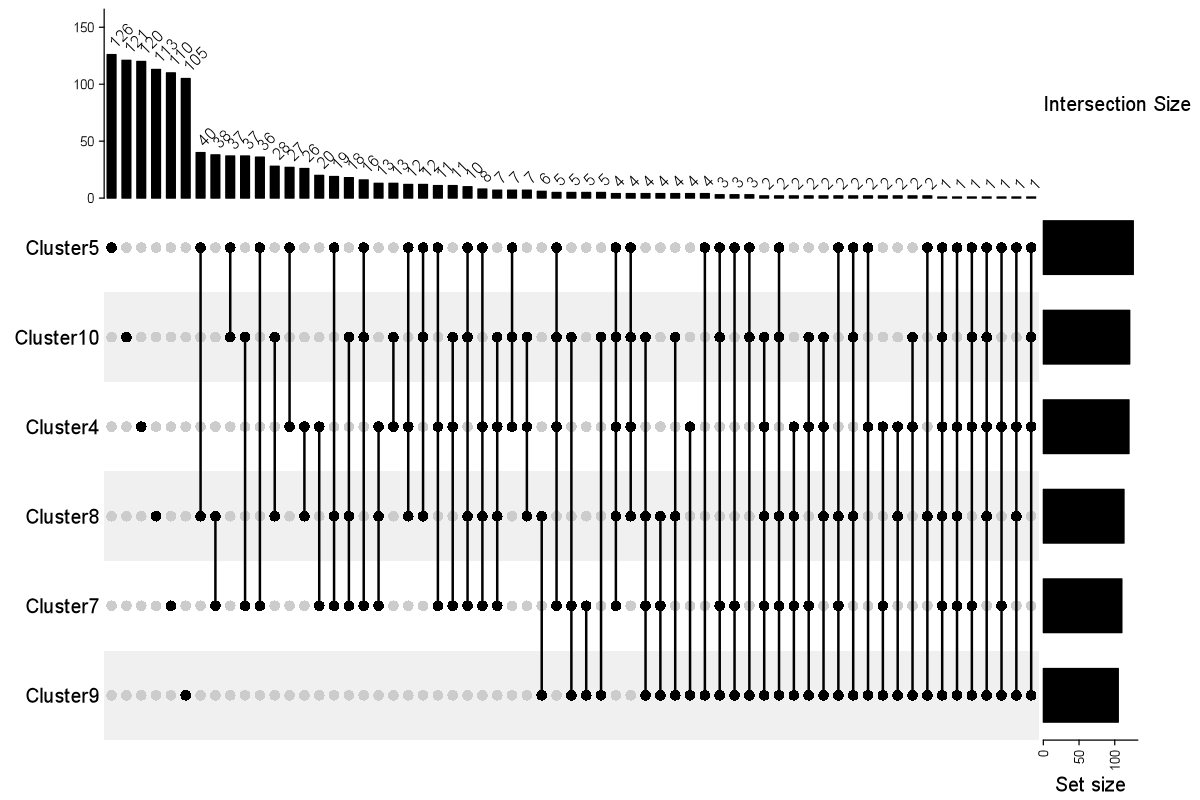}
    \caption{Inclusive Upset Plot of genes with 80\% probability of being ranked top 100 for the 6 non-empty clusters from lowBM3 with $C = 10$. Each intersection bar counts every gene present in the specified combination, regardless of whether it also appears in other clusters or intersections.}
 \label{fig:genes_upset_inclusive}
\end{figure}

\end{document}